\documentclass[12pt]{article}
\pdfoutput=1

\usepackage{graphicx}
\usepackage{dsfont}
\usepackage{amssymb}
\usepackage{amsmath}
\usepackage{mathrsfs}
\usepackage{bm}

\numberwithin{equation}{section}

\begin{document}

\title{Transformation Optics and the Geometry of Light}
\author{Ulf Leonhardt$^1$ and Thomas G. Philbin$^2$\\
$^1$School of Physics and Astronomy, University of St Andrews,\\
North Haugh, St Andrews KY16 9SS, UK\\
$^2$Max Planck Research Group 
Optics, Information and Photonics,\\
G\"{u}nther-Scharowsky Str.\ 1, 91058 Erlangen, Germany
}
\date{\today}
\maketitle

\begin{center}
To appear in Progress in Optics (edited by Emil Wolf).
\end{center}

\newpage

\section{Introduction}

Metamaterials are beginning to transform optics and microwave technology
thanks to their versatile properties that, in many cases, can be tailored 
according to practical needs and desires
\cite{Caloz,Eleftheriades,Engheta,Halpern,Krowne,Marques,Milton,Sarychev}.
Although metamaterials are surely not the answer to all engineering problems,
they have inspired a series of significant technological developments 
and also some imaginative research,
because they invite researchers and inventors to dream. 
Imagine there were no practical limits 
on the electromagnetic properties of materials.
What is possible? And what is not? If there are no practical limits, what are the fundamental limits?
Such questions inspire taking a fresh look at the foundations 
of optics \cite{BornWolf} and at connections between optics
and other areas of physics. 
In this article we discuss such a connection, 
the relationship between optics and general relativity, or, expressed more precisely, between geometrical ideas 
normally applied in general relativity 
and the propagation of light, or electromagnetic waves in general,
in materials \cite{GREE}.
Farfetched as it may appear, general relativity turns out \cite{GREE} to have been put to practical use in the first working prototype
of an electromagnetic cloaking device \cite{Schurig},
it gives perhaps the most elegant approach to achieving invisibility
\cite{LeoConform,PSS},
and \cite{GREE}, 
general relativity even works behind the scenes of 
perfect lenses \cite{Pendry,Veselago1}.

The practical use of 
general relativity in electrical and optical engineering
may seem surprisingly unorthodox:
traditionally, relativity has been associated with 
the physics of gravitation \cite{Telephone}
and cosmology \cite{Peacock}
or, in engineering \cite{VanBladel}, 
has been considered a complication,
not a simplification.
For example, the Global Positioning System
would not be as accurate as it is without taking 
relativistic corrections into account 
that are due to gravity and the motion of the 
navigation satellites. 
However, here we are not concerned with the influence of the 
natural geometry of space and time on optics, 
the space-time curvature due to gravity,
but rather we show how optical materials create 
artificial geometries for light and how such geometries 
can be exploited in designing novel optical devices. 

Connections between geometry and optics are nothing new;
the ideas we explain here are rather, to borrow a phrase of
Sir Michael Berry, ``new things in old things''.
These ideas are based on Fermat's principle \cite{BornWolf}
formulated in 1662 by Pierre de Fermat,
but anticipated nearly a millennium ago by the 
Arab scientist Ibn al-Haytham and inspired by the 
Greek polymath Hero of Alexandria's reflections on light
almost two millennia ago.
According to Fermat's principle,
light rays follow extremal optical paths in materials
(shortest or longest, mostly shortest)
where the length measure is given by the refractive index.
Media change the measure of length.
This means that any optical medium establishes a geometry
\cite{Bortolotti,Rytov,SchleichScully}:
the glass in a lens, 
the water in a river 
or the air creating a mirage in the desert.  
General relativity has cultivated the 
theoretical tools for fields in curved geometries
\cite{LL2,Telephone}.
In this article we show how to use these tools for applications in 
electromagnetic or optical metamaterials.

\begin{figure}[t]
\begin{center}
\includegraphics[width=23.0pc]{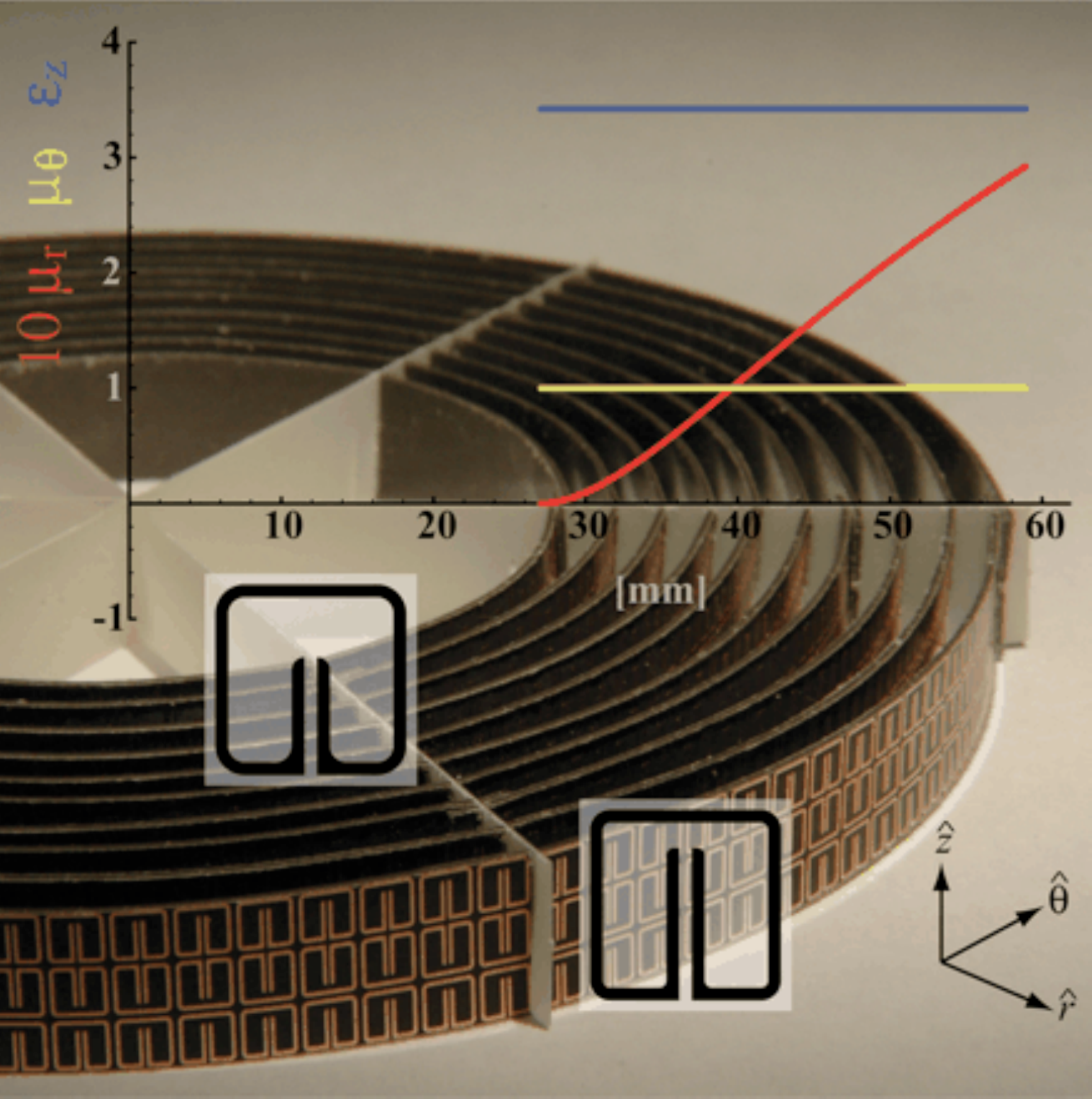}
\caption{
\small{
Cloaking device. 
(From Ref.\ \cite{Schurig}. Reprinted with permission from AAAS.)
Two-dimensional microwave cloaking structure (background image) with a plot of the material parameters that are implemented. The cloaking device is made of circuit-board with structures that are about an order of magnitude smaller than the wavelength. The structures are split-ring resonators with tunable magnetic response. The split-ring resonators of the inner and outer rings are shown in expanded schematic form (transparent square insets).
}
\label{fig:cloak}}
\end{center}
\end{figure}

\begin{figure}[t]
\begin{center}
\includegraphics[width=23.0pc]{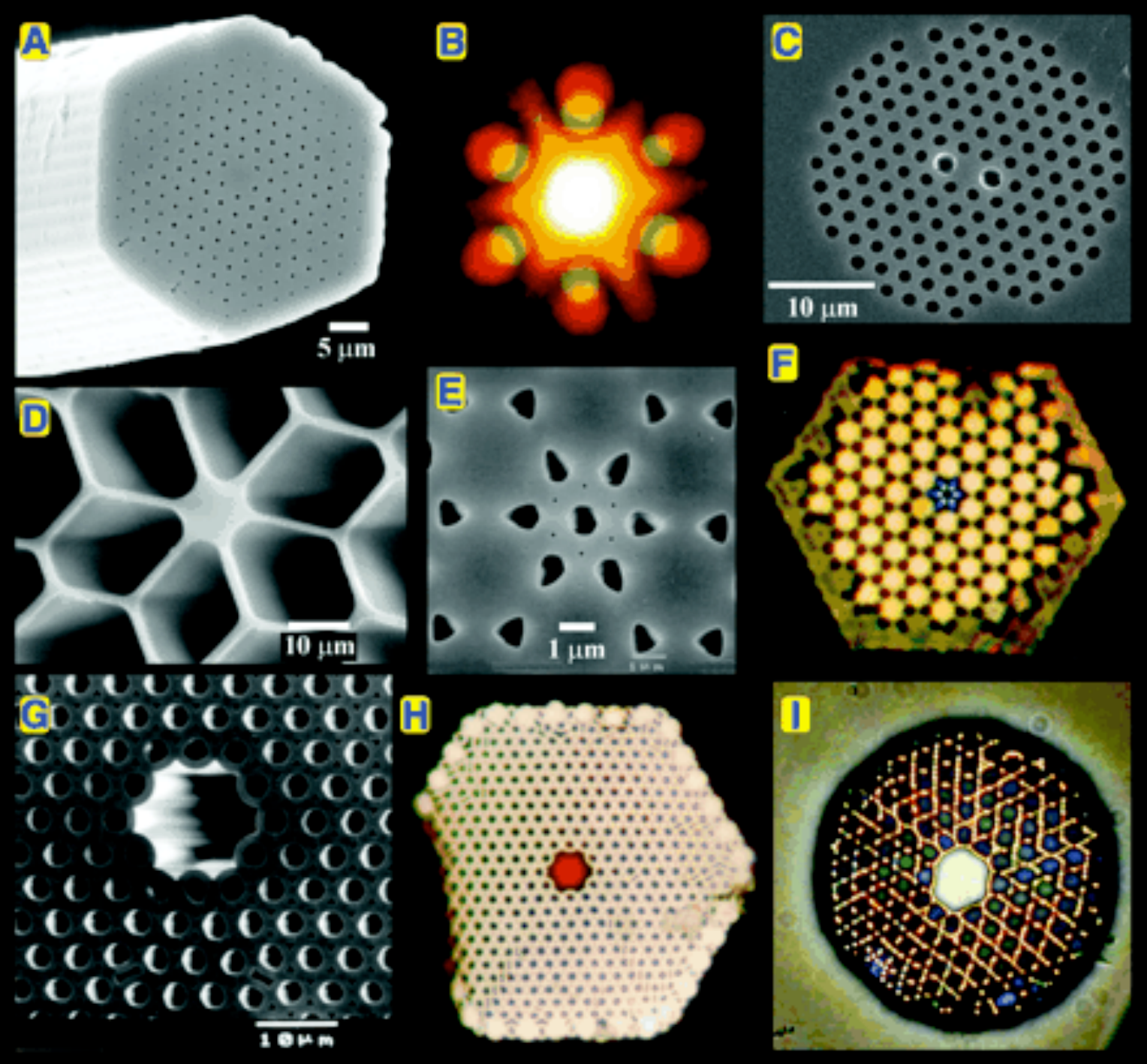}
\caption{
\small{
Photonic-crystal fibres.
(From Ref.\ \cite{Russell}. Reprinted with permission from AAAS.)
An assortment of optical (OM) and scanning electron (SEM) micrographs of photonic-crystal fibre (PCF) structures. (A) SEM of an endlessly single-mode solid core PCF. (B) Far-field optical pattern produced by (A) when excited by red and green laser light. (C) SEM of a recent birefringent PCF. (D) SEM of a small (800 nm) core PCF with ultrahigh nonlinearity and a zero chromatic dispersion at $560$-nm wavelength. (E) SEM of the first photonic band gap PCF, its core formed by an additional air hole in a graphite lattice of air holes. (F) Near-field OM of the six-leaved blue mode that appears when (E) is excited by white light. (G) SEM of a hollow-core photonic band gap fiber. (H) Near-field OM of a red mode in hollow-core PCF (white light is launched into the core). (I) OM of a hollow-core PCF with a Kagom\'e cladding lattice, guiding white light.
}
\label{fig:pcf}}
\end{center}
\end{figure}

Metamaterials are materials with electromagnetic properties
that originate from man-made sub-wavelength structures
\cite{Marques,Milton,Sarychev}.
Perhaps the best known metamaterials are 
the materials used in the pioneering demonstrations 
of negative refraction \cite{Shelby} 
or invisibility cloaking \cite{Schurig} of microwaves, 
see Fig.\ \ref{fig:cloak},
or for negative refraction of near-visible light \cite{SLW}.
These materials consist of metallic cells that are smaller 
than the relevant electromagnetic wavelength.
Each cell acts like an artificial atom that can be tuned
by changing the shape and the dimensions
of the metallic structure.
It is probably fair to regard microstructured 
or photonic-crystal fibres \cite{Russell} 
as metamaterials as well, see Fig.\ \ref{fig:pcf}.
Here sub-wavelength structures --- airholes along the fibre ---
significantly influence the optical properties of the fused silica
the fibres are made of. 
Metamaterials have a long history:
the ancient Romans invented ruby glass, 
which is a metamaterial,
although the Romans presumably did not know this concept. 
Ruby glass \cite{Wagner}
contains nano-scale gold colloids that render
the glass neither golden nor transparent, but ruby,
depending on the size and concentration of the gold droplets.
The color originates from a resonance of the surface plasmons
\cite{Barnes} on the metallic droplets.
Metamaterials {\it per se} are nothing new:
what is new is the degree of control over the structures
in the material that generate the desired properties. 

The specific starting point of our theory is not new either.
In the early 1920's
Gordon \cite{Gordon} noticed that moving isotropic media
appear to electromagnetic fields
as certain effective space-time geometries.
Bortolotti \cite{Bortolotti} and Rytov \cite{Rytov}
pointed out that ordinary isotropic media establish
non-Euclidean geometries for light.
Tamm \cite{Tamm1,Tamm2} 
generalized the geometric approach
to anisotropic media
and briefly applied this theory \cite{Tamm2}
to the propagation of light in curved geometries.
In 1960 Plebanski \cite{Plebanski}
formulated the electromagnetic effect 
of curved space-time or curved coordinates 
in concise constitutive equations.
Electromagnetic fields perceive media as geometries
and geometries act as effective media.
Furthermore, in 2000 it was understood \cite{LeoGeometry}
that media perceive electromagnetic fields as geometries as well.
Light acts on dielectric media via dipole forces
(forces that have been applied in optical trapping and tweezing
\cite{Dholakia,Neuman}).
These forces turn out to appear like the inertial forces
in a specific space-time geometry. 
This geometric approach \cite{LeoMomentum} 
was used to shed light on the 
Abraham-Minkowski controversy
about the electromagnetic momentum in media
\cite{Abraham,LeoNature,Minkowski,Peierls}.
Geometrical ideas have been applied to construct
conductivities that are undetectable by static electric fields 
\cite{Greenleaf1,Greenleaf2},
which was the precursor of invisibility devices
\cite{Alu,Gbur,LeoConform,LeoNotes,MN,PSS,SPS}
based on implementations of coordinate transformations.
From these recent developments grew the subject
of transformation optics. 
Here media, possibly made of metamaterials,
are designed such that they appear to perform 
a coordinate transformation from physical space to some
virtual electromagnetic space.
As we describe in this article,
the concept of transformation optics embraces 
some of the spectacular recent applications
of metamaterials.

Transformation optics is beginning to transform optics. 
We would do injustice to this emerging field 
if we attempted to record every recent result. 
By the time this article goes to press, it would be outdated already.
Instead we focus on the ``old things in new things'',
because those are the ones that are guaranteed to last 
and to remain inspiring for a long time to come.
This article rather is a primer, not a typical literature review.
We try to give an introduction into
connections between geometry and electromagnetism in media
that is as consistent and elementary as possible,
without assuming much prior knowledge.
We begin in \S 2 with a brief section on Fermat's principle 
and the concept of transformation optics.
In \S 3  we develop in detail 
the mathematical machinery of geometry. Although this is textbook material, many readers will appreciate a (hopefully) readable introduction. We do not assume any prior knowledge of differential geometry;
readers familiar with this subject may skim through most of \S 3.
After having honed the mathematical tools, we 
apply them to Maxwell's electromagnetism in \S 4
where we develop the concept of transformation optics. 
In \S 5 we discuss some examples of transformation media:
perfect invisibility devices, perfect lenses, 
the Aharonov-Bohm effect in moving media 
and analogues of the event horizon.
Let's begin at the beginning, Fermat's principle.

\section{Fermat's principle}

\begin{figure}[h]
\begin{center}
\includegraphics[width=25.0pc]{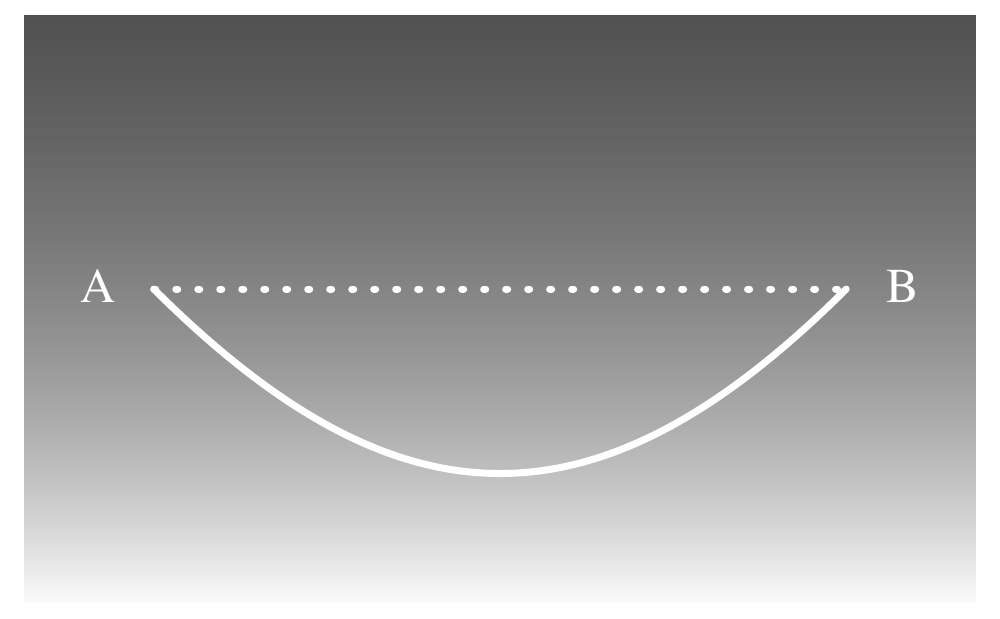}
\caption{
\small{
Fermat's principle.
Light takes the shortest optical path from A to B (solid line) which is not
a straight line (dotted line) in general. 
The optical path length is measured in terms of the refractive index $n$
integrated along the trajectory. 
The greylevel of the background indicates the refractive index.
The figure illustrates the ray trajectories involved in forming a mirage.
}
\label{fig:fermat}}
\end{center}
\end{figure}

\begin{figure}[h]
\begin{center}
\includegraphics[width=25.0pc]{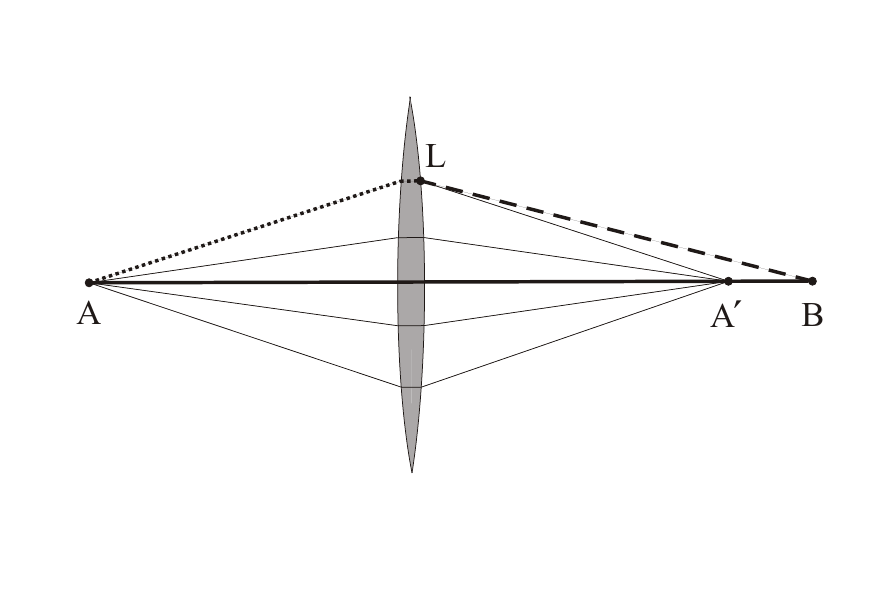}
\caption{
\small{
Fermat's principle: longest optical path. The figure shows a simple example where light takes the longest optical path: light beyond the focal point of a lens, traveling along a straight line from $\mathrm{A}$ to $\mathrm{B}$. To see why the optical path from $\mathrm{A}$ to $\mathrm{B}$ is the longest, compare the solid line of the actual path with an example of a virtual path, with the dashed and dotted paths, and note that the optical path length between the focal points $\mathrm{A}$ and $\mathrm{A}'$ of the lens is always the same, regardless whether light travels along a straight line or is refracted in the lens. Therefore the optical path length taken differs from the virtual path length by the difference between the two short sides and the long side of the triangle from $\mathrm{L}$ to $\mathrm{A}'$ and $\mathrm{B}$. The sum of the two short sides of a triangle is always longer than the long side: light has taken the longest optical path.  
}
\label{fig:longestpath}}
\end{center}
\end{figure}

In a letter dated January 1st, 1662, Pierre de Fermat formulated a physical principle that was destined to shape geometrical optics, to give rise to Lagrangian and Hamiltonian dynamics and to inspire Schr\"odinger's quantum mechanics and Feynman's form of quantum field theory and statistical mechanics.
Fermat's principle is the principle of the shortest optical path: light rays passing between two spatial points A and B chose the optically shortest path, see Fig.\ \ref{fig:fermat}.
In some cases, however, light takes the longest path; in any case, light rays follow extremal optical paths, see Fig.\ \ref{fig:longestpath}.
The optical path length $s$ is defined in terms of the refractive index $n$ as
\begin{equation}
\label{eq:fermat}
s = \int n\,\mathrm{d}l = 
\int_\mathrm{A}^\mathrm{B} n \,
\sqrt{\mathrm{d}x^2+\mathrm{d}y^2+\mathrm{d}z^2}
\end{equation}
in Cartesian coordinates.
If the refractive index varies in space --- for non-uniform media --- the shortest optical path is not a straight line, but is curved. This bending of light is the cause of many optical illusions. For example, picture a mirage in the desert \cite{Feynman}. The tremulous air above the hot sand conjures up images of water in the distance, but it would be foolish to follow these deceptions; they are not water, but images of the sky. The hot air above the sand bends light from the sky, because hot air is thin with low refractive index and so light prefers to propagate there.

Fermat's principle has profoundly influenced modern physics, and like most if not all profound discoveries it has deep roots in the history of science. Fermat was inspired  by the Greek polymath Hero of Alexandria's theory of light reflection in mirrors. The Arab scientist Ibn al-Haytham anticipated Fermat's principle in his {\it Book of Optics} (written under house arrest in Cairo from 1011 to 1021). Fermat's principle was instantly greeted with objections, because it appears to violate causality --- it presumes an idea of destiny. The principle governs the path between A and B if it is known that light travels from A to B; Fermat's principle shows how the ray's destiny is fulfilled, but it does not explain why the light ray arrives at B and not at some other end point.
Wave optics resolves this problem, because a wave emitted at A propagates in all directions (but possibly with greatly varying amplitude). The path of extremal optical length (\ref{eq:fermat}) is the place of constructive interference between all possible paths.

\begin{figure}[h]
\begin{center}
\includegraphics[width=30.0pc]{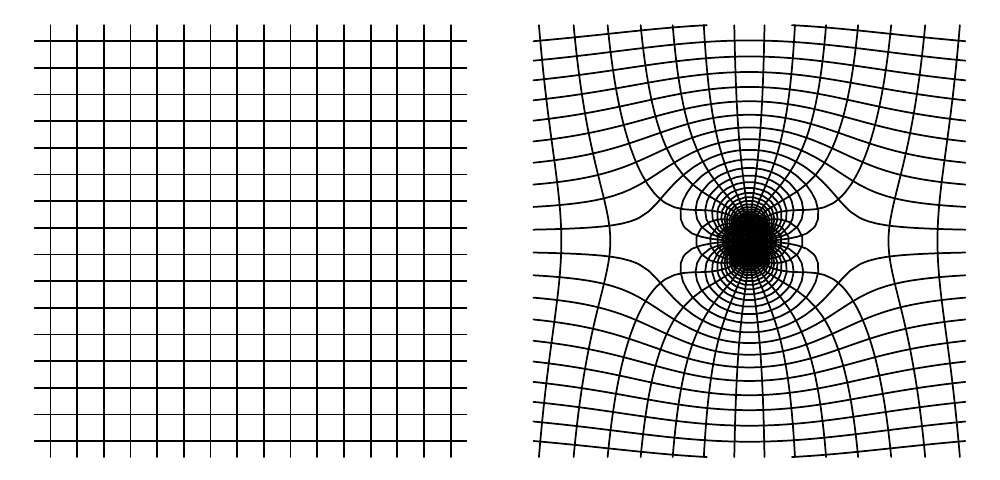}
\caption{
\small{
Optical conformal mapping \cite{LeoConform}.
Suppose that an optical medium performs a coordinate transformation
from a straight Cartesian grid (left) to curved coordinates in 
physical space (right).
The trajectories of light rays follow the curved coordinates, 
but this apparent curvature is an illusion and may be used to 
create optical illusions, for example invisibility.
}
\label{fig:grid}}
\end{center}
\end{figure}

We will derive Fermat's principle later, in Sec.\ 4.3. The question we pose here is this: imagine one illuminates a non-uniform medium with a grid of light rays, see Fig.\ \ref{fig:grid}. Each ray is curved according to Fermat's principle. Is it possible to transform away the curvature of the grid? 
In this case the curved path of each ray would appear as a straight line in some transformed space. So, in other words, is it possible that the bending of light is an illusion of choosing the wrong coordinates? Curvature would be an illusion of Cartesian linear thinking. Such transformable media may create optical illusions by themselves, in fact they may create the ultimate illusion, invisibility: if the transformable grid contains a hole, anything inside the hole is invisible. The transformation medium acts as a cloaking device.

For simplicity, imagine a two-dimensional situation where the refractive index varies in $x$ and $y$, and light is confined to the $x,y$ plane. Consider a coordinate transformation from $x$ and $y$ to $x'$ and $y'$. Is Fermat's principle obeyed in transformed coordinates,
\begin{equation}
s =  \int n \, \sqrt{\mathrm{d}x^2+\mathrm{d}y^2} = 
\int n' \, \sqrt{\mathrm{d}x'^2+\mathrm{d}y'^2}
\quad ?
\end{equation}
The reader easily sees from 
\begin{equation}
\label{eq:difftrans}
\mathrm{d}x' = \frac{\partial x'}{\partial x} \, \mathrm{d}x 
+ \frac{\partial x'}{\partial y} \, \mathrm{d}y \, ,  \quad
\mathrm{d}y' = \frac{\partial y'}{\partial x} \, \mathrm{d}x 
+ \frac{\partial y'}{\partial y} \, \mathrm{d}y 
\end{equation}
that $\mathrm{d}x'^2+\mathrm{d}y'^2$ is proportional to $\mathrm{d}x^2+\mathrm{d}y^2$ if 
\begin{equation}
\label{eq:cr}
\frac{\partial x'}{\partial x} = \frac{\partial y'}{\partial y} \, ,  \quad
\frac{\partial y'}{\partial x} = - \frac{\partial x'}{\partial y} \, .
\end{equation}
In this case the coordinate transformation changes $ n^2 \, (\mathrm{d}x^2+\mathrm{d}y^2$) into $n'^2 \, (\mathrm{d}x'^2+\mathrm{d}y'^2)$, thus preserving the form (\ref{eq:fermat}) of Fermat's principle. 
One obtains 
\begin{equation}
\label{eq:n}
n'^2 \left( \left( \frac{\partial x'}{\partial x} \right)^2 + \left( \frac{\partial x'}{\partial y} \right)^2 \right) = 
n'^2 \left( \left( \frac{\partial y'}{\partial x} \right)^2 + \left( \frac{\partial y'}{\partial y} \right)^2 \right) =  n^2 \, .
\end{equation}
A transformation with the property (\ref{eq:cr}) is known as a conformal transformation in two-dimensional space \cite{Nehari}. Conformal transformations leave Fermat's principle 
(\ref{eq:fermat}) intact, they correspond to materials with an isotropic refractive index profile. 
If $n'=1$, the transformed space is empty; light would propagate along a straight line there: the refractive-index profile acts as a transformation medium.

So far we have discussed light rays. How does the conformal transformation (\ref{eq:cr}) act on light waves? Suppose that both amplitudes $\psi$ of the optical polarization satisfy the Helmholtz equation
\begin{equation}
\label{eq:helm}
\left(  \bm{\nabla}^2 + \frac{\omega^2}{c^2} n^2 \right) \, \psi = 0
\end{equation}
where $\omega$ denotes the frequency and $c$ the speed of light in vacuum.
It is convenient to write the Laplacian $ \bm{\nabla}^2$ as
\begin{equation}
 \bm{\nabla}^2 = \frac{\partial^2 }{\partial x^2} + \frac{\partial^2 }{\partial y^2} =
\left( \frac{\partial}{\partial x} + \mathrm{i} \, \frac{\partial}{\partial y}\right)
\, \left( \frac{\partial}{\partial x} - \mathrm{i} \, \frac{\partial}{\partial y}\right)   .
\end{equation}
We obtain from the differential equations (\ref{eq:cr}) of the conformal map the transformation
\begin{eqnarray}
\left( \frac{\partial}{\partial x} - \mathrm{i} \, \frac{\partial}{\partial y}\right) & = & \left( \frac{\partial x'}{\partial x} \frac{\partial}{\partial x'} + \frac{\partial y'}{\partial x} \frac{\partial}{\partial y'} 
- \mathrm{i} \, \frac{\partial x'}{\partial y} \frac{\partial}{\partial x'}
- \mathrm{i} \, \frac{\partial y'}{\partial y} \frac{\partial}{\partial y'}  
\right) \nonumber \\
& = & \left( \frac{\partial x'}{\partial x} + \mathrm{i} \, \frac{\partial y'}{\partial x}\right) \, \left( \frac{\partial}{\partial x'} - \mathrm{i} \, \frac{\partial}{\partial y'}\right) \nonumber \\
& = & \left( \frac{\partial y'}{\partial y} - \mathrm{i} \, \frac{\partial x'}{\partial y}\right) \, \left( \frac{\partial}{\partial x'} - \mathrm{i} \, \frac{\partial}{\partial y'}\right) 
\end{eqnarray}
and, by exchanging $\mathrm{i}$ with $- \mathrm{i}$,
\begin{equation}
\left( \frac{\partial}{\partial x} + \mathrm{i} \, \frac{\partial}{\partial y}\right)  = \left( \frac{\partial x'}{\partial x} - \mathrm{i} \, \frac{\partial y'}{\partial x}\right) \, \left( \frac{\partial}{\partial x'} + \mathrm{i} \, \frac{\partial}{\partial y'}\right)
= \left( \frac{\partial y'}{\partial y} + \mathrm{i} \, \frac{\partial x'}{\partial y}\right) \, \left( \frac{\partial}{\partial x'} + \mathrm{i} \, \frac{\partial}{\partial y'}\right)   .
\end{equation}
Furthermore, since
\begin{equation}
\left( \frac{\partial}{\partial x} + \mathrm{i} \, \frac{\partial}{\partial y}\right)  \left( \frac{\partial x'}{\partial x} + \mathrm{i} \, \frac{\partial y'}{\partial x}\right) =  
\frac{\partial^2 x'}{\partial x^2} - \frac{\partial^2 y'}{\partial x \partial y} 
+ \mathrm{i} \, \frac{\partial^2 y'}{\partial x^2}
+ \mathrm{i} \, \frac{\partial^2 x'}{\partial x \partial y} = 0   ,
\end{equation}
the Laplacian $ \bm{\nabla}^2$ is transformed into 
\begin{eqnarray}
 \bm{\nabla}^2 & = & 
\left( \frac{\partial x'}{\partial x} + \mathrm{i} \, \frac{\partial y'}{\partial x}\right) 
\left( \frac{\partial x'}{\partial x} - \mathrm{i} \, \frac{\partial y'}{\partial y}\right)
\left( \frac{\partial}{\partial x'} + \mathrm{i} \, \frac{\partial}{\partial y'}\right)
\left( \frac{\partial}{\partial x'} - \mathrm{i} \, \frac{\partial}{\partial y'}\right) \nonumber \\
& = & \left( \left( \frac{\partial x'}{\partial x} \right)^2 + \left( \frac{\partial y'}{\partial x} \right)^2 \right) \, { \bm{\nabla}'}^2 \nonumber \\
& = & \left( \left( \frac{\partial x'}{\partial y} \right)^2 + \left( \frac{\partial y'}{\partial y} \right)^2 \right) \, { \bm{\nabla}'}^2
\, .
\end{eqnarray}
Consequently, the Helmholtz equation (\ref{eq:helm}) is invariant under conformal transformations if the refractive index is transformed according to Eq.\ (\ref{eq:n}).
Waves are transformed in precisely the same way as rays.

There is an elegant short-cut to the theory of this optical conformal mapping \cite{LeoConform} that allows us to condense the previous calculations in a few lines: complex analysis \cite{Ablowitz,Needham}.
Suppose we denote the two-dimensional coordinates by complex numbers
\begin{equation}
z = x + \mathrm{i} \, y , \quad   z^* = x - \mathrm{i} \, y   .
\end{equation}
From 
\begin{equation}
\frac{\partial}{\partial x} = 
\frac{\partial z}{\partial x} \frac{\partial}{\partial z} +
\frac{\partial z}{\partial x}\frac{\partial}{\partial z^*} =
\frac{\partial}{\partial z} + \frac{\partial}{\partial z^*} \, ,  \quad
\frac{\partial}{\partial y} = \mathrm{i} \,\frac{\partial}{\partial z} - \mathrm{i} \,\frac{\partial}{\partial z^*}
\end{equation}
we obtain
\begin{equation}
\frac{\partial}{\partial z} = \frac{1}{2} \, \left( \frac{\partial}{\partial x} -   \mathrm{i} \,\frac{\partial}{\partial y} \right) \, ,   \quad
\frac{\partial}{\partial z^*} = \frac{1}{2} \, \left( \frac{\partial}{\partial x} +  \mathrm{i} \,\frac{\partial}{\partial y} \right)
\end{equation}
and hence
\begin{equation}
 \bm{\nabla}^2 = 4 \, \frac{\partial^2}{\partial z \partial z^*}   .
\end{equation}
Consider a coordinate transformation described by a function $w (z)$ that depends on $z$, but not on $z^*$,
\begin{equation}
x' + \mathrm{i} \, y' = w (z) \quad \mathrm{with} \quad \frac{\partial w}{\partial z^*} = 0   .
\end{equation}
Since 
\begin{equation}
2 \, \frac{\partial w}{\partial z^*} = \left( \frac{\partial}{\partial x} +   \mathrm{i} \, \frac{\partial}{\partial y} \right) \, \left( x' + \mathrm{i} \, y' \right) = \frac{\partial x'}{\partial x} - \frac{\partial y'}{\partial y} +
\mathrm{i} \, \frac{\partial x'}{\partial y} + \mathrm{i} \, \frac{\partial y'}{\partial x}
\end{equation}
the differential equation (\ref{eq:cr}) of conformal maps are naturally satisfied; they are the Cauchy-Riemann differential equations of analytic functions \cite{Ablowitz,Needham,Nehari}.
Finally we obtain from
\begin{equation}
 \bm{\nabla}^2 = 4 \, \frac{\partial^2}{\partial z \partial z^*} = 
4 \, \frac{\mathrm{d}w}{\mathrm{d}z} \, \frac{\mathrm{d}w^*}{\mathrm{d}z^*} \,\frac{\partial^2}{\partial w \partial w^*}  
= \left| \frac{\mathrm{d}w}{\mathrm{d}z} \right|^2 \, { \bm{\nabla}'}^2
\end{equation}
the relationship (\ref{eq:n}) between the original and the transformed refractive-index profile in the Helmholtz equation (\ref{eq:helm}) as
\begin{equation}
n = \left| \frac{\mathrm{d}w}{\mathrm{d}z} \right| \, n'   .
\end{equation}
Complex analysis not only simplifies the theory, it provides optical conformal mapping \cite{LeoConform} with a vast resource in calculational tools and geometrical insights \cite{Ablowitz,Needham,Nehari}.

Conformal coordinate transformations represent a special case; most spatial transformations are non-conformal, and we could also envision transformations that mix space and time. Consequently, media that implement such transformations are not subject to Fermat's principle in the form (\ref{eq:fermat}). Furthermore, the Helmholtz equation (\ref{eq:helm}) is only approximately valid \cite{BornWolf}. Light should be described as an electromagnetic wave subject to Maxwell's equations. There are various ways of developing the concepts of transformation media for the general case. In \S 4 we discuss a theory that perhaps plays a similar role to that played by complex analysis in optical conformal mapping. It will equip the reader with calculational short-cuts and geometrical insights. For this theory we borrow concepts from general relativity, but we do not assume that the reader is familiar with them. The necessary ingredients from differential geometry are derived by elementary means in the following section.


\section{Arbitrary coordinates} \label{maths}
The theory of transformation media requires consideration of Maxwell's equations in arbitrary coordinates. This means that the natural mathematical language of transformation media is {\it differential geometry}, the mathematics that also describes curved spaces and Einstein's general relativity. Here we introduce the reader to the mathematics of arbitrary coordinates, to the extent necessary to deal with coordinate transformations of Maxwell's equations. The reader will see that this formalism provides the most transparent way of describing non-Cartesian coordinate systems: even something as apparently familiar as electromagnetism in spherical polar coordinates is much simpler in the language of differential geometry than in the standard treatment found in the  textbooks.\footnote{The introduction to curved coordinates given here is broadly similar to that in Schutz's excellent text \cite{Schutz}.} 


\subsection{Coordinate transformations}

We deal first with spatial coordinates; the extension to space-time coordinates will then be straightforward. Our interest is in writing equations that are valid in an arbitrary spatial coordinate system $\{x^i,\ i=1,2,3\}$ and in performing an arbitrary transformation to another set of coordinates that we distinguish from the original by a prime on the index: $\{x^{i'},\ i'=1,2,3\}$. Throughout, we take as concrete examples Cartesian coordinates $\{x,y,z\}$ and spherical polar coordinates $\{r,\theta,\phi\}$, related by
\begin{equation}
\begin{array}{ll}
\{x^i\}=\{x,y,z\},  & \quad \{x^{i'}\}=\{r,\theta,\phi\},  \\[5pt] 
x=r\sin\theta\cos\phi, & \quad r=\sqrt{x^2+y^2+z^2}, \\[5pt]
y=r\sin\theta\sin\phi, & \quad  \theta=\tan^{-1}(\sqrt{x^2+y^2}/z), \\[5pt]
z=r\cos\theta & \quad  \phi=\tan^{-1}(y/z). \label{coords}
\end{array}
\end{equation}
At this point we introduce the {\it Einstein summation convention} in which a summation is implied over repeated indices; for example 
\begin{equation} \label{sum}
A^iB_i\equiv\sum_iA^iB_i=A^1B_1+A^2B_2+A^3B_3.
\end{equation}
This convention allows us to dispense with writing summation signs, which are completely unnecessary. For reasons that will become clear later on, our summations will generally be over a pair of indices in which one index is a subscript and one is a superscript, as in Eq.\ (\ref{sum}). We also introduce the {\it Einstein range convention} by which a free index ({\it i.e.}\ an index that is not summed over) is understood to range over all possible values of the index, for example
\begin{equation}
A^i\equiv\{A^i,\ i=1,2,3\},
\end{equation}
Together, the summation and range conventions allow an economy of notation such as the following:
\begin{equation}
R^i_{\ jik}\equiv\left\{\sum_iR^i_{\ jik},\ \ j,k=1,2,3\right\}.
\end{equation}
The differentials of our two sets of coordinates, $x^i$ and $x^{i'}$, are related by the chain rule:
\begin{equation} \label{ds}
\mathrm{d}x^i=\frac{\partial x^i}{\partial x^{i'}}\, \mathrm{d}x^{i'}, \quad \mathrm{d}x^{i'}=\frac{\partial x^{i'}}{\partial x^i}\, \mathrm{d}x^i,
\end{equation}
with similar relations holding for the differential operators:
\begin{equation} \label{dels}
\frac{\partial}{\partial x^{i'}}=\frac{\partial x^i}{\partial x^{i'}}\,\frac{\partial}{\partial x^i},  \quad \frac{\partial}{\partial x^i}=\frac{\partial x^{i'}}{\partial x^i}\,\frac{\partial}{\partial x^{i'}}.
\end{equation}
We denote the transformation matrices in Eqs.\ (\ref{ds}) and (\ref{dels}) by
\begin{equation} \label{Lambdas}
\Lambda^i_{\ i'}=\frac{\partial x^i}{\partial x^{i'}}, \quad \Lambda^{i'}_{\ i}=\frac{\partial x^{i'}}{\partial x^i}.
\end{equation}
Note that primes or unprimed indices in $\Lambda^i_{\ i'}$ and $\Lambda^{i'}_{\ i}$ do not mean that we simply use different indices:  $\Lambda^i_{\ i'}$ and $\Lambda^{i'}_{\ i}$ are different matrices where we differentiate with respect to different sets of coordinates.
The reader may verify that for the example (\ref{coords}) the transformation matrices (\ref{Lambdas}) are
\begin{align}
\Lambda^i_{\ i'}=&\left(\begin{array}{ccc} \sin\theta\cos\phi & r\cos\theta\cos\phi & -r\sin\theta\sin\phi \\ \sin\theta\sin\phi & r\cos\theta\sin\phi & r\sin\theta\cos\phi \\ \cos\theta & -r\sin\theta & 0 \end{array}\right) \label{Lam1a} \\[10pt]
=& \left(\begin{array}{ccc} {\displaystyle \frac{x}{r} }& {\displaystyle\frac{xz}{\sqrt{x^2+y^2}} }& -y \\[15pt] {\displaystyle\frac{y}{r} }& {\displaystyle\frac{yz}{\sqrt{x^2+y^2}}} & x \\[15pt] {\displaystyle\frac{z}{r}} & -\sqrt{x^2+y^2} & 0  \end{array}\right), \label{Lam1b}  \\[10pt]
\Lambda^{i'}_{\ i}=&\left(\begin{array}{ccc} \sin\theta\cos\phi & \sin\theta\sin\phi & \cos\theta \\[5pt] {\displaystyle \frac{1}{r} }\cos\theta\cos\phi & {\displaystyle \frac{1}{r} }\cos\theta\sin\phi & {\displaystyle -\frac{1}{r} }\sin\theta \\[15pt] {\displaystyle -\frac{1}{r} }\csc\theta\sin\phi & {\displaystyle \frac{1}{r} }\csc\theta\cos\phi & 0 \end{array}\right),  \label{Lam2a} \\[10pt] 
=&\left(\begin{array}{ccc} {\displaystyle \frac{x}{r} }& {\displaystyle \frac{y}{r}}& {\displaystyle \frac{z}{r} } \\[10pt] {\displaystyle \frac{xz}{r^2\sqrt{x^2+y^2}} }& {\displaystyle \frac{yz}{r^2\sqrt{x^2+y^2}} } & {\displaystyle -\frac{\sqrt{x^2+y^2}}{r^2} } \\[15pt]  {\displaystyle -\frac{y}{x^2+y^2} } &  {\displaystyle \frac{x}{x^2+y^2} } & 0  \end{array}\right).  \label{Lam2b}
\end{align}
From Eqs.\ (\ref{ds}) and (\ref{Lambdas}) we find
$
\mathrm{d}x^i=\Lambda^i_{\ i'}\mathrm{d}x^{i'}=\Lambda^i_{\ i'}\Lambda^{i'}_{\ j}\mathrm{d}x^j
$
and
$
\mathrm{d}x^{i'}=\Lambda^{i'}_{\ i}\mathrm{d}x^{i}=\Lambda^{i'}_{\ i}\Lambda^{i}_{\ j'}\mathrm{d}x^{j'},
$
which imply
\begin{equation}  \label{inv}
\Lambda^i_{\ i'}\Lambda^{i'}_{\ j}=\delta^i_{\ j}, \quad \Lambda^{i'}_{\ i}\Lambda^{i}_{\ j'}=\delta^{i'}_{\ j'},
\end{equation}
where $\delta^i_{\ j}$ and $\delta^{i'}_{\ j'}$ are the Kronecker delta,
the matrix elements of the unity matrix. Equations (\ref{inv}) state that the matrices $\Lambda^i_{\ i'}$ and $\Lambda^{i'}_{\ i}$ are the inverses of each other. This property can be deduced directly from the definitions (\ref{Lambdas}) and the chain rule. 
The reader may verify the relations (\ref{inv}) for the example (\ref{Lam1a})-(\ref{Lam2b}).


\subsection{The metric tensor}

Although we have awarded ourselves the freedom of covering space with any coordinate system we wish, the distances between points in space are {\it invariant} --- they are the same no matter which coordinates we use to calculate them. The basic quantity is the square of the infinitesimal distance $\mathrm{d}s$ between the points $x^i$ and $x^i+\mathrm{d}x^i$. For Cartesian coordinates $x^i=\{x,y,z\}$ this is given by the 3-dimensional Pythagoras theorem:
\begin{equation} \label{dscar}
\mathrm{d}s^2=\mathrm{d}x^2+\mathrm{d}y^2+\mathrm{d}z^2=\delta_{ij}\mathrm{d}x^i\mathrm{d}x^j,
\end{equation}
where $\delta_{ij}$ is again the Kronecker delta. For general coordinates $x^i$, the square of the line element $\mathrm{d}s^2$ is given by an expression quadratic in the coordinate differentials $\mathrm{d}x^i$:
\begin{equation} \label{metric}
\mathrm{d}s^2=g_{ij}\mathrm{d}x^i\mathrm{d}x^j.
\end{equation}
In Eq.\ (\ref{metric}) we have introduced the {\it metric tensor} $g_{ij}$, the quantity that allows us to calculate distances in space. The metric tensor is always symmetric in its indices, 
\begin{equation} \label{gsym}
g_{ij}=g_{ji},
\end{equation}
for the following reason: a matrix can always be written as a sum of its symmetric and antisymmetric parts, and the reader can verify that an antisymmetric part of $g_{ij}$ would not contribute to the distance (\ref{metric}). So it only makes sense to consider a symmetric metric tensor. 
In Cartesian coordinates (\ref{dscar}) the metric tensor is the {\it Euclidean metric} $\delta_{ij}$. 

We can write the relation (\ref{metric}) also in the coordinate system $x^{i'}$, denoting the metric tensor in this system by $g_{i'j'}$; from the invariance of $\mathrm{d}s$ we have
\begin{equation} \label{gs}
\mathrm{d}s^2=g_{i'j'}\mathrm{d}x^{i'}\mathrm{d}x^{j'}=g_{ij}\mathrm{d}x^i\mathrm{d}x^j 
=g_{ij}\Lambda^i_{\ i'}\Lambda^j_{\ j'}\mathrm{d}x^{i'}\mathrm{d}x^{j'},
\end{equation}
where we have used Eqs.\ (\ref{ds}) and (\ref{Lambdas}) in the second line. Equation (\ref{gs}) reveals how the metric tensor changes under a coordinate transformation:
\begin{equation} \label{gtrans}
g_{i'j'}=\Lambda^i_{\ i'}\Lambda^j_{\ j'}g_{ij}.
\end{equation}
Writing the metric tensors $g_{i'j'}$ and $g_{ij}$ as matrices $G'$ and $G$ we can display the transformation procedure in the matrix form
\begin{equation} \label{gtransmatrix}
G'=\Lambda^T G \Lambda
\end{equation}
where $\Lambda$ denotes the transformation matrix $\Lambda^i_{\ i'}$
defined in Eq.\ (\ref{Lambdas}), whereas $\Lambda^{i'}_{\ i}$ is the inverse matrix $\Lambda^{-1}$.

Consider a transformation from a Cartesian coordinate system, in which $g_{ij}=\delta_{ij}$, to another Cartesian system, by means of a rotation. As the new coordinates are Cartesian, the transformed metric must also be Euclidean and Eq.\ (\ref{gtrans}) shows that this is the case:
\begin{equation} \label{orth}
g_{i'j'}=\Lambda^i_{\ i'}\Lambda^j_{\ j'}\delta_{ij}=\delta_{i'j'} \quad \text{(Cartesian $\longrightarrow$ Cartesian)}
\end{equation}
where the second equality follows from the fact that rotations are performed by {\it orthogonal matrices} with $\Lambda^T=\Lambda^{-1}$. Rotations thus preserve the Euclidean metric.

The metric tensor not only characterizes the measure of length in arbitrary coordinates, it turns out to describe the volume element as well. To see this, we represent the Cartesian volume element, $\mathrm{d}V=\mathrm{d}x\,\mathrm{d}y\,\mathrm{d}z$, in arbitrary coordinates according to the standard rule
\begin{equation} \label{voltrans}
\mathrm{d}V = \left|\mathrm{det}\Lambda\right|\,\mathrm{d}V'.
\end{equation}
From the matrix representation (\ref{gtransmatrix}) follows 
\begin{equation} \label{gdet}
g' = (\mathrm{det}\Lambda)^2 g
\end{equation}
where $g$ and $g'$ denote the determinants of the metric tensors.  
Note that $g'$ is always positive, because the determinant $g$ of the Euclidian metric is unity.
Consequently, we obtain the volume element
\begin{equation} \label{volelement}
\mathrm{d}V = \sqrt{g'}\,\mathrm{d}V'.
\end{equation}
Dropping the primes, we note that $\sqrt{g}\,\mathrm{d}V$ always describes the volume element, in Cartesian or curved coordinates. 

Returning to our example of spherical coordinates, we can use the transformation procedure (\ref{gtrans}) to compute the metric tensor and the volume element: the metric tensor in Cartesian coordinates is $g_{ij}=\delta_{ij}$ and the required transformation matrix is expressed in Eq.\ (\ref{Lam1a}); so we obtain
\begin{gather}
g_{i'j'}=\left(\begin{array}{ccc} 1 & 0 & 0 \\ 0 & r^2 & 0 \\ 0 & 0 & r^2\sin^2\theta \end{array}\right),  \label{gsph} \\[5pt]
\mathrm{d}s^2=g_{i'j'}\mathrm{d}x^{i'}\mathrm{d}x^{j'}=\mathrm{d}r^2+r^2(\mathrm{d}\theta^2+\sin^2\theta\, \mathrm{d}\phi^2). \label{dssph}
\end{gather}
The volume element (\ref{volelement}) is given by the square root of the determinant of the matrix (\ref{gsph}); we arrive at the familiar spherical volume element $\mathrm{d}V=r^2\sin\theta\,\mathrm{d}r\,\mathrm{d}\theta\,\mathrm{d}\phi$.


\subsection{Vectors and bases}

The transformation relations (\ref{ds})-(\ref{Lambdas}) determine the transformation properties of vectors, and of more general objects. The coordinate displacements $\mathrm{d}x^i$ are the components of a vector in space, therefore the components of a general vector $\bm{V}$ will transform in the same way under a change of coordinates:
\begin{equation} \label{vtrans}
V^{i'}=\Lambda^{i'}_{\ i}V^i, \quad V^i=\Lambda^i_{\ i'}V^{i'}.
\end{equation}
The components $V^{i}$ and $V^{i'}$ in (\ref{vtrans}) refer to an expansion of $\bm{V}$ in terms of the basis vectors $\bm{e}_i$ and $\bm{e}_{i'}$ associated with each coordinate system:
\begin{equation} \label{expan}
\begin{split}
\bm{V}=&\,V^{i}\bm{e}_i=V^{i'}\bm{e}_{i'} \\
=&\,\Lambda^{i}_{\ i'}V^{i'}\bm{e}_i.
\end{split}
\end{equation}
The second line in Eq.\ (\ref{expan}) was obtained by use of Eq.\ (\ref{vtrans}) and comparison with the first line gives the transformation of the basis vectors:
\begin{equation} \label{etrans}
\bm{e}_{i'}=\Lambda^{i}_{\ i'}\bm{e}_i.
\end{equation}
In Cartesian coordinates $x^i=\{x,y,z\}$ the basis vectors are the familiar unit vectors in the $x$-, $y$- and $z$ directions:
\begin{equation} \label{ecar}
\bm{e}_i=\{\bm{e}_x,\bm{e}_y,\bm{e}_z\}=\{\bm{i},\bm{j},\bm{k}\}.
\end{equation}
From (\ref{etrans}) and (\ref{Lam1a}) we then obtain the basis vectors in spherical polar coordinates as $\bm{e}_{i'}=\{\bm{e}_r,\bm{e}_\theta,\bm{e}_\phi\}$
with
\begin{eqnarray}
\bm{e}_r&=&\,\sin\theta\cos\phi\,\bm{i}+\sin\theta\sin\phi\,\bm{j}+\cos\theta\,\bm{k}, \nonumber\\
\bm{e}_\theta&=&\,r\cos\theta\cos\phi\,\bm{i}+r\cos\theta\sin\phi\,\bm{j}-r\sin\theta\,\bm{k}, \nonumber \\
\bm{e}_\phi&=&-\!r\sin\theta\sin\phi\,\bm{i}+r\sin\theta\cos\phi\,\bm{j}. \label{es}
\end{eqnarray}


\subsection{One-forms and general tensors} 

The expression (\ref{metric}) is the squared length of the vector $\mathrm{d}x^i$, and the metric tensor similarly gives the squared length of a general vector:
\begin{equation} \label{vsqd}
\left|\bm{V}\right|^2=\bm{V}\cdot\bm{V}=g_{ij}V^iV^j=g_{i'j'}V^{i'}V^{j'}.
\end{equation}
In Eq.\ (\ref{vsqd}) we have used the fact that the length of a vector is an invariant quantity, and the equality of the expressions evaluated in the coordinate systems $x^i$ and $x^{i'}$ can be explicitly shown from the transformation rules (\ref{gtrans}), (\ref{vtrans}) together with the inverse relations (\ref{inv}). Note that the invariance of Eq.\ (\ref{vsqd}) under coordinate transformations works because the vector $V^i$ and metric $g_{ij}$ are transformed by matrices that are inverse to each other. This inverse relationship of their transformations is to be associated with the fact that $V^i$ is an upper-index object whereas $g_{ij}$ is a lower-index object. Each summation in (\ref{vsqd}) over an upper and lower index, called a {\it contraction}, is a coordinate-invariant operation because of this inverse property. It is useful to construct from $V^i$ and $g_{ij}$ a lower-index quantity $V_i$ that transforms in the manner of $g_{ij}$, as follows:
\begin{equation} \label{vcov}
V_i=g_{ij}V^j.
\end{equation}
The transformation rule for $V_i$ is easily established from those of $V^i$ and $g_{ij}$:
\begin{equation} \label{covectrans}
V_{i'}=g_{i'j'}V^{j'}=\Lambda^{i}_{\ i'}g_{ij}V^j=\Lambda^{i}_{\ i'}V_i,
\end{equation}
so that $V_i$ does indeed transform similarly to $g_{ij}$. The quantities $V_i$ are the components of a {\it covariant vector} or {\it one-form}. We can view Eq.\ (\ref{vcov}) as lowering the index on the vector $V^i$ using the metric tensor, producing the associated one-form $V_i$. Defining the {\it inverse metric tensor} $g^{ij}$ by
\begin{equation} \label{ginv}
g^{ij}g_{jk}=\delta^i_{\ k},
\end{equation}
we obtain from relation (\ref{vcov})
\begin{equation} \label{vraise}
V^i=g^{ij}V_j.
\end{equation}
Equation (\ref{vraise}) can be regarded as raising the index on the one-form $V_i$ using the inverse metric tensor, producing the associated vector $V^i$. Note that the vector $V^i$ and one-form $V_i$ are the same only in Cartesian coordinates where $g_{ij}=\delta_{ij}$. The expression (\ref{vsqd}) for the squared length of a vector is more compactly written using the one-form $V_i$:
$
\left|\bm{V}\right|^2=\bm{V}\cdot\bm{V}=V_iV^i
$.
The scalar product of two vectors $\bm{V}$ and $\bm{U}$ is
\begin{equation} \label{dot}
\bm{V}\cdot\bm{U}=g_{ij}V^iU^i=V_iU^i=V^iU_i=V_{i'}U^{i'}=V^{i'}U_{i'},
\end{equation}
where the coordinate invariance of the scalar product follow from the transformation properties of vectors and one-forms.
From the position of the indices on the inverse metric tensor $g^{ik}$ we expect it to transform in a vector-like fashion. We verify this by writing the vector and one-form in Eq.\ (\ref{vraise}) in terms of their transformed values,
$
\Lambda^{i}_{\ i'}V^{i'}=g^{ij}\Lambda^{j'}_{\ j}V_{j'}$
from which follows
$
V^{i'}=\Lambda^{i'}_{\ i}g^{ij}\Lambda^{j'}_{\ j}V_{j'}=g^{i'j'}V_{j'}
$
so that
\begin{equation}\label{ggtrans}
g^{i'j'}=\Lambda^{i'}_{\ i}\Lambda^{j'}_{\ j}g^{ij}.
\end{equation}

The transformation rule for a tensor with an arbitrary collection of indices is now clear; we give as an example a four-index tensor:
\begin{equation} \label{ttrans}
R^{i'}_{\ j'k'l'} = 
\Lambda^{i'}_{\ i}\Lambda^{j}_{\ j'}\Lambda^{k}_{\ k'}\Lambda^{l}_{\ l'}
R^i_{\ jkl}.
\end{equation}
One can also raise and lower the indices of a general tensor in complete analogy to Eqs.\ (\ref{vraise}) and (\ref{vcov}). In this manner Eq.\ (\ref{ginv}) may be regarded as a raising or lowering operation so that the Kronecker delta $\delta^i_{\ j}$ is the metric tensor with one index raised, or the inverse metric tensor with one index lowered.

It is straightforward to introduce bases for general tensors, but we will not pursue this here. The reader should, however, be able to deduce the index positions and transformation properties of any tensor basis, starting with the one-form basis.


\subsection{Coordinate and non-coordinate bases}

The scalar product of two basis vectors is seen from the definition (\ref{dot}) to be
\begin{equation} \label{edot}
\bm{e}_i\cdot\bm{e}_j=g_{ij},
\end{equation}
since the components $(\bm{e}_i)^j$ of a basis vector $\bm{e}_i$ are $\delta^j_{\ i}$.
The basis vectors (\ref{ecar}) in Cartesian coordinates constitute an orthonormal basis. For the spherical polar basis (\ref{es}) we can compute the dot products using the right-hand sides of Eqs.\ (\ref{es}) or, much more simply, by using the scalar product (\ref{edot}) and the metric (\ref{gsph}) in spherical polar coordinates. We see that the basis vectors are orthogonal to each other, but they are not all unit vectors:
\begin{equation}
\left|\bm{e}_r\right|^2=1, \quad
\left|\bm{e}_\theta\right|^2=r^2, \quad
\left|\bm{e}_\phi\right|^2=r^2\sin^2\theta. \label{eesph}
\end{equation}
One can of course easily construct an orthonormal basis $\hat{\bm{e}}_i$ by rescaling $\bm{e}_\theta$ and $\bm{e}_\phi$:
\begin{equation}
\hat{\bm{e}}_r=\bm{e}_r, \quad
\hat{\bm{e}}_\theta=\frac{1}{r}\,\bm{e}_\theta, \quad
\hat{\bm{e}}_\phi=\frac{1}{r\sin\theta}\,\bm{e}_\phi. \label{ensph}
\end{equation}
The reader unfamiliar with the material of this section will only have encountered spherical polar coordinates in combination with the orthonormal basis (\ref{ensph}). How did we end up with the non-orthonormal basis? The answer is that we let the coordinates induce our basis through their differentiable structure. Recall that the components of a vector were introduced by analogy with the coordinate differentials $\mathrm{d}x^i$. The discussion of vector components then immediately specifies the basis as in Eqs.\ (\ref{vtrans})-(\ref{etrans}). Such a basis, induced naturally by the coordinates, is called a {\it coordinate basis}. The fact that the differentiable properties of the coordinates completely determine the coordinate basis is the reason why the coordinate bases (\ref{etrans}) behave exactly like the partial derivative operators in (\ref{dels}).\footnote{This is not just a pleasant correspondence; in modern differential geometry the partial derivative operators {\it are} the coordinate basis vectors.} 
The orthonormal basis (\ref{ensph}), by contrast, is not induced in a similar manner by {\it any} coordinate system --- it is a {\it non-coordinate basis} (see Ref.\ \cite{Schutz} for more detail). The only coordinate system that induces an orthonormal coordinate basis is the Cartesian system. It might be suspected that it is always simpler to work in an orthonormal basis; in fact, for most purposes a coordinate basis is much simpler, in particular for the manipulations in curvilinear coordinates performed in electromagnetism textbooks. The reason these texts use the more complicated non-coordinate bases is that to exploit the simplicity of coordinate bases requires a little knowledge of tensor analysis. 


\subsection{Vector products and Levi-Civita tensor} 

For describing electromagnetism in arbitrary coordinates, we  need to define the notion of vector products in three-dimensional space.  Vector products ${\bm V}\times {\bm U}$ are antisymmetric, ${\bm V}\times {\bm U}=-{\bm U}\times {\bm V}$, and the vector products of the Cartesian basis vectors are cyclic: ${\bm i}\times {\bm j}= {\bm k}$, ${\bm j}\times {\bm k}= {\bm i}$ and ${\bm k}\times {\bm i}= {\bm j}$. For implementing these properties,  we introduce the permutation symbol $[ijk]$ defined by
\begin{equation}\label{ijk}
[ijk]=\left\{\begin{array}{l} +1\ \text{if $ijk$ is an even permutation of 123,} \\
-1\ \text{if $ijk$ is an odd permutation of 123,} \\
0\ \text{otherwise.} \end{array}\right.
\end{equation}
We define the {\it Levi-Civita tensor} $\epsilon^{ijk}$ as the tensor whose components in some particular right-handed Cartesian coordinate system are given by the permutation symbol:
\begin{equation} \label{LC}
\epsilon^{ijk}=[ijk], \quad \text{(Right-handed Cartesian coordinates).}
\end{equation}
We can find the components of the Levi-Civita tensor in any other coordinate system, Cartesian or otherwise, by transforming the expression (\ref{LC}) according to the rule (\ref{ttrans}); we constructed a tensor by definition. Specifically, the Levi-Civita tensor in an arbitrary coordinate system is
\begin{equation} \label{LCtrans}
\epsilon^{i'j'k'}=\Lambda^{i'}_{\ i}\Lambda^{j'}_{\ j}\Lambda^{k'}_{\ k}[ijk] =\det (\Lambda^{l'}_{\ l})[i'j'k'] = \frac{[i'j'k']}{\det \Lambda}
\end{equation}
using the Leibniz formula for the determinant of $\Lambda^{l'}_{\ l}$ \cite{Stoll}. As in Sec.\ 3.2, $\det \Lambda$ is the determinant of the transformation matrix $\Lambda^{l}_{\ l'}$, the matrix inverse of $\Lambda^{l'}_{\ l}$. We obtain from Eq.\ (\ref{gdet})
\begin{equation} \label{detL}
\det \Lambda=\pm\sqrt{g'}.
\end{equation}
Which sign should we take here? Clearly the sign in question is the sign of $\det\Lambda$. If $\det\Lambda$ is negative the transformation changes the handedness of the coordinate system, so the new system is left-handed \cite{Goldstein}. For example, a transformation that changes the sign of one or all three of the coordinates in the right-handed Cartesian system has $\det\Lambda=-1$ and results in a left-handed Cartesian system. A general transformation between Cartesian coordinate systems consists of a rotation and a translation, together with possible reflections of the coordinates. For any such transformation the Euclidean metric is preserved so Eq.\ (\ref{orth}) holds and the transformation matrix is orthogonal; it then follows from Eq.\ (\ref{detL}) that
\begin{equation} \label{detorth}
\det\Lambda=\pm 1 \quad \text{(Cartesian $\longrightarrow$ Cartesian)},
\end{equation}
where the sign is negative if the transformation includes a handedness change. Using the relationship (\ref{detL}) we can now write the Levi-Civita tensor in arbitrary coordinates (\ref{LCtrans}) in terms of the metric; we drop the prime on the indices of the arbitrary coordinate system and obtain
\begin{equation} \label{LCgen}
\epsilon^{ijk}=\pm\frac{1}{\sqrt{g}}[ijk],
\end{equation}
where it is understood that the plus (minus) sign obtains if the system is right-handed (left-handed). It is now clear that Eq.\ (\ref{LC}) holds in {\it all} right-handed Cartesian coordinate systems, not just in the one we started with.

Some or all of the indices of $\epsilon^{ijk}$ can be lowered according to the prescription (\ref{vcov}); if all are lowered we obtain another simple expression to complement the representation (\ref{LCgen}):
\begin{equation} \label{LClow}
\epsilon_{ijk}=g_{il}g_{jm}g_{kn}\epsilon^{lmn}=\pm\frac{1}{\sqrt{g}}g_{il}g_{jm}g_{kn}[lmn]=\pm\frac{1}{\sqrt{g}}g[ijk]=\pm\sqrt{g}[ijk].
\end{equation}
The Levi-Civita tensor is {\it completely antisymmetric}; this means that when its components are taken with all indices in the upper or lower position they are antisymmetric under interchange of two adjacent indices, e.g.\ $\epsilon^{ijk}=-\epsilon^{jik}$.

The Levi-Civita tensor is required to compute vector products in an arbitrary coordinate system:
\begin{equation} \label{vecprod}
\bm{U}\times\bm{V}=\epsilon^{ijk}U_jV_k\bm{e}_i.
\end{equation}
The reader can verify that (\ref{vecprod}) is the standard vector product in right-handed Cartesian coordinates; the definitions we have given show that it maintains the same form when transformed to an arbitrary coordinate system. The components of the vector product can be written in terms of Eqs.\ (\ref{LCgen}) or (\ref{LClow}):
\begin{equation} \label{vecprodcpts}
(\bm{U}\times\bm{V})^i=\epsilon^{ijk}U_jV_k, \quad (\bm{U}\times\bm{V})_i=\epsilon_{ijk}U^jV^k.
\end{equation}
We can apply the Levi-Civita tensor for expressing the well-known double vector product in arbitrary coordinates,
\begin{equation}\label{baccab}
\bm{A}\times(\bm{B}\times\bm{C}) = 
\bm{B} (\bm{A}\cdot\bm{C}) -
\bm{C} (\bm{A}\cdot\bm{B}).
\end{equation}
We obtain from formulas (\ref{LCgen}) and (\ref{LClow}) and the defining properties (\ref{ijk}) of the permutation symbol:
\begin{equation}\label{baccabgen}
\epsilon^{ijk}\epsilon_{klm} = \sum_k [ijk]\,[klm] = 
\delta^i_l\delta^j_m-\delta^i_m\delta^j_l.
\end{equation}
Contracted with the components of three vectors $A_j$, $B^l$ and $C^m$, this identity generates the double vector product (\ref{baccab}).
Curls are also computed using the Levi-Civita tensor, but they contain a differentiation and we must learn how to differentiate in arbitrary coordinates.


\subsection{The covariant derivative of a vector} 

A scalar field in space is a function of the coordinates and we can take its partial derivative with respect to $x^i$; in writing the partial derivative we can introduce another ink-saving device, as follows:
\begin{equation} \label{par}
\frac{\partial}{\partial x^i}\psi\equiv\psi_{,i}.
\end{equation}
Thus a comma means partial differentiation, with the following index giving the coordinate with respect to which the derivative is taken. In Cartesian coordinates the derivatives (\ref{par}) are of course the components of the gradient vector $ \bm{\nabla}\psi$. It is easy to see, however, from Eqs.\ (\ref{dels}), (\ref{Lambdas}) and (\ref{covectrans}), that the expression (\ref{par}) transforms as a one-form. Consistent with the index being in the lower position, the derivatives (\ref{par}) are in fact the components of the one-form associated with the gradient vector, and this distinction can only be ignored in Cartesian coordinates where the vector and the one-form have the same components. In a general coordinate system we must raise the index in Eq.\ (\ref{par}) using the inverse metric tensor to obtain the components of the gradient vector, {\it i.e.}
\begin{equation}
( \bm{\nabla}\psi)_i=\psi_{,i},  \quad ( \bm{\nabla}\psi)^i=g^{ij}\psi_{,j},  \quad  \bm{\nabla}\psi=g^{ij}\psi_{,j}\bm{e}_i. \label{grad} 
\end{equation}
$ \bm{\nabla}\psi$ is a vector and so it is a coordinate-independent object; the reader may verify that $g^{ij}\psi_{,j}\bm{e}_i$ is the same in every coordinate system using the transformation rules for the quantities involved, $
g^{ij}\psi_{,j}\bm{e}_i=g^{i'j'}\psi_{,j'}\bm{e}_{i'}$.
Note that if we were to use an orthonormal frame the components of the gradient one-form and vector would not be given by Eq.\ (\ref{grad}); this is easily seen in the case of spherical polar coordinates, where if we replace the coordinate basis (\ref{es}) in Eq.\ (\ref{grad}) by the orthonormal, non-coordinate basis (\ref{ensph}), the components (\ref{grad}) are rescaled. Thus in a non-coordinate basis the partial derivatives $\psi_{,i}$ are not the components of the gradient one-form and this makes such a basis unsuitable for our tensor calculus. Partial derivatives with respect to the coordinates only have a simple tensorial meaning if we use the basis induced by those coordinates.
\begin{figure}[t]
\includegraphics[width=35.0pc]{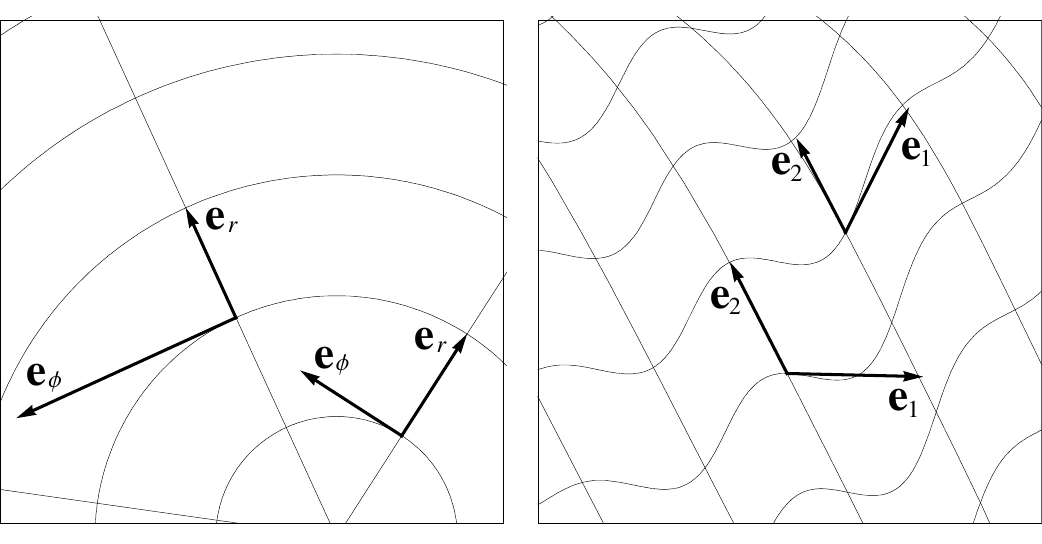}
\caption{\small{Basis vectors change with position. Polar coordinates in two dimensions (spherical polar coordinates with $\theta=\pi/2$) are depicted on the left. The basis vectors are everywhere orthogonal but they rotate as one moves through the plane. While $\bm{e}_r$ remains a unit vector, the length of $\bm{e}_\phi$, equal to $r$, varies from zero to infinity. A general two-dimensional coordinate system is shown on the right. Here the orientations and lengths of both basis vectors vary, as does the angle between them. Since the 
dot product $\bm{e}_1\cdot\bm{e}_2$ does not vanish, the metric tensor is not diagonal 
in these coordinates ($g_{12}\neq 0$).}}   \label{grids}
\end{figure}

A vector field in space $\bm{V}=V^i\bm{e}_i$ consists of a sum of products of scalar fields $V^i$ and basis vector fields $\bm{e}_i$; to differentiate $\bm{V}$ we must of course use the Leibniz rule:
\begin{equation} \label{vder}
\frac{\partial}{\partial x^i}\bm{V}=\frac{\partial V^j}{\partial x^i}\bm{e}_j+V^j\frac{\partial\bm{e}_j}{\partial x^i}.
\end{equation}
This simple relation represents the most important fact about curvilinear coordinates. In Cartesian coordinates the basis vectors are constant and to differentiate a vector we need only to differentiate its components. The coordinate basis vectors for any other coordinate system, however, change in orientation and magnitude as one moves through space, see Fig.\ \ref{grids}. The rate of change of a vector is itself a vector and so can be expanded in terms of the basis $\bm{e}_i$; we therefore have
\begin{equation} \label{eder}
\frac{\partial\bm{e}_j}{\partial x^i}=\Gamma^k_{\ ji}\bm{e}_k.
\end{equation}
The 27 quantities $\Gamma^k_{\ ji}$ are called the {\it Christoffel symbols}; $\Gamma^k_{\ ji}$ is the $k$th component of the derivative of $\bm{e}_j$ with respect to $x^i$. 

Let us pause to consider how $\Gamma^k_{\ ji}$ change under a coordinate transformation. In the transformed system the Christoffel symbols are defined by 
$
{\partial\bm{e}_{j'}}/{\partial x^{i'}}=\Gamma^{k'}_{\ j'i'}\bm{e}_{k'}
$
and since we know how the basis vectors and the partial derivative operators transform, Eqs.\ (\ref{etrans}) and (\ref{dels}), we can deduce the transformation law for the Christoffel symbols. We leave it as an exercise for the reader to show that
\begin{equation} \label{Gammatrans}
\Gamma^{k'}_{\ j'i'}=\Lambda^{k'}_{\ k}\Lambda^{j}_{\ j'}\Lambda^{i}_{\ i'}\Gamma^{k}_{\ ji}-\Lambda^{k'}_{\ j,i}\Lambda^{j}_{\ j'}\Lambda^{i}_{\ i'}.
\end{equation}
Note that $\Gamma^{k}_{\ ji}$ do not obey the transformation law (\ref{ttrans}) of tensor components; they do not therefore constitute a tensor.

The transformation rule (\ref{Gammatrans}) reveals an important property of the Christoffel symbols: they are symmetric in their lower indices, {\it i.e.}
\begin{equation} \label{Gammasym}
\Gamma^{i}_{\ jk}=\Gamma^{i}_{\ kj}.
\end{equation}
To prove this, first write the second term in the transformation rule (\ref{Gammatrans}) explicitly in term of the partial derivatives (\ref{Lambdas}) and show that it is symmetric in $j'$ and $i'$. Now, the Christoffel symbols in any coordinate system can be obtained by transforming from Cartesian coordinates according to the rule (\ref{Gammatrans}), but in Cartesian coordinates the $\Gamma^{k}_{\ ji}$ vanish. This shows that in the new coordinate system $\Gamma^{k'}_{\ j'i'}=\Gamma^{k'}_{\ i'j'}$; but this coordinate system is arbitrary, so (\ref{Gammasym}) holds for any coordinates.

Using the Christoffel symbols we write down the derivative (\ref{vder}) of a vector. Insertion of Eq.\ (\ref{eder}) in Eq.\ (\ref{vder}) gives
\begin{equation}
\frac{\partial}{\partial x^i}\bm{V}=V^j_{\ \,,i}\bm{e}_j+\Gamma^k_{\ ji}V^j\bm{e}_k = V^j_{\ \,;i}\bm{e}_j, \label{derv}
\end{equation}
where we have defined the quantities
\begin{equation} \label{dvcov}
V^j_{\ \,;i}\equiv V^j_{\ \,,i}+\Gamma^j_{\ ki}V^k
\end{equation}
which give the components of the derivative of $\bm{V}$ with respect to $x^i$. 
Just as the derivative of a scalar (a zero-index tensor) gives a one-form (a one-index tensor), the derivative (\ref{derv}) of a vector $\bm{V}$ gives a two-index tensor $V^j_{\ \,;i}$ called the {\it covariant derivative} of $\bm{V}$. The semi-colon in the definition (\ref{dvcov}) thus means covariant differentiation of the vector, in which both the components and the basis are differentiated, whereas the comma means differentiation of the components. ``Covariant derivative'' therefore just means ``correct derivative''! 

Suppose we transport the vector $V^j$ from a point $x^i$ to its infinitesimally close neighbor $x^i+\mathrm{d}x^i$ without rotating it or changing its length; this is called {\it parallel transport}. At the new point the coordinate basis has changed, but the vector has not changed; so the vector components must vary as well as the coordinate basis. Here it is important that we differentiate correctly: the covariant derivative of the vector along $\mathrm{d}x^i$ vanishes, but the ordinary derivative of its components will not, unless we are in Cartesian coordinates. Imagine $V^j$ and its parallel-transported neighbour as being part of a vector field. We use $\mathrm{D}V^j$ to denote the increment of $V^j$ along $\mathrm{d}x^i$ and introduce an alternative notation for the covariant derivative, as the differential operator
\begin{equation} \label{Dv}
\nabla_i V^j \equiv V^j_{\ \,;i}= V^j_{\ \,,i}+\Gamma^j_{\ ki}V^k.
\end{equation}
The increment of $V^i$ along $\mathrm{d}x^i$ is then
\begin{equation} \label{D}
\mathrm{D}V^j =
(\nabla_i V^j)\,\mathrm{d}x^i.
\end{equation}
In the case of parallel transport the vector $V^j$ simply remains the same:
\begin{equation} \label{parallel}
\mathrm{D}V^j = 0. \qquad \text{(Parallel transport.)}
\end{equation}
A nontrivial example of parallel transport is given by Foucault's pendulum \cite{Hart}: consider a pendulum that is attached to the Earth, but free to oscillate (Foucault suspended one from the dome of the Pantheon in Paris in 1851). The direction in which a pendulum swings follows the rotation of the Earth, it is parallel-transported on the surface of a sphere. The pendulum slowly turns due to the non-Euclidean curvature of the sphere, as we discuss in Sec.\ 3.11. 

The covariant derivative of a vector $V^j$ is the tensor $V^j_{\ \,;i}$; so it must transform according to the tensor prescription (\ref{ttrans}). Note that neither of the two terms on the right-hand side of Eq.\ (\ref{dvcov}) separately constitutes a tensor. From the transformation rule of the Christoffel symbols (\ref{Gammatrans}) we see that the transformation of $\Gamma^j_{\ ki}V^k$ will not adhere to the tensor transformation (\ref{ttrans}), and the same is true of $V^j_{\ \,,i}$; but when they are added together, however, the result does obey the transformation rule (\ref{ttrans}) of a tensor, as the reader should verify. 

In standard vector calculus one is confined to scalars and vectors, and so one does not encounter the two-index tensor $V^j_{\ \,;i}$ in its full glory, but only some aspects of covariant differentiation. For example, the divergence $ \bm{\nabla}\cdot\bm{V}$ is constructed from the covariant derivative $V^j_{\ \,;i}$ by contraction of its two indices. To see this note that in Cartesian coordinates the divergence is $V^i_{\ \,,i}$, which (only) in these coordinates is equal to $V^i_{\ \,;i}$. Now, since $V^j_{\ \,;i}$ is a tensor, the contraction $V^i_{\ \,;i}$ is a scalar, the same in all coordinate systems, just like the dot product (\ref{dot}). We therefore have
\begin{equation} \label{divv}
 \bm{\nabla}\cdot\bm{V}= \nabla_i V^i = V^i_{\ \,;i}=V^{i'}_{\ \,;i'}.
\end{equation}

Let us return to the example of spherical polar coordinates to see what a set of Christoffel symbols looks like. We can compute the Christoffel symbols from Eqs.\ (\ref{eder}) and (\ref{es}), or by transforming from Cartesian coordinates, in which $\Gamma^i_{\ jk}$ are zero, using the rules (\ref{Gammatrans}) and (\ref{Lam1a})-(\ref{Lam2b}). Clearly, the recipe (\ref{eder}) presents the easier path and we find
\begin{gather}
\Gamma^\theta_{\ r\theta}=\frac{1}{r}, \quad 
\Gamma^\phi_{\ r\phi}=\frac{1}{r}, \quad 
\Gamma^\theta_{\ \theta r}=\frac{1}{r}, \quad 
\Gamma^r_{\ \theta\theta}=-r,  \quad
\Gamma^\phi_{\ \theta\phi}=\cot\theta, \nonumber \\[5pt]
\Gamma^\phi_{\ \phi r}=\frac{1}{r}, \quad 
\Gamma^\phi_{\ \phi\theta}=\cot\theta, \quad
\Gamma^r_{\ \phi\phi}=-r\sin^2\theta, \quad
\Gamma^\theta_{\ \phi\phi}=-\sin\theta\cos\theta, \label{Gammasph}
\end{gather}
all the other Christoffel symbols vanishing. We can now compute the divergence (\ref{divv}) in spherical polar coordinates, applying the covariant derivative (\ref{dvcov}):
\begin{align}
 \bm{\nabla}\cdot\bm{V}&=\frac{\partial}{\partial r}V^r+\frac{\partial}{\partial \theta}V^\theta+\frac{\partial}{\partial \phi}V^\phi+\frac{2}{r}\,V^r+\cot\theta\, V^\theta   \nonumber \\[5pt]
&=\frac{1}{r^2}\frac{\partial}{\partial r}\left(r^2V^r\right)+\frac{1}{\sin\theta}\frac{\partial}{\partial \theta}\left(\sin\theta\, V^\theta\right)+\frac{\partial}{\partial \phi}V^\phi. \label{divsph}
\end{align}
Remember that in Eq.\ (\ref{divsph}) we are using the coordinate basis (\ref{es}); to find the expression in the orthonormal, non-coordinate basis (\ref{ensph}) requires an obvious rescaling of the vector components.


\subsection{Covariant derivatives of tensors and of the metric} 

Our knowledge of how to differentiate scalars and vectors leads directly to the expressions for the covariant derivatives of more general tensors. Consider the scalar product (\ref{dot}), written in terms of a one-form and a vector: $U_iV^i$. Since this is a scalar, the correct derivative is the ordinary partial derivative of a scalar field:
\begin{equation} \label{derdot}
(U_iV^i)_{,j}=U_{i}V^i_{\ ,j}+U_{i,j}V^i,
\end{equation}
where we have employed the Leibniz rule. Let us rewrite this equation in terms of the covariant derivative (\ref{dvcov}) of the vector:
\begin{equation}
(U_iV^i)_{,j}=U_{i}V^i_{\ ;j}-U_i\Gamma^i_{\ kj}V^k+U_{i,j}V^i
=U_{i}V^i_{\ ;j}+(U_{i,j}-\Gamma^k_{\ ij}U_k)V^i.  \label{duv}
\end{equation}
The left-hand side of Eq.\ (\ref{duv}) is a tensor, the gradient one-form of the scalar $U_iV^i$, so the right-hand side is also a tensor. Now everything here, except the quantity in brackets, has already been shown to be a tensor; therefore the quantity in brackets is also a tensor, it is the covariant derivative of the one-form $U_i$:
\begin{equation} \label{ducov}
U_{i;j}=U_{i,j}-\Gamma^k_{\ ij}U_k,
\end{equation}
or, using the notion of the covariant derivative as a differential operator,
\begin{equation} \label{Du}
\nabla_jU_i=U_{i,j}-\Gamma^k_{\ ij}U_k.
\end{equation}
The covariant derivatives depends on the character of the object that is differentiated, vector components (\ref{dvcov}) are differentiated differently than the components of one-forms  (\ref{ducov}). The index position of the semi-colon in the definitions (\ref{dvcov}) and (\ref{ducov}) indicates this better than the differential operators (\ref{Dv}) and (\ref{Du}). Note that the covariant derivative obeys the Leibniz rule:
\begin{equation} \label{Leibnizuv}
(U_iV^i)_{,j}=(U_iV^i)_{;j}=U_{i}V^i_{\ ;j}+U_{i;j}V^i.
\end{equation}
The fact that $U_{i;j}$ is a tensor can be also shown directly by proving it transforms according the the tensor rule (\ref{ttrans}), using the known transformation properties of the objects on the right-hand side of (\ref{ducov}).

One can deduce the expression for the covariant derivative of any tensor by constructing a scalar from it with vectors and one-forms and applying the above procedure. For example, expanding the derivative of $A_j^{\ i}U_iV^j$ one finds the covariant derivative of a mixed tensor,
\begin{equation} \label{dmix}
\nabla_kA_j^{\ i} 
= A_{j \ ,k}^{\ i} -\Gamma^m_{\ jk}A_m^{\ i}+\Gamma^i_{\ mk}A_j^{\ m}.
\end{equation} 
The general rule is simple: high indices get a positive sign in front of the Christoffel symbols and low indices a negative sign. An important case is the covariant derivative of the metric tensor itself:
\begin{eqnarray}
g_{ij;k}&=&g_{ij,k}-\Gamma^l_{\ ik}g_{lj}-\Gamma^l_{\ jk}g_{il}, 
\label{dg}\\
g^{ij}_{\ \ ;k}&=&g^{ij}_{\ \ ,k}+\Gamma^i_{\ lk}g^{lj}+\Gamma^j_{\ lk}g^{il}.\label{dgup}
\end{eqnarray}
A highly significant property of the metric tensor now emerges if we consider the fact that, as a tensor, it transforms as
\begin{equation} \label{dgtrans}
g_{i'j';k'}=\Lambda^i_{\ i'}\Lambda^j_{\ j'}\Lambda^k_{\ k'}g_{ij;k}.
\end{equation}
It is clear from Eq.\ (\ref{dg}) that the covariant derivative of the metric vanishes in Cartesian coordinates, where $g_{ij}=\delta_{ij}$ and the Christoffel symbols are all zero. But Eq.\ (\ref{dgtrans}) shows that if $g_{ij;k}$ is zero in one coordinate system it is zero in all coordinate systems\footnote{This is a general property of tensors as a consequence of their transformation rule (\ref{ttrans}): if a tensor vanishes in one coordinate system it vanishes in  all of them.}; so we have 
\begin{equation} \label{dg=0}
g_{ij;k}=0.
\end{equation}
Similar reasoning starting from Eq.\ (\ref{dgup}) shows that
\begin{equation} \label{dgup=0}
g^{ij}_{\ \ ;k}=0.
\end{equation}
It is instructive to verify the property (\ref{dg=0}) explicitly for spherical polar coordinates using the expressions (\ref{dg}) and (\ref{gsph}) and the Christoffel symbols (\ref{Gammasph}).

It follows from Eqs.\ (\ref{dg=0}) and (\ref{dgup=0}) that 
$
V_{i;j}=(g_{ik}V^k)_{;j}=g_{ik;j}V^k+g_{ik}V^i_{\ ;j}=g_{ik}V^i_{\ ;j}
$ and
$ 
V^i_{\ ;j}=(g^{ik}V_k)_{;j}=g^{ik}_{\ \ ;j}V_k+g^{ik}V_{i;j}=g^{ik}V_{i;j}
$,
which are further examples of the general index lowering and raising operations (\ref{vcov}) and (\ref{vraise}). Since $V_{i;j}$ is a tensor, we can in fact raise either of its indices, so we have
\begin{equation} \label{raise;}
V_{i}^{\; ;j}=g^{jk}V_{i;k}.
\end{equation}
Equation (\ref{raise;}) defines what it means to have a covariant-derivative index in the upper position.

It is very important that Eqs.\ (\ref{dg=0}) and (\ref{dg}) serve to determine the Christoffel symbols in terms of the metric tensor. To see this, we insert the expression (\ref{dg=0}) in Eq.\ (\ref{dg}) and write it three times, with different permutations of the indices:
\begin{align}
g_{ij,k}&=\Gamma^l_{\ ik}g_{lj}+\Gamma^l_{\ jk}g_{il}, \nonumber \\
g_{ik,j}&=\Gamma^l_{\ ij}g_{lk}+\Gamma^l_{\ kj}g_{il}, \nonumber \\
-g_{jk,i}&=-\Gamma^l_{\ ji}g_{lk}-\Gamma^l_{\ ki}g_{jl},
\end{align}
In view of the symmetry (\ref{gsym}) of the metric tensor and the symmetry (\ref{Gammasym}) of the Christoffel symbols, the sum of the three lines gives
$
2g_{il}\Gamma^l_{\ jk}=g_{ij,k}+g_{ik,j}-g_{jk,i}
$.
Employing the inverse metric tensor $g^{ij}$ we finally find
\begin{equation} \label{Gammag}
\Gamma^i_{\ jk}=\frac{1}{2}g^{il}(g_{lj,k}+g_{lk,j}-g_{jk,l}).
\end{equation}
Equation (\ref{Gammag}) represents the most economic way of computing the Christoffel symbols in general. Again, it is highly instructive to take the spherical-polar metric (\ref{gsph}) and re-calculate the Christoffel symbols (\ref{Gammasph}) using the recipe (\ref{Gammag}). 


\subsection{Divergence and curl} 

We can utilize the expression (\ref{Gammag}) for the Christoffel symbols to find a very simple formula for the divergence of a vector in arbitrary coordinates. From the definitions (\ref{divv}) and (\ref{dvcov}) the divergence is
\begin{equation} \label{divvi}
 \bm{\nabla}\cdot\bm{V}=V^i_{\ ;i}=V^i_{\ ,i}+\Gamma^i_{\ ji}V^j,
\end{equation}
and inserting (\ref{Gammag}) gives
\begin{equation} \label{divvg}
 \bm{\nabla}\cdot\bm{V}=V^i_{\ ,i}+\frac{1}{2}\,g^{il}(g_{lj,i}-g_{ji,l})V^j+\frac{1}{2}\,g^{il}g_{li,j}V^j = 
V^i_{\ ,i}+\frac{1}{2}\,g^{il}g_{il,j}V^j,
\end{equation}
as is seen by relabeling summation indices and employing the symmetry of the metric tensor $g_{ij}$ and its inverse $g^{ij}$. Then we use a general property of the determinant of a matrix: the derivative of the determinant $g$ with respect to the matrix element $g_{ij}$ gives $g\,g^{ij}$ where $g^{ij}$ is the inverse matrix. (This property follows from Laplace's formula of expressing the determinant in terms of cofactors 
and Cramer's rule for the inverse matrix \cite{Stoll}.) Consequently,
\begin{equation}
\frac{1}{2}\,g^{il}g_{il,j}=
\frac{1}{2g}\,\frac{\partial g}{\partial g_{il}}\,g_{il,j}=
\frac{g_{,j}}{2g} = \frac{1}{\sqrt{g}}\left(\sqrt{g}\right)_{,j}
\end{equation}
and so we arrive at the simplified formula for the divergence
\begin{equation}\label{divvf}
 \bm{\nabla}\cdot\bm{V}=\frac{1}{\sqrt{g}}\left(\sqrt{g}\,V^i\right)_{,i}.
\end{equation}
The advantage of this formula compared to the definition (\ref{divvi}) is clear, because it only contains an ordinary partial derivative.
It is easy to see from the metric (\ref{gsph}) that in spherical polar coordinates the divergence formula (\ref{divvf}) gives the previous result (\ref{divsph}).

Another derivative operation familiar from Maxwell's equations is the curl of a vector. Like the vector product (\ref{vecprod})--(\ref{vecprodcpts}), the curl $ \bm{\nabla}\times\bm{V}$ is formed using the Levi--Civita tensor. Since the partial derivatives of Cartesian coordinates are covariant derivatives in general coordinates, we have
\begin{equation} \label{curl}
( \bm{\nabla}\times\bm{V})^i=\epsilon^{ijk}V_{k;j}.
\end{equation}
A simplification occurs in formula (\ref{curl}), however: from the covariant derivative (\ref{ducov}) one finds that the terms containing the Christoffel symbols cancel. We can therefore, in the case of the curl, use partial derivatives:
\begin{equation} \label{curlcpts}
( \bm{\nabla}\times\bm{V})^i=\epsilon^{ijk}V_{k,j}, \quad ( \bm{\nabla}\times\bm{V})_i=\epsilon_i^{\ jk}V_{k,j}.
\end{equation}
In this way we have found convenient expressions for the mathematical ingredients of Maxwell's equations, the divergence (\ref{divvf}) and the curl (\ref{curlcpts}).


\subsection{The Laplacian} 

We can use formula (\ref{divvf}) to find the divergence of the gradient vector $( \bm{\nabla}\psi)^i$ of a scalar $\psi$, which will give us the general expression for the Laplacian of a scalar. Note that we must use the gradient vector $( \bm{\nabla}\psi)^i$, rather than the gradient one-form $( \bm{\nabla}\psi)_i$, since the divergence is defined for vectors. From Eqs.\ (\ref{grad}) and (\ref{divvf}) we get
\begin{equation} \label{Laplacian}
 \bm{\nabla}^2\psi= \bm{\nabla}\cdot( \bm{\nabla}\psi)=( \bm{\nabla}\psi)^i_{\ ;i}=(g^{ij}\psi_{,j})_{;i}=\frac{1}{\sqrt{g}}\left(\sqrt{g}\,g^{ij}\psi_{,\,j}\right)_{,i}.
\end{equation}
For spherical polar coordinates we apply Eq.\  (\ref{gsph}) and easily obtain the well-known result
\begin{equation}    \label{Laplaciansph}
 \bm{\nabla}^2\psi=\frac{1}{r^2}\,\frac{\partial}{\partial r}\left(r^2\frac{\partial\psi}{\partial r}\right)+\frac{1}{r^2\sin\theta}\,\frac{\partial}{\partial \theta}\left(\sin\theta\frac{\partial\psi}{\partial\theta}\right)+\frac{1}{r^2\sin^2\theta}\,\frac{\partial^2\psi}{\partial\phi^2}. 
\end{equation}
Any reader who has had the misfortune of having to work out this expression without using formula (\ref{Laplacian}) from differential geometry should now appreciate the power of the machinery we have developed. As a further salutary example, consider the monochromatic wave equation for the electric field
\begin{equation}  \label{Ewave}
 \bm{\nabla}^2\bm{E}+\frac{\omega^2}{c^2}\bm{E}=0.
\end{equation}
This is the equation as it is usually written in the textbooks, but the notation in the first term is treacherous for the student. When the wave equation (\ref{Ewave}) is considered in curvilinear coordinates, for example to find the radiation modes in a waveguide \cite{Jackson}, the student is apt to think that the components $( \bm{\nabla}^2\bm{E})^i$ of the vector $ \bm{\nabla}^2\bm{E}$ are the Laplacians $ \bm{\nabla}^2(E^i)$ of the electric field components $E^i$ --- this is {\it not} true. The partial derivatives of Cartesian coordinates get replaced by covariant derivatives in a general system, so the vector $ \bm{\nabla}^2\bm{E}$ has in fact components $g^{jk}E^i_{\ ;j;k}=E^{i;j}_{\ \ \,;j}$, which are not the Laplacians $(g^{jk}E^i_{\ ,j})_{;k}$ of $E^i$. The three Laplacians $(g^{jk}E^i_{\ ,j})_{;k}$ are not the components of a vector. The correct wave equation (\ref{Ewave}) is thus
\begin{equation} 
E^{i;j}_{\ \ \,;j}+\frac{\omega^2}{c^2}E^i=0.
\end{equation}
Expressed in curvilinear coordinates the wave equation (\ref{Ewave}) provides three {\it coupled} equations for the components $E^i$. In practice, the textbooks ensure that they only deal with those components of the wave equation (\ref{Ewave}) for which $( \bm{\nabla}^2\bm{E})^i$ {\it is} equal to $ \bm{\nabla}^2(E^i)$, such as the $z$-component in cylindrical coordinates. But this only confirms in the student's mind the false believe that $( \bm{\nabla}^2\bm{E})^i$ is the same as $ \bm{\nabla}^2(E^i)$, and in the future he or she may come a cropper as a consequence.


\subsection{Geodesics and curvature} 

We are now acquainted with the mathematics of arbitrary coordinates in three-dimensional Euclidean space. But there is an extra bonus for our efforts --- we can also deal with {\it curved} space. A notion of curvature on a space is naturally induced by the metric tensor: if coordinates exist in which the metric has the Euclidean form (\ref{dscar}) the space is {\it flat}, otherwise it is {\it curved}. Curvature of a three-dimensional space is difficult to visualize, but a familiar curved two-dimensional space is provided by the surface of a sphere. If the sphere has radius $a$, then from Eq.\ (\ref{dssph}) the metric on the surface is
\begin{equation} \label{dssphere}
\mathrm{d}s^2=a^2(\mathrm{d}\theta^2+\sin^2\theta\, \mathrm{d}\phi^2).
\end{equation}
There is no transformation to coordinates $\{x^1,x^2\}$ in which this metric takes the Euclidean form
\begin{equation} \label{2Dcar}
\mathrm{d}s^2=(\mathrm{d}x^1)^2+(\mathrm{d}x^2)^2.
\end{equation}
Note that the crucial feature of the sphere that prevents its metric being transformed to the Euclidean (\ref{2Dcar}) is not that it is a closed space with a finite area; we can consider any finite patch of the sphere, ignoring its global structure, and we would still be unable to find a coordinate transformation to Eq.\ (\ref{2Dcar}). The crucial fact that makes the sphere a curved space is that we cannot form a patch of the sphere from a flat piece of paper without stretching the paper. Consider, in contrast, the surface of a cylinder. A cylinder can be formed by rolling up a flat piece of paper, so this space must be flat. The metric on a cylinder
of radius $a$, in cylindrical polar coordinates with $r=a$, is
\begin{equation} \label{cyl}
\mathrm{d}s^2=\mathrm{d}z^2+a^2\, \mathrm{d}\phi^2=\mathrm{d}z^2+(\mathrm{d}(a\phi))^2,
\end{equation}
which is of the Eulidean form (\ref{2Dcar}), proving that it is flat. It is not important that the coordinate $\phi$ in (\ref{cyl}) is periodic; curvature is a local property of a space, in contrast to its topology. A cylinder and a plane have different topologies but they have the same curvature, namely zero.

How can we quantify the curvature of a space? We first need to generalize the notion of a straight line to the case of curved spaces like the sphere. The key property of a straight line joining two points in flat space is that it is the shortest path between those points. In a curved space we can still construct the shortest line between two points and this is called a {\it geodesic}. For a sphere, the geodesics are the great circles, {\it i.e.}\ the circles whose centres are the centre of the sphere. A geodesic is a curve $x^i(\xi)$ in space, with parameter $\xi$. The length $s$ of the curve between two points $\xi=\xi_1$ and $\xi=\xi_2$ is the integral of the line element  (\ref{metric}),
\begin{equation} \label{s}
s=\int^{\xi_2}_{\xi_1}ds=\int^{\xi_2}_{\xi_1}\sqrt{g_{ij}\mathrm{d}x^i(\xi)\,\mathrm{d}x^j(\xi)}=\int^{\xi_2}_{\xi_1}\sqrt{g_{ij}\frac{\mathrm{d}x^i(\xi)}{\mathrm{d}\xi}\frac{\mathrm{d}x^j(\xi)}{\mathrm{d}\xi}}\,\mathrm{d}\xi.
\end{equation}
For the curve $x^i(\xi)$ to be a geodesic the length (\ref{s}) must be a minimum, so the variation $\delta s$ must vanish when we perform a variation $\delta x^i(\xi)$ of the curve (maintaining the parameter values $\xi_1$ and $\xi_2$ at the endpoints). But this is completely equivalent to the principle of least action  in mechanics \cite{LL1} with ``Lagrangian''
\begin{equation} \label{L}
L=\sqrt{g_{ij}\dot{x}^i\dot{x}^j}, \quad
\dot{x}^i \equiv \frac{\mathrm{d}x^i(\xi)}{\mathrm{d}\xi}.
\end{equation}
The geodesic is therefore given by the Euler-Lagrange equations:
\begin{equation} \label{geo1}
0=
\frac{\mathrm{d}}{\mathrm{d} \xi}\,
\frac{\partial L}{\partial \dot{x}^i} -  
\frac{\partial L}{\partial x^i} =
\frac{\mathrm{d}}{\mathrm{d} \xi}\,
\left(\frac{1}{L}\,g_{ij} \dot{x}^j\right) - 
\frac{1}{2L}\,g_{lj,i} \dot{x}^l\dot{x}^j.
\end{equation}
Equation (\ref{geo1}) determines a geodesic curve once an initial tangent vector (direction of the geodesic) $\mathrm{d}x^i(\xi)/\mathrm{d}\xi$ is specified. The parameter $\xi$ is arbitrary, but we obtain a much simpler equation if we choose $\xi$ to be the distance $s$ along the geodesic, because in this case $L=1$. We calculate
\begin{equation}
\frac{\mathrm{d}}{\mathrm{d} \xi}\,
\frac{\partial L}{\partial \dot{x}^i} =
\frac{\mathrm{d}}{\mathrm{d} \xi}\, g_{ij}
\frac{\mathrm{d}x^i}{\mathrm{d}s} =
g_{ij}\frac{\mathrm{d}^2x^i}{\mathrm{d}s^2} +
g_{lj,i}\frac{\mathrm{d}x^l}{\mathrm{d}s}
\frac{\mathrm{d}x^j}{\mathrm{d}s}
\end{equation}
and insert this result in the Euler-Lagrangian equations (\ref{geo1}) that
we matrix-multiply with $g^{ij}$ from the left.
Recalling the expression (\ref{Gammag}) for the Christoffel symbols we obtain the simple geodesic equation
\begin{equation} \label{geo3}
\frac{\mathrm{d}^{\,2}x^i(s)}{\mathrm{d}s^2}+\Gamma^i_{\ jk}\frac{\mathrm{d}x^j(s)}{\mathrm{d}s}\frac{\mathrm{d}x^k(s)}{\mathrm{d}s}=0.
\end{equation}
In Euclidean space the geodesic equation (\ref{geo3}) always gives the equation of a straight line, whether this line is expressed in Cartesian or curved coordinates, as the reader may verify for the case of the spherical polar coordinates using the Christoffel symbols (\ref{Gammasph}). The reader may also show that the geodesics for the sphere, with metric (\ref{dssphere}), are the great circles.

We find a simple interpretation for the geodesic equation (\ref{geo3}) using the concept of parallel transport discussed in Sec.\ 3.7. Note that the vector $\mathrm{d}x^i/\mathrm{d}s$ appearing in the geodesic equation is the tangent vector to the geodesic curve $x^i(s)$. Equation (\ref{geo3}) thus means that the covariant increment of the tangent vector along $\mathrm{d}x^i(s)$ is zero: the tangent vector is parallel-transported along the geodesic curve:  
\begin{equation} \label{geo4}
\mathrm{D}U^i=\left(\nabla_jU^i\right)\mathrm{d}x^j(s)=0, \quad
U^i\equiv \frac{\mathrm{d}x^i(s)}{\mathrm{d}s}.
\end{equation}
Thus, a geodesic parallel-transports its tangent vector. In flat space, as one moves along a geodesic (straight line), the tangent vectors at all points are parallel. Equation (\ref{geo4}) generalizes this property for geodesics in curved space. Geodesics are lines of inertia.

\begin{figure}[t] 
\vspace*{-15mm}
\includegraphics[width=30.0pc]{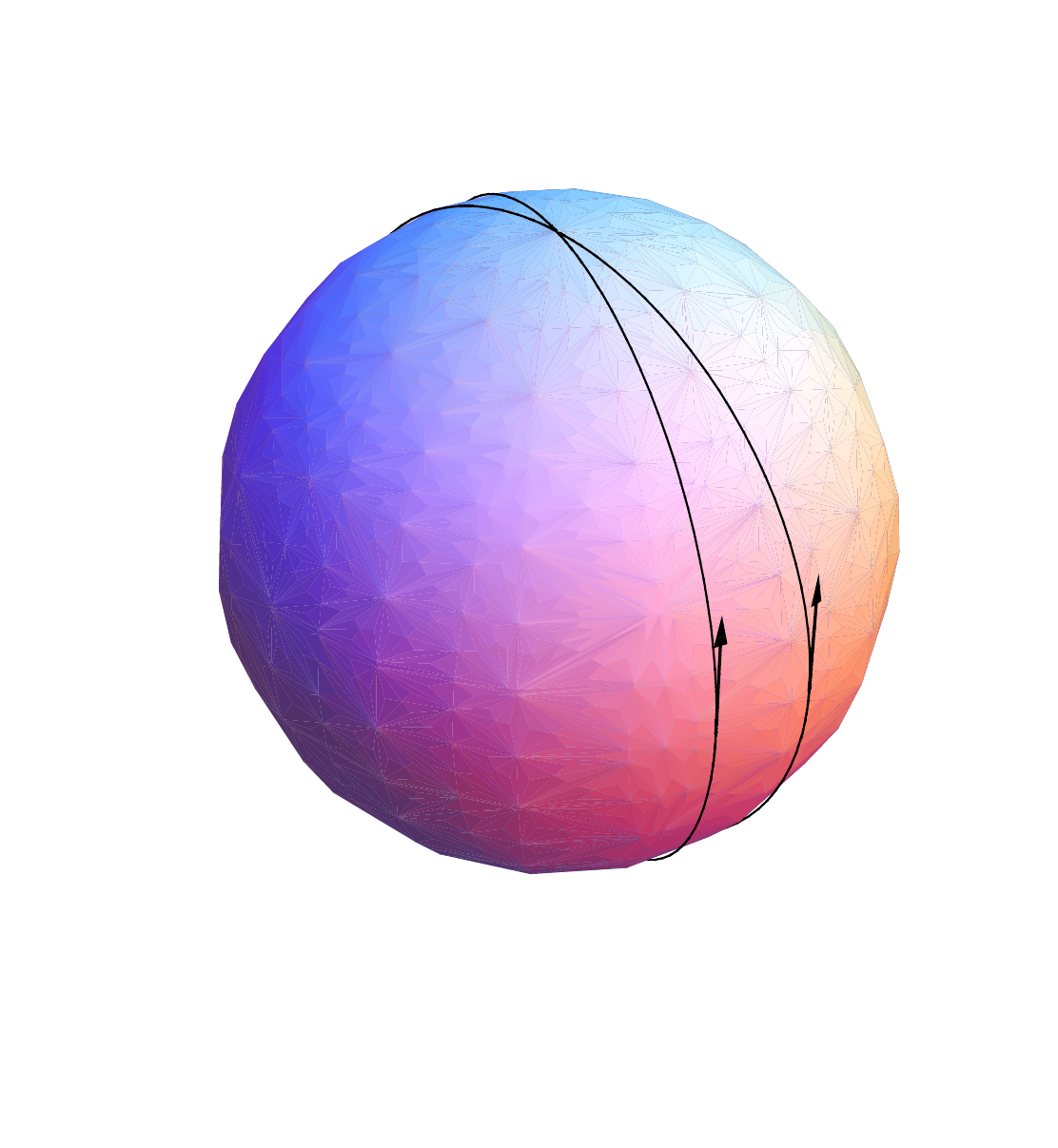}
\vspace*{-30mm}
\caption{\small{Two geodesics (great circles) on the sphere.  The separation as a function of distance from a meeting point is not linear; as one moves along, the geodesics have an ``acceleration'' towards each other. Tangent vectors to the geodesics are drawn at the points where the separation has reached a maximum. These vectors are parallel in three-dimensional space.}}  \label{fig:sphere}
\end{figure}

The behaviour of geodesics in a space can be used to quantify the amount of curvature. In flat space two geodesics (straight lines) either have a constant separation (parallel lines) or the separation distance changes at a constant rate, {\it i.e.}\ it changes linearly with distance along the lines. The second derivative of the separation with respect to distance along the geodesics is therefore zero. In curved space by contrast, this second derivative does not vanish, a phenomenon called {\it geodesic deviation}; it is used to measure the curvature. As an example, consider again the surface of a sphere, depicted in Fig.~\ref{fig:sphere}.
We see clearly that the separation of any two geodesics (great circles) does not change linearly with distance along the geodesics: for, the separation increases from zero to a maximum and then decreases to zero again. At the points where the tangent vectors to the geodesics are drawn in Fig.\ \ref{fig:sphere}, the separation is a maximum and the tangent vectors are parallel as viewed in the three-dimensional space in which the sphere is embedded. The geodesics are parallel at these points, inasmuch as a notion of parallelism can be introduced on the sphere, but these ``parallel lines'' meet! Thus the postulates of Euclidean geometry do not hold in this space: it is curved. As another example, consider the geodesics on the surface of a cylinder. These are drawn by simply cutting open the cylinder into a flat sheet, drawing straight lines on the sheet and rolling it up again to reform the cylinder. Clearly the geodesics behave just as in the plane: there is no geodesic deviation (second derivative of separation with respect to distance along geodesics is zero), showing again that a cylinder is a flat space.

\begin{figure}[t] 
\vspace*{-15mm}
\includegraphics[width=30.0pc]{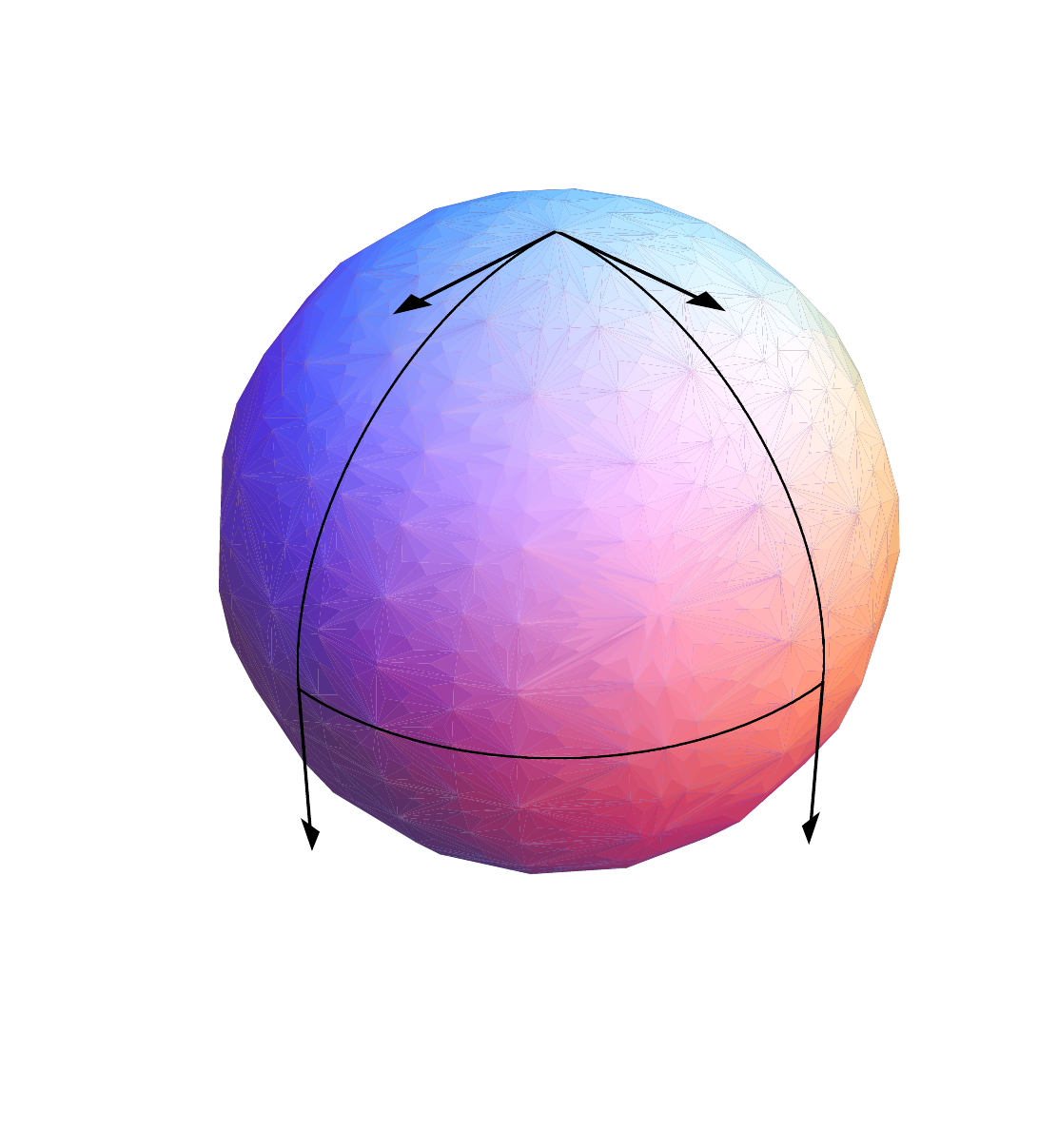}
\vspace*{-30mm}
\caption{\small{Parallel transport around a closed loop on the sphere. A vector at the North Pole is parallel transported to the equator along the geodesic to which it is tangent. Since a geodesic parallel transports its tangent vector, the vector is still tangent to the geodesic at the equator. The vector is then parallel transported a quarter of the distance around the equator (also a geodesic). Parallel transport preserves the vector, as much as the curvature allows, so it stays at the same angle to the equator. Finally the vector is parallel transported back to the North Pole along a geodesic. On return to the North Pole the vector is rotated by 90 degrees with respect to the initial vector. This rotation is completely specified by the Riemann curvature tensor. Note that on a plane (zero Riemann tensor) this kind of operation does not rotate the vector.}}  \label{fig:sphere2}
\end{figure}

How are we to compute the geodesic deviation? We need to consider a family of geodesics $x^i(s,\delta)$, where the parameter $\delta$ labels (continuously) the geodesics in the family and for fixed $\delta$ the parameter $s$ is the distance along a geodesic. Note that $\delta$ is not the distance between geodesics since two fixed values of $\delta$ determine two geodesics that may be moving apart. The vector joining two infinitesimally separated geodesics with equal parameter value $s$ is, however, given by 
$V^i\,\mathrm{d}\delta$ with
\begin{equation} \label{vsdef}
V^i=\frac{\partial x^i(s,\delta)}{\partial \delta}.
\end{equation}
Now, what we are interested in is the rate of change of this joining vector as $s$ increases, specifically its second covariant increment along the geodesic, $\mathrm{D}^2V^i$. To compute the geodesic deviation we use the notation (\ref{Dv}) of $\mathrm{D}$ and utilize the relationship
\begin{equation} 
U^k\nabla_kV^i
= \frac{\partial x^k}{\partial s}\nabla_k
\frac{\partial x^i}{\partial \delta}
= \frac{\partial^2x^i}{\partial s\, \partial\delta} + 
\Gamma^i_{\ jk} \frac{\partial x^k}{\partial s}\,
\frac{\partial x^j}{\partial \delta}
= \frac{\partial x^k}{\partial \delta}\nabla_k
\frac{\partial x^i}{\partial s}
= V^k\nabla_kU^i.
\end{equation}
Then we calculate $\mathrm{D}^2V^i = U^j \nabla_j (U^k \nabla_k V^i) \mathrm{d}s^2$, regarding the $\nabla_j$ as operators obeying the Leibniz rule of derivatives (\ref{Leibnizuv}),
\begin{eqnarray}
U^j \nabla_j (U^k \nabla_k V^i)
& = & U^j \nabla_j (V^k \nabla_k U^i)
\nonumber\\
& = & U^jV^k\nabla_j\nabla_kU^i + 
\left(U^j\nabla_jV^k\right)\left(\nabla_kU^i\right)
\nonumber\\
& = & U^jV^k\nabla_j\nabla_kU^i + 
V^j\left(\nabla_jU^k\right)\left(\nabla_kU^i\right)
\nonumber\\
& = & U^jV^k 
\left(\nabla_j\nabla_k - \nabla_k\nabla_j\right) U^i
+ V^k\nabla_k \left(U^j \nabla_j U^i\right)
\nonumber\\
& = & U^jV^k 
\left(\nabla_j\nabla_k - \nabla_k\nabla_j\right) U^i
\end{eqnarray}
where we applied the geodesic equation (\ref{geo4}) in the last step.
If the commutator of the covariant derivatives vanishes the geodesic deviation  is zero and the space is flat. What is the meaning of this commutator? Recall the concept of parallel transport. Imagine we move a vector $A^i$ from the point $x^i$ to the infinitesimally close neighbor $x^i+\mathrm{d}x^i$ and then again to the next neighbor by another increment $\mathrm{d}y^i$;
finally we close a loop in moving the vector back by $-\mathrm{d}x^i$ followed by $-\mathrm{d}y^i$: the vector would change as 
$(\nabla_j\nabla_k - \nabla_k\nabla_j)A^i\,
\mathrm{d}x^j \mathrm{d}y^k$. Any closed loop we can imagine as consisting of patches of infinitesimal loops; so, in a non-Euclidean geometry, a vector transported along a closed loop does not return to itself, see Fig.~\ref{fig:sphere2}. Foucault's pendulum \cite{Hart}, transported with the rotating Earth on a sphere, does not return to its original oscillation direction after one loop (24 hours). 

To calculate the commutator $(\nabla_j\nabla_k - \nabla_k\nabla_j)A^i$ we use the covariant derivative (\ref{dmix}) of the mixed tensors $\nabla_kA^i$ and $\nabla_jA^i$ first and apply then the formula (\ref{Dv}) for the covariant derivative of a vector. We get, utilizing the symmetry (\ref{Gammasym}) of the Christoffel symbols, 
\begin{eqnarray}
\left(\nabla_j\nabla_k - \nabla_k\nabla_j\right)A^i 
&=& 
\left(\nabla_k A^i\right)_{,j} + \Gamma^i_{\ lj}\left(\nabla_k A^l\right) - 
\left(\nabla_j A^i\right)_{,k} - \Gamma^i_{\ lk}\left(\nabla_j A^l\right) 
\nonumber\\
& = &
\left(\Gamma^i_{\ lk}A^l\right)_{,j} + \Gamma^i_{\ lj}\left(A^l_{\ ,k} + \Gamma^l_{\ km} A^m\right) 
\nonumber\\
& & \quad\quad -
\left(\Gamma^i_{\ lj}A^l\right)_{,k} - \Gamma^i_{\ lk}\left(A^l_{\ ,j} + \Gamma^l_{\ jm} A^m\right)
\nonumber\\
& = &
\left(\Gamma^i_{\ lk,j}-\Gamma^i_{\ lj,k}
+ \Gamma^i_{\ mj} \Gamma^m_{\ kl}
- \Gamma^i_{\ mk} \Gamma^m_{\ jl}
\right)A^l.
\label{ccc}
\end{eqnarray}
The quantity in brackets on the right-hand side of this formula depends only on the geometry of the space (the metric) and determines the amount of curvature. It transforms as a tensor with four indices (\ref{ttrans}), because the left-hand side of Eq.\ (\ref{ccc}) is a tensor with three indices and the right-hand side contains a contraction over a vector. The space is flat if and only if this quantity is zero, giving zero geodesic deviation. It is one of the most important objects in geometry, the {\it Riemann curvature tensor}\footnote{The tensor character (\ref{ttrans}) of $R^i_{\ jkl}$ can be verified using the transformation properties of the Christoffel symbols.}:
\begin{equation} \label{riemann}
R^i_{\ jkl}\equiv\Gamma^i_{\ jl,k}-\Gamma^i_{\ jk,l}+\Gamma^i_{\ mk}\Gamma^m_{\ jl}-\Gamma^i_{\ ml}\Gamma^m_{\ jk}.
\end{equation}
The Riemann tensor describes both the geodesic deviation 
\begin{equation}
\mathrm{D}^2 V^i = R^i_{\ jkl}V^l \mathrm{d}x^j\, \mathrm{d}x^k
\end{equation}
and the result of loops in parallel transport
\begin{equation}\label{commutator}
\left(\nabla_j\nabla_k - \nabla_k\nabla_j\right)A^i 
= R^i_{\ ljk} A^l.
\end{equation}
In Euclidean space, no matter how complicated the coordinate system we use, the Riemann tensor will always turn out to be zero (the reader may check this for spherical polar coordinates). On the other hand, for a curved space like the sphere, the Riemann tensor will not vanish in any coordinate system. In arbitrary coordinates $x^A=\{x^1,x^2\}$ a sphere of radius $a$ has a Riemann tensor
\begin{equation}
R^A_{\ BCD}=\frac{1}{a^2}(\delta^A_{\ C}g_{BD}-\delta^A_{\ D}g_{BC}),
\end{equation}
as can be checked for the case of the coordinates (\ref{dssphere}). Incidentally, besides the sphere, all readers are familiar with the physical effect of another curved space: the space-time geometry in which they reside, the geometry of gravity \cite{LL2,Telephone}. Here space-time is curved, not only three-dimensional space. This space-time has a Riemann tensor whose largest components at the surface of the Earth are of the order of $10^{-23}\text{m}^{-2}$. This space-time curvature is the reason the reader does not float off into space.


\section{Maxwell's equations}

After having discussed the mathematical machinery of differential geometry we are now well-prepared to formulate the foundations of electromagnetism, Maxwell's equations \cite{Jackson}.  In empty space, the Maxwell equations for the electric field strength $\bm{E}$ and the magnetic induction $\bm{B}$ are
\begin{gather}
 \bm{\nabla}\cdot\bm{E}=\frac{\rho}{\varepsilon_0}, \quad  \bm{\nabla}\cdot\bm{B}=0, 
\nonumber\\[5pt]
 \bm{\nabla}\times\bm{E} = -\frac{\partial \bm{B}}{\partial t}, \quad  \bm{\nabla}\times\bm{B} = \frac{1}{c^2}\,\frac{\partial \bm{E}}{\partial t}
+ \mu_0\bm{j}.
\label{maxwell}
\end{gather}
We use SI units with electric permittivity $\varepsilon_0$, magnetic permeability $\mu_0$ and speed of light $c$ in vacuum, $\varepsilon_0\mu_0=c^{-2}$. Charge and current densities are denoted by $\rho$ and $\bm{j}$.
In the following we express Maxwell's equations in arbitrary coordinates and arbitrary geometries. We show how a geometry appears as a medium and how a medium appears as a geometry. We develop the concept of transformation optics where we use the freedom of coordinates to describe transformation media as elegantly as possible. Furthermore, we generalize transformation optics to space-time geometries. We also return to our starting point, Fermat's principle. 


\subsection{Geometries and media}

Maxwell's equations (\ref{maxwell}) contain curls and divergences. 
Using the expressions (\ref{divvf}) and (\ref{curlcpts}) from differential geometry we can now write these in arbitrary coordinates:
\begin{gather}
\frac{1}{\sqrt{g}}\left(\sqrt{g}\,E^i\right)_{,i}=\frac{\rho}{\varepsilon_0}, \quad \frac{1}{\sqrt{g}}\left(\sqrt{g}\,B^i\right)_{,i}=0, \nonumber \\[5pt]
\epsilon^{ijk}E_{k,j}= -\frac{\partial B^i}{\partial t}, \quad \epsilon^{ijk}B_{k,j}= \frac{1}{c^2}\,\frac{\partial E^i}{\partial t}+\mu_0j^i. \label{max1}
\end{gather}
This form of Maxwell's equation is also valid in arbitrary geometries, {\it i.e.} in curved space, for the following reason: any geometry, no matter how curved, is locally flat --- at each spatial point we can always construct an {\it infinitesimal} patch of a Cartesian coordinate system, although these local systems do not constitute a single global grid. For each locally flat piece we postulate Maxwell's equations (\ref{maxwell}), and in writing these equations in arbitrary coordinates we naturally express them in a global frame.

Let us rewrite the form (\ref{max1}) of Maxwell's equations with all the vector indices in the lower position and the Levi-Civita tensor expressed in terms of the permutation symbol according to formula (\ref{LCgen}):
\begin{gather}
\left(\sqrt{g}\,g^{ij}E_j\right)_{,i}=\frac{\sqrt{g}\,\rho}{\varepsilon_0}, \quad \left(\sqrt{g}\,g^{ij}B_j\right)_{,i}=0, \nonumber\\[5pt]
[ijk]E_{k,j}= -\frac{\partial (\pm\sqrt{g}\,g^{ij}B_j)}{\partial t}, \quad [ijk]B_{k,j}= \frac{1}{c^2}\,\frac{\partial (\pm\sqrt{g}\,g^{ij}E_j)}{\partial t}+\mu_0\sqrt{g}\,j^i. \label{max2}
\end{gather}
In this form, Maxwell's equations in empty space, but in curved coordinates or curved geometries, resemble the {\it macroscopic} Maxwell equations in dielectric media \cite{Jackson},
\begin{gather}
 \bm{\nabla}\cdot\bm{D}=\rho, \quad  \bm{\nabla}\cdot\bm{B}=0, \nonumber\\[5pt]
 \bm{\nabla}\times\bm{E} = -\frac{\partial \bm{B}}{\partial t}, \quad  \bm{\nabla}\times\bm{H} = \frac{\partial \bm{D}}{\partial t} \label{macbf2}
+ \mu_0\bm{j},
\end{gather}
written in right-handed Cartesian coordinates:
\begin{gather}
D^i_{\ ,i}=\rho, \quad B^i_{\ ,i}=0, \nonumber \\[5pt]
[ijk]E_{k,j}= -\frac{\partial B^i}{\partial t}, \quad [ijk]H_{k,j}= \frac{\partial D^i}{\partial t}+j^i. \label{mac2}
\end{gather}
In fact, the empty-space equations (\ref{max2}) can be expressed exactly in the macroscopic form (\ref{mac2}) if we replace $B_i$ in the free-space equations by $H_i/\mu_0$, rescale the charge and current densities, and take the constitutive equations
\begin{gather} 
D^i=\varepsilon_0\varepsilon^{ij}E_j, \quad B^i=\mu_0\mu^{ij}H_j.  \label{cons1} \\
\varepsilon^{ij}=\mu^{ij}=\pm\sqrt{g}\,g^{ij}. \label{cons2}
\end{gather}
Consequently, the {\it empty-space} Maxwell equations in {\it arbitrary} coordinates and geometries are equivalent to the {\it macroscopic} Maxwell equations in {\it right-handed Cartesian} coordinates. Geometries appear as dielectric media. The electric permittivities $\varepsilon^{ij}$ and magnetic permeabilities $\mu^{ij}$ are identical --- these media are {\it impedance-matched} \cite{Jackson}, and the $\varepsilon^{ij}$ and $\mu^{ij}$ are matrices --- the media are anisotropic. Spatial geometries appear as anisotropic impedance-matched media. 

The converse is also true: anisotropic impedance-matched media appear as geometries. We easily derive this statement from the constitutive equations (\ref{cons2}): calculate the determinant $\det \epsilon$ of $\varepsilon^{ij}$. The result is $\det \epsilon = \pm \sqrt{g}$, the factor in front of the metric in the constitutive equations (\ref{cons2}). So we obtain from the constitutive equations of a geometry the metric tensor of a medium
\begin{equation}\label{epsg}
g^{ij} = \frac{\varepsilon^{ij}}{\det \epsilon}.
\end{equation}
For general $\varepsilon^{ij}=\mu^{ij}$, this geometry is curved. But, if and only if the Riemann tensor (\ref{riemann}) vanishes the spatial geometry is flat. In such a case there exists a coordinate transformation of physical space where Maxwell's equations are purely Cartesian, where space is flat and empty. The electromagnetic fields in real, physical space are transformed fields --- the results of coordinate transformations. Media that perform such a feat are called {\it transformation media}. 


\subsection{Transformation media}

\begin{figure}[h]
\begin{center}
\includegraphics[width=30.0pc]{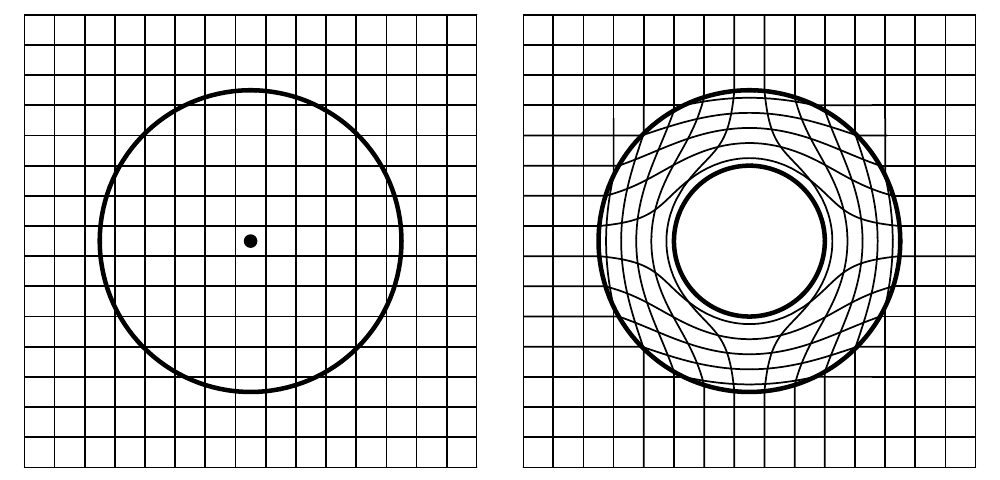}
\caption{
\small{
Transformation media implement coordinate transformations.
The left figure shows the Cartesian grid of electromagnetic space
that is mapped to the curved grid of physical space shown in the right figure.
The physical coordinates enclose a hole that is made invisible
in electromagnetic space (where it shrinks to the point indicated there).
Consequently, a medium that performs this transformation 
acts as an invisibility device.  
The case illustrated in the figure corresponds to 
the transformation
$r=R_1 + r'(R_2-R_1)/R_2$ in cylindrical coordinates
where the prime refers to the radius in electromagnetic space.
The region with radius $R_1$ is invisible; $R_2-R_1$
describes the thickness of the cloaking layer.
}
\label{fig:twospaces}
}
\end{center}
\end{figure}

Transformation media implement coordinate transformations in Maxwell's equations. Note carefully how this interpretation of Maxwell's equations (\ref{max1}) works: we write the free-space equations in coordinates that are not right-handed Cartesian, but we then interpret these equations as being in a right-handed Cartesian system with an effective medium (\ref{cons2}). This sounds a bit paradoxical, but the way to think of it is to imagine {\it two} different spaces as well as two different coordinate systems, see Fig.\ \ref{fig:twospaces}. In the first space, which we call {\it electromagnetic space}, we have no medium and we write the empty-space Maxwell equations in right-handed Cartesian coordinates. We then perform a transformation that gives us a non-trivial effective medium (\ref{cons2}) and we interpret the transformed coordinates as being right-handed Cartesian in a {\it new} space, physical space, which contains the medium (\ref{cons2}). The Cartesian grid in electromagnetic space will deform under the transformation and this deformed grid shows ray trajectories in physical space.  

There are two aspects of transformation media that make them highly significant. First, we know a good deal about solutions of Maxwell's equation in vacuum (light rays travel in straight lines, etc.) and to find the effect of the medium we can just take a vacuum solution in electromagnetic space and transform to physical space using the coordinate transformation that defines the medium: the transformed fields are a solution of the macroscopic Maxwell equations in physical space. Second, since a transformation medium is defined by a coordinate transformation, we can use this as a design tool to find materials with remarkable electromagnetic properties.

Some readers may have nagging doubts about the juxtaposition of the mathematical tools of general relativity with the attribution of a physical significance to coordinate transformations. For a relativist coordinate systems have no physical meaning; the geometry of the space is the important thing, and that is independent of the coordinate grid one chooses to cover the space. But here we wish to consider materials that, as far as electromagnetism is concerned, perform active coordinate transformations. In this theory, the coordinate transformation {\it is} physically significant, it describes completely the macroscopic electromagnetic properties of the material, and differential geometry is just as useful for these purposes as it is in general relativity. 

\begin{figure}[h]
\begin{center}
\includegraphics[width=25.0pc]{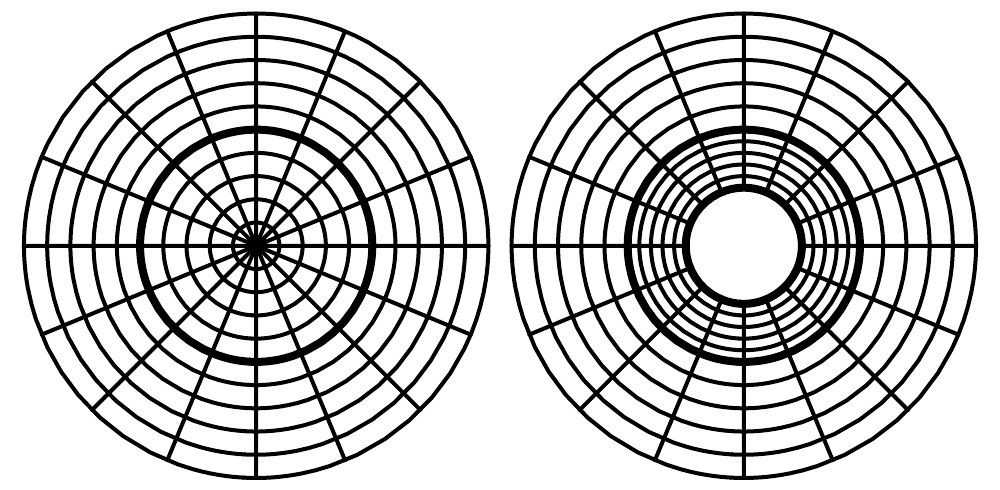}
\caption{
\small{
Transformation media in cylindrical coordinates.
The figure shows the transformation of Fig.\ \ref{fig:twospaces} in cylindrical coordinates. These are the coordinates best adapted to this case.
}
\label{fig:spider}
}
\end{center}
\end{figure}

In our description of transformation media, the starting point of the theory was a right-handed Cartesian system in electromagnetic space; any non-trivial transformation from this system gives an effective medium in physical space. It is, however, often convenient to adjust coordinates to the particular situation under investigation, see Fig.\ \ref{fig:spider}. We should be allowed to use any coordinates we wish in electromagnetic space. In order to implement this freedom of coordinates, we generalize the theory. Suppose that we describe electromagnetic space by a curvilinear system such as cylindrical or spherical polar coordinates; any deformation of this system through a coordinate transformation is to be interpreted as a medium, but to describe the electromagnetism in the presence of this medium we employ the original curvilinear grid. 

Let $x'^{i}$ be the curvilinear system in electromagnetic space and we denote its metric tensor by $\gamma_{ij}$. Then in electromagnetic space the empty-space Maxwell equations (\ref{max1}) are:
\begin{gather}
\frac{1}{\sqrt{\gamma}}(\sqrt{\gamma}\,E^{i})_{,i}=\frac{\rho}{\varepsilon_0}, \quad \frac{1}{\sqrt{\gamma}}(\sqrt{\gamma}\,B^{i})_{,i}=0, \nonumber \\[5pt]
[ijk]E_{k,j}= -\frac{\partial (\sqrt{\gamma}\, B^{i})}{\partial t}, \quad [ijk]B_{k,j}= \frac{1}{c^2}\,\frac{\partial (\sqrt{\gamma}\,E^i)}{\partial t}+\mu_0j^{i}, \label{maxem2}
\end{gather}
where we have made the metric dependence of the Levi-Civita tensor explicit. Now we perform a coordinate transformation and, as before, we interpret the resulting equations as being macroscopic Maxwell equations written in the {\it same} (curvilinear) system we started with, but in physical space. We cast the equations (\ref{max2}) in physical space as the macroscopic equations (\ref{macbf2}) in the curvilinear system with the metric $\gamma_{ij}$:
\begin{gather}
\left(\sqrt{\gamma}\,D^i\right)_{,i}=\sqrt{\gamma}\,\rho, \quad \left(\sqrt{\gamma}\,B^i\right)_{,i}=0, \nonumber \\[5pt]
[ijk]E_{k,j}= -\frac{\partial (\sqrt{\gamma}\,B^i)}{\partial t}, \quad [ijk]H_{k,j}= \frac{\partial (\sqrt{\gamma}\,D^i)}{\partial t}+\mu_0\sqrt{\gamma}\,j^i. \label{maccur2}
\end{gather}
By the same reasoning as before, we can interpret the free-space equations (\ref{max2}) as macroscopic equations (\ref{maccur2}) written in the curvilinear system if we rescale the charge and current densities and take the constitutive equations (\ref{cons1}) with
\begin{equation} 
\varepsilon^{ij}=\mu^{ij}=\pm\frac{\sqrt{g}}{\sqrt{\gamma}}\,g^{ij}. \label{constspace}
\end{equation}
If we wish to implement a certain coordinate transformation, formula (\ref{constspace}) gives a simple and efficient recipe for calculating the required material properties in arbitrary coordinates
\cite{GREE}.


\subsection{Wave equation and Fermat's principle}

Impedance-matched media establish geometries and geometries appear as impedance-matched media.  In general these media are anisotropic, but we still expect that light rays follow a version of Fermat's principle (\ref{eq:fermat}) of the extremal optical path. The metric $g_{ij}$ should set the measure of optical path length. So we anticipate that light rays follow a geodesic with respect to the geometry given by the material properties (\ref{epsg}). How do we deduce Fermat's principle from Maxwell's equations? First we write down the wave equation for monochromatic fields with frequency $\omega$, the refined version of the Helmholtz equation (\ref{eq:helm}). Any electromagnetic field consists of a superposition of monochromatic fields,  if  $g_{ij}$ does not change in time, which we assume, and also that there are no external charges and currents in the region we consider. We obtain from Maxwell's equations (\ref{max1}):
\begin{equation}
\epsilon^{ijk}\left(\epsilon_{klm} E^{m;l}\right)_{;k} = 
\frac{\omega^2}{c^2}\, E^i.
\end{equation}
Note that we expressed the derivatives as covariant derivatives (semi-colons instead of commas) in anticipation of simplifications to come. The covariant derivative of the Levi-Civita tensor $\epsilon_{klm}$ is zero, because it is determined by the metric tensor which has vanishing covariant derivative \cite{Telephone}. Consequently, using the notation (\ref{Dv}) for the covariant derivatives
\begin{equation}\label{wave0}
\frac{\omega^2}{c^2}\, E^i = 
\epsilon^{ijk} \epsilon_{klm} \nabla_k \nabla^l E^m =
\left(\delta^i_l\delta^j_m-\delta^i_m\delta^j_l\right)
\nabla_k \nabla^l E^m =
\nabla_j \nabla^i E^j - \nabla^j \nabla_j E^i,
\end{equation}
where we applied formula (\ref{baccabgen}) for the double vector product. The first term in the right-hand side of Eq.\ (\ref{wave0}) resembles the covariant derivative of the divergence, $\nabla_j E^j$, that would vanish since the charge density is zero. But covariant derivatives do not commute in general: their commutator is given by the Riemann tensor (\ref{commutator}). So we obtain, lowering the index $i$,
\begin{equation}
\frac{\omega^2}{c^2}\, E_i = 
\left(\nabla_k \nabla_i - \nabla_k \nabla_i\right) E^k - \nabla^j \nabla_j E_i,
\end{equation}
and arrive at the {\it wave equation} \cite{Piwnicki1,Piwnicki2}
\begin{equation}\label{wave}
\nabla^j \nabla_j E_i + R_{ij} g^{jk} E_k + \frac{\omega^2}{c^2}\, E_i = 0
\end{equation}
where $R_{ij}$ denotes the Ricci tensor
\cite{LL2,Telephone}
\begin{equation}\label{ricci}
R_{ij}= R^k_{\ ikj}.
\end{equation}
The non-Euclidean geometry established by the medium may scatter light, even in the case of impedance matching. Impedance matching \cite{Jackson} may significantly reduce scattering, but not completely, unless the Ricci tensor (\ref{ricci}) vanishes. In this case the Riemann tensor vanishes as well in three-dimensional space (see exercise 1 in \S 92 of Ref.\ \cite{LL2}). The Riemann tensor quantifies the measure of curvature, irrespective of the coordinates. If the Riemann tensor vanishes the geometry is flat --- the apparent curvature is an illusion where curved coordinates disguise a straight system: the material is a transformation medium. So only transformation media cause absolutely no local scattering.\footnote{In one dimension, all impendance-matched non-moving dielectrics are transformation media, so impedance-matched waveguides are reflectionless \cite{Jackson}, but some non-impedance-matched materials (with soliton index-profiles) are reflectionless too  \cite{Gupta,LL3}.}\ Only they guide electromagnetic waves without disturbing them; they merely transform fields from electromagnetic to real space.  However, some transformation media still cause scattering in situations where the topologies of the two spaces differ from each other. In \S 5 we discuss two examples of topological scattering (though for space-time transformations), the optical Aharonov-Bohm effect and analogues of the event horizons.

Let us return to Fermat's principle. In the regime of geometrical optics \cite{BornWolf} the phase of electromagnetic waves advances more rapidly than the variations of the dielectric properties of the material. The medium determines the wavelength $\lambda$, and so $\lambda$ varies with the variations of the material properties. For geometrical optics to be a valid approximation, the gradient of the wavelength should be small, $| \bm{\nabla}\lambda|\ll1$ \cite{LL3}. 
We represent the electric-field components as
\begin{equation}\label{ansatz}
E_i = {\cal E}_i\, \mathrm{e}^{\mathrm{i}\varphi}
\end{equation}
where ${\cal E}_i$ is a slowly varying envelope and $\varphi$ the rapidly advancing phase. The gradient of the phase $\bm{\nabla}\varphi$ describes the wave vector $\bm{k}$ and the wavelength is given by $\lambda=2\pi/|\bm{k}|$. 
We substitute the ansatz (\ref{ansatz}) into the wave equation (\ref{wave}) and take only the dominant terms into account, terms that contain products of the first derivatives of the phase or $\omega^2$, ignoring all other terms, including the curvature contribution. The amplitude ${\cal E}_i$ is common to all remaining terms, and so the wave equation reduces to the {\it dispersion relation} 
\begin{equation}\label{dispersion}
g^{ij} \varphi_{,i} \varphi_{,j} = \frac{\omega^2}{c^2}.
\end{equation}
The dispersion relation contains the speed of light in the medium, $c'$. For example, for an isotropic medium with refractive index $n=\varepsilon=\mu$ we have the inverse metric tensor $g^{ij}=n^{-2}\mathds{1}$ and $c'=c/n$.
For anisotropic media, the eigenvalues of the matrix $g^{ij}$ characterize $c'$ for light rays propagating in the directions of its eigenvectors; $c'/c$ is the square root of the corresponding eigenvalue of $g^{ij}$.

In order to obtain Fermat's principle, we take advantage of the connection between optics and classical mechanics (the connection that inspired Schr\"odinger's quantum mechanics in analogy to wave optics). 
We read the dispersion relation (\ref{dispersion}) as the Hamilton-Jacobi equation of a fictitious point particle that draws the spatial trajectory of the light ray  \cite{LL1}. For this we need to identify the wave vector $\varphi_{,i}$ as the canonical momentum, as the derivative of some Lagrangian $L'$ with respect to the velocity $\dot{x}^i$. We might be tempted to employ the Lagrangian (\ref{L}) of the geodesic equation, but it is wise to use 
\begin{equation}\label{LL}
L' = \frac{L^2}{2} = \frac{1}{2}\, g_{ij} \dot{x}^i \dot{x}^j.
\end{equation}
In this case we obtain
\begin{equation}
\varphi_{,i} = \frac{\partial L'}{\partial \dot{x}^i} = g_{ij} \dot{x}^j
\end{equation}
and the Hamiltonian 
\begin{equation}
H' = \frac{\partial L'}{\partial \dot{x}^i} \, \dot{x}^i - L' =
\frac{1}{2}\, g_{ij} \dot{x}^i \dot{x}^j =
\frac{1}{2}\, g^{ij} \varphi_{,i}\varphi_{,j}.
\end{equation}
Since the metric does not depend on time the Hamiltonian is conserved \cite{LL1}. The conserved quantity is the ``energy'' of the fictitious particle. Our choice (\ref{LL}) of the Lagrangian is consistent with the dispersion relation (\ref{dispersion}) if we put
\begin{equation}
H' = \frac{\omega^2}{2c^2}.
\end{equation}
This proves that $L'$ is a suitable Lagrangian for light rays. The resulting Euler-Lagrange equations give the geodesic equation (\ref{geo3}). So light rays follow a geodesic of the metric $g_{ij}$, they take an extremal optical path given by the properties of the medium (\ref{epsg}): light follows Fermat's principle.


\subsection{Space-time geometry} 

So far we considered spatial geometries or coordinate transformations in space, but there are also important examples of transformation media that mix space and time \cite{GREE}. In \S 5 we discuss two cases in detail, the optical Aharonov-Bohm effect of a vortex and optical analogues of the event horizon. Here we write down the foundations for the theory of space-time transformation media, Maxwell's equations in a space-time geometry.

First, let us introduce space-time coordinates. The role of the Cartesian coordinates $\{x,y,z\}$ is played by the Galilean system $\{ct,x,y,z\}$ where $t$ denotes time. In \S 3 we developed differential geometry in three-dimensional space, but it is easy to extend the treatment to four dimensions --- the indices just take one more value and the Levi-Civita tensor has one more index. In this manner one can treat arbitrary coordinates in space-time. The appropriate distance in Galilean coordinates $\{ct,x,y,z\}$ in flat space-time is, however, not given by the four-dimensional Euclidean metric, but rather by the Minkowski metric \cite{LL2}:
\begin{equation} \label{mink}
\mathrm{d}s^2=c^2\mathrm{d}t^2-\mathrm{d}x^2-\mathrm{d}y^2-\mathrm{d}z^2.
\end{equation}
Here we use the Landau-Lifshitz convention for the metric where spatial distances are counted as negative contributions (one can also use the opposite metric $-\mathrm{d}s^2$ where space counts as positive). The metric (\ref{mink}) rather describes a measure of time. It is customary to denote the time coordinate by $x^0$ and the three spatial coordinates by $x^i$ where the Latin indices run over $\{1,2,3\}$. The metric in a general space-time coordinate system (or in curved space-time) is
\begin{equation}\label{spacetime}
\mathrm{d}s^2=g_{\alpha\beta}\,\mathrm{d}x^\alpha \mathrm{d}x^\beta,
\end{equation}
where the Greek indices run over the four values $\{0,1,2,3\}$. The component $g_{00}$ is usually positive throughout space-time, as in the Minkowski metric (\ref{mink}), and the determinant $g$ is always negative. 

It turns out that the theory of transformation media and materials that mimic curvature also works in four-dimensional space-time. This is proved in the Appendix, which provides a more challenging example of tensor algebra. There we show that the free-space Maxwell equations in arbitrary right-handed space-time coordinates can be written as the macroscopic Maxwell equations in right-handed Cartesian coordinates $\{ct,x,y,z\}$ with Plebanski's constitutive equations \cite{Plebanski}
\begin{gather}
D^i=\varepsilon_0\varepsilon^{ij}E_j+\frac{1}{c}[ijk]w_jH^k, \label{eq:constitution} \quad
B_i=\mu_0\mu^{ij}H^j-\frac{1}{c}[ijk]w_jE_k, \\[5pt]
\varepsilon^{ij}=\mu^{ij}=-\frac{\sqrt{-g}}{g_{00}}\,g^{ij},\quad 
w_i=\frac{g_{0i}}{g_{00}}. \label{eq:constrel}
\end{gather}
Space-time geometries appear as media.

The constitutive equations (\ref{eq:constrel}) turn out to reveal an important hidden property of electromagnetism: electromagnetic fields are {\it conformally invariant} in space-time. Suppose we compare two geometries, one with the metric (\ref{spacetime}) and one with a metric where we re-scale equally space and time at each point, but the scaling factor $\Omega$ may vary over space-time,
\begin{equation}\label{stc}
g_{\alpha\beta} \rightarrow \Omega(x^\nu)\, g_{\alpha\beta}.
\end{equation}
This is not a coordinate transformation in general: we  compare two different geometries. They measure space-time distances differently, but the angles between world lines are the same. As a result of the conformal scaling (\ref{stc}), the inverse metric tensor $g^{\alpha\beta}$ scales with $\Omega^{-1}$ and the determinant $g$ with $\Omega^4$. So $\varepsilon^{ij}$, $\mu^{ij}$ and $w_i$ do not change; but these are the only quantities that depend on the geometry. Consequently, the electromagnetic field does not notice a conformal space-time transformation, electromagnetism is invariant if space and time are re-scaled equally.  However, if we only altered the measure of space by a spatially dependent factor $n$, the new geometry would behave like a dielectric medium with refractive index profile $n$. Conformal invariance is the basis of Penrose diagrams \cite{Wald} where the entire causal structure of infinitely extended space-time is condensed, by a conformal factor, into a finite map one can draw and discuss. In Sec.\ 5.4 we apply the conformal invariance of electromagnetism to discuss the space-time geometry generated by moving media \cite{Gordon,LeoGeometry}.  

For transformation media, we can generalize the constitutive equations (\ref{eq:constitution})-(\ref{eq:constrel}) to allow for a handedness change in the spatial part of the space-time coordinate transformation, and also to allow for a curvilinear spatial coordinate system in electromagnetic space. Our previous result (\ref{constspace}) shows how to incorporate these possibilities in the permittivity and permeability in the constitutive equations (\ref{eq:constrel}). In addition, we can express the constitutive equations (\ref{eq:constitution})-(\ref{eq:constrel}) in index-free form if we denote the permittivity and permeability matrices by $\varepsilon$ and $\mu$, respectively, and understand $\varepsilon\bm{E}$, etc., as a matrix product. Our final constitutive relations are then \cite{GREE}
\begin{gather}
\bm{D}=\varepsilon_0\varepsilon\bm{E}+\frac{\bm{w}}{c}\times\bm{H},   \quad
\bm{B}=\mu_0\mu\bm{H}-\frac{\bm{w}}{c}\times\bm{E},  \\[5pt]
\varepsilon^{ij}=\mu^{ij}=\mp\frac{\sqrt{-g}}{\sqrt{\gamma}\,g_{00}}\,g^{ij},\quad 
w_i=\frac{g_{0i}}{\,g_{00}}. 
\label{eq:const}
\end{gather}
In addition to the familiar impedance-matched electric permittivity $\varepsilon$ and magnetic permeability $\mu$, a transformation that mixes space and time mixes electric and magnetic fields. A space-time geometry appears as a magneto-electric medium, also called a bi-anisotropic medium  \cite{SSTS,SVLT}. The mixing of electric and magnetic fields is brought about  by the bi-anisotropy vector ${\bm w}$ that has the physical dimension of a velocity. In Sec.\ 5.4 we show that ${\bm w}$ is closely related to the velocity of the medium (for slow media, ${\bm w}$ is proportional to the velocity). Moving media are naturally magneto-electric \cite{LL8} --- a moving dielectric responds to the electromagnetic field in its local frame, but this frame is moving and motion mixes electric and magnetic fields by Lorentz transformations \cite{Jackson,LL2}. Such phenomena have been observed before special relativity was discovered, for example in the R\"ontgen effect \cite{LPRoentgen,Roentgen} in 1888 or, indirectly, in Fizeau's 1851 demonstration \cite{Fizeau} of the Fresnel drag \cite{Fresnel}. 
More recently, moving optical media have been shown to generate analogues of the event horizon \cite{Philbin}. 


\section{Transformation media}

The concept of transformation media has been the key idea for the design of invisibility devices \cite{Greenleaf1,Greenleaf2,LeoConform,PSS}, an idea that was put into practice using electromagnetic metamaterials \cite{Schurig}.
Moreover, the idea that inspired the surge of interest in metamaterials in the first place, the perfect lens \cite{Pendry}, turned out to represent an example of transformation optics as well \cite{GREE}. Furthermore, the optical Aharonov-Bohm effect \cite{CFM,Hannay,Stor,Liten} and optical analogues of the event horizon \cite{LeoInstruments,Philbin} can be understood as cases of media that perform space-time transformations \cite{GREE}. In this section we discuss these examples in some detail.

\subsection{Spatial transformation media}

Let us first focus on some general properties of spatial transformation media. These are media that perform purely spatial coordinate transformations of electromagnetic fields. We develop an economic form of the theory that allows quick calculations and rapid judgment on the physical properties of spatial transformation media.
For this, we distinguish two different coordinate systems and three matrices of metric tensors: the coordinates $x^{i'}$ of electromagnetic space with metric tensor 
$g_{i' j'}$, which in physical coordinates $x^i$ appears as the tensor components 
$g_{i j}$. The physical coordinates $x^i$ are not necessarily Cartesian, but characterized by the metric $\gamma_{i j}$ of physical space, and neither are the electromagnetic coordinates required to be Cartesian,
because it is often wise in physics to adopt the coordinates to the particular physical problem under investigation.

Transformation media are made of anisotropic materials, in general, that are characterized by $\varepsilon$ and $\mu$ tensors. In curvilinear coordinates, the components of the dielectric tensors are given by Eqs.\ (\ref{ggtrans}) and (\ref{constspace}) as 
\begin{equation}
\varepsilon^{i j} = \mu^{i j} = \pm \frac{\sqrt{g}}{\sqrt{\gamma}}   g^{i' j'} \,  {\Lambda^i}_{i'} \, {\Lambda^j}_{j'}
\label{eq:dtensors}
\end{equation}
where $\pm$ indicates the handedness of the physical coordinates with respect to the electromagnetic coordinates, $+$ for right-handed and $-$ for left-handed coordinate systems; $g$ and $\gamma$ denote the determinants of $g_{i j}$ and $\gamma_{i j}$.
We obtain from the transformation (\ref{gtrans}) of the metric tensor
\begin{equation}
g = \mathrm{det} \, ( g_{i j}^{'}) \, \mathrm{det}^2 ( {\Lambda^{i'}}_{i} ) = 
\frac{ \mathrm{det} \, ( g_{i j}^{'})}{\mathrm{det}^2 ( {\Lambda^i}_{i'} )} =
\frac{g'}{\mathrm{det}^2 ( {\Lambda^i}_{i'} )}   .
\end{equation}
We employ a convenient matrix form for the dielectric tensors (\ref{eq:dtensors}) by defining the matrices
\begin{equation}
G' \equiv \left( g_{i' j'} \right) \, , \quad  \Gamma \equiv \left( \gamma_{i j} \right) \, , \quad \Lambda \equiv \left( {\Lambda^i}_{i'} \right) = 
\left( \frac{\partial x^i}{\partial x^{i'}} \right)   .
\label{eq:matrix}
\end{equation}
The first index labels the rows and the second index the columns of the matrices. In terms of the matrix representation (\ref{eq:matrix}) we obtain for the dielectric tensors (\ref{eq:dtensors})
\begin{equation}
\varepsilon = \mu = \frac{\sqrt{\mathrm{det} \, G'}}{\sqrt{\mathrm{det} \, \Gamma}} \, \frac{\Lambda \, G^{' -1} \, \Lambda^{T}}{\mathrm{det} \, \Lambda}   .
\label{eq:epsmu}
\end{equation}
The equality of the electric permitivity $\varepsilon$ and the magnetic permeability $\mu$ means that transformation media are impedance-matched \cite{Jackson} to the vacuum, but in practice impedance-matching can be relaxed at the expense of introducing some slight additional scattering  \cite{Cai,Cai2,Schurig}.

For an anisotropic medium, the eigenvalues of the dielectric tensors $\varepsilon_c$ and $\mu_c$ in Cartesian coordinates describe the dielectric responses along the three axes of the medium at each point in space; the axes are given by the eigenvectors. For transformation media, the tensor $\varepsilon_c=\mu_c$ is symmetric, because it is constructed from the symmetric inverse metric tensor 
$ g^{i' j'} $ in Eq.\ (\ref{eq:dtensors}). Consequently, the matrix $\varepsilon_c$ has three real eigenvalues $\epsilon_a$ with the orthogonal eigenvectors $\bm{a}$, the three orthogonal axes of dielectric response. Here we show how to calculate the eigenvalues $\epsilon_a$ from the dielectric tensor (\ref{eq:epsmu}) in curvilinear coordinates. 
Consider the mixed tensor $\varepsilon \, \Gamma = \left( \varepsilon^{il} \gamma_{lj} \right)$. Suppose we transform this tensor from curvilinear to Cartesian coordinates by the transformation matrix $\Lambda_c$.
According to the rules of tensor transformations conveniently indicated by the index positions in ${\varepsilon^i}_{j}$, the curvilinear $\varepsilon \, \Gamma$ turns into the Cartesian $\varepsilon_c = \Lambda_c \, \varepsilon \, \Gamma \, \Lambda_c^{-1}$, and vice versa $\varepsilon \, \Gamma = \Lambda_c^{-1} \, \varepsilon_c \, \Lambda_c$. Consequently, $\Lambda_c^{-1} {\bm a}$ is an eigenvector of $\varepsilon \, \Gamma$ with eigenvalue $\epsilon_a$; the eigenvalues of $\varepsilon \, \Gamma$ are those of the dielectric tensor in Cartesian coordinates. Note that the eigenvectors of $\varepsilon \, \Gamma$ are not orthogonal in general (and ${\varepsilon^i}_{j}$ is not symmetric). Nevertheless, the eigenvalues of $\varepsilon \, \Gamma$ directly give the dielectric functions of the transformation medium.

\begin{figure}[h]
\begin{center}
\includegraphics[width=16.0pc]{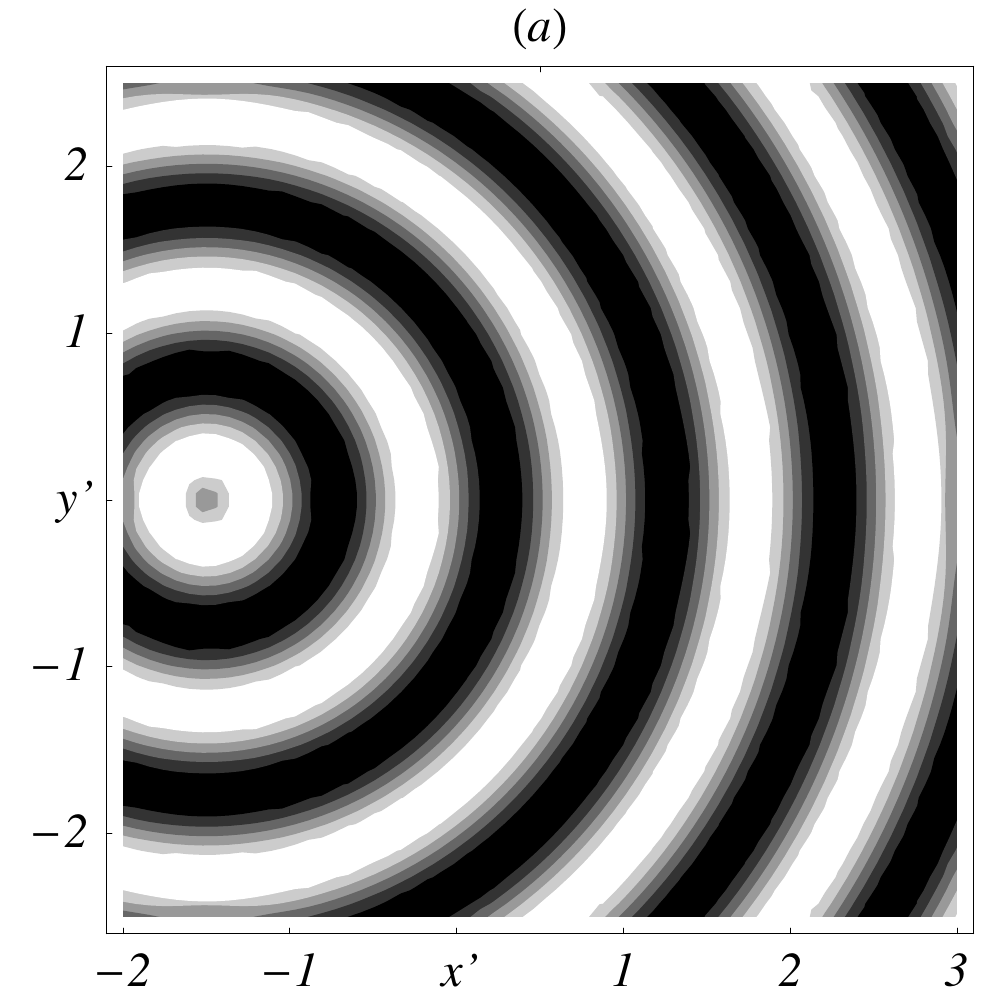}
\includegraphics[width=16.0pc]{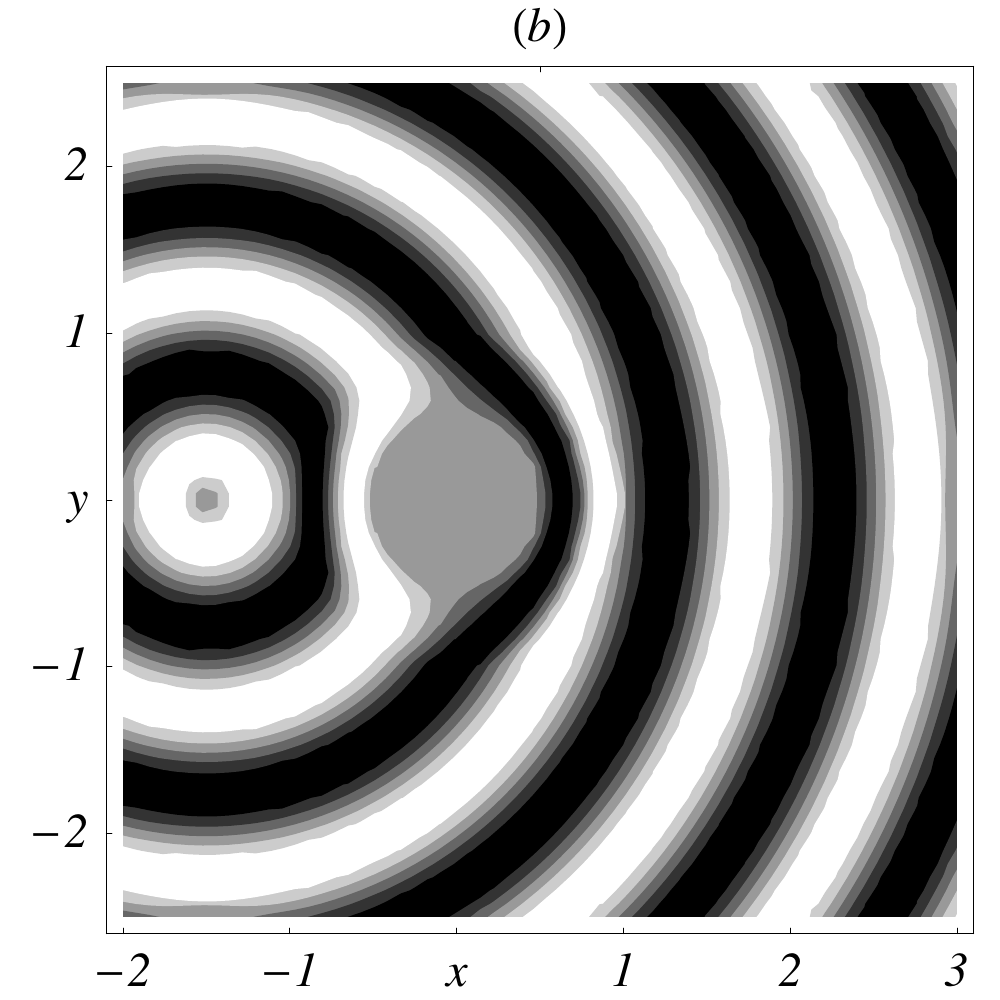}
\caption{
\small{
Invisibility for waves.
The device transforms 
waves emitted in physical space as if they propagate 
through empty space in transformed coordinates.
(a) Transformed space.
The figure shows the propagation of a
spherical wave emitted at the point $(-1.5,0,0)$.
(b) Physical space.
The invisibility device 
turns the wave of figure (a) into physical space
by the coordinate transformation illustrated in Fig.\ \ref{fig:spider}.
The cloaking layer with radii $R_1=0.5$ and $R_2=1.0$
deforms electromagnetic waves such that they
propagate around the invisible region
and leave without carrying any trace of the interior of the device.}
\label{fig:waves}
}
\end{center}
\end{figure}

To illustrate why this procedure offers considerable economy in calculations, we discuss the example of a cylindrical transformation. Suppose that the medium transforms the radius $r$ in cylindrical coordinates, but preserves the angle $\phi$ and the vertical coordinate $z$,
\begin{eqnarray}
r = r (r') \, ,   & r'=\sqrt{x'^2+y'^2} \, ,  &  \nonumber \\
x=\frac{r \, x'}{\sqrt{x'^2+y'^2}} \, ,   & y=\displaystyle \frac{r \, y'}{\sqrt{x'^2+y'^2}} \, ,   & z = z' \, .
\end{eqnarray}
We calculate the transformation matrix $\Lambda$ from $\{x,y,z\}$ to $\{x',y',z'\}$ and express the result as
\begin{equation}
\Lambda = \left(
\begin{array}{c@{\quad}c@{\quad}c}
R \cos^2 \phi + \displaystyle\frac{r}{r'} \sin^2 \phi & \left( R-\displaystyle\frac{r}{r'}\right) \cos \phi \sin \phi & 0  \\
\left( R- \displaystyle\frac{r}{r'}\right) \cos \phi \sin \phi & \displaystyle\frac{r}{r'} \cos^2 \phi + R \sin^2 \phi & 0  \\
0 & 0 & 1 
\end{array}
\right) 
\end{equation}
with 
\begin{equation}
R = \frac{\mathrm{d}r}{\mathrm{d}r'} \, , \quad  \mathrm{det} \, \Lambda = \frac{r}{r'} \, R   .
\end{equation}
We obtain the dielectric tensor in Cartesian coordinates
\begin{equation}
\varepsilon_c = \frac{r'}{r R} \left(
\begin{array}{c@{\quad}c@{\quad}c}
R^2 \cos^2 \phi + \displaystyle\frac{r^2}{r'^2} \sin^2 \phi & \left( R^2 -\displaystyle\frac{r^2}{r'^2}\right) \cos \phi \sin \phi & 0  \\
\left( R^2 -\displaystyle\frac{r^2}{r'^2}\right) \cos \phi \sin \phi & \displaystyle\frac{r^2}{r'^2} \cos^2 \phi + R^2 \sin^2 \phi & 0 \\
0 & 0 & 1 
\end{array}
\right) 
\end{equation}
calculate the eigenvalues, and obtain after some algebra
\begin{equation}
\epsilon_a = R \, \frac{r'}{r} \, ,  \frac{r}{R \, r'} \, ,   \frac{r'}{r R}   .
\end{equation}
Alternatively, we calculate $\varepsilon$ in the coordinate system that is best adapted to cylindrical transformations: both in electromagnetic and in physical space we use cylindrical coordinates. From the line element in cylindrical coordinates $ \mathrm{d}s^2 = \mathrm{d}r^2 + r^2 \mathrm{d}\phi^2 + \mathrm{d}z^2 $ we read off the metric tensors $\Gamma = \mathrm{diag} \, (1, r^2, 1)$ and $G' = \mathrm{diag} \, (1, r'^2, 1)$. For transformations of the radius $\Lambda = \mathrm{diag} \, (R, 1, 1)$.  
Consequently, 
\begin{equation}
\varepsilon \, \Gamma = \mathrm{diag} \left( R \, \frac{r'}{r} \, ,  \frac{r}{R \, r'} \, ,   \frac{r'}{r R} \right)   ;
\end{equation}
this elementary calculation reveals the dielectric properties of the cylindrical transformation medium. 
If the transformation opens a hole in physical space, anything inside this hole is decoupled from the electromagnetic field: it has become invisible, as Fig.\ \ref{fig:waves} shows.\footnote{The corresponding figure in Ref.\ \cite{LPQuant} is not correct. We thank T.\ Tyc for pointing this out.}

\subsection{Perfect invisibility devices}

\begin{figure}[h]
\begin{center}
\includegraphics[width=15.0pc]{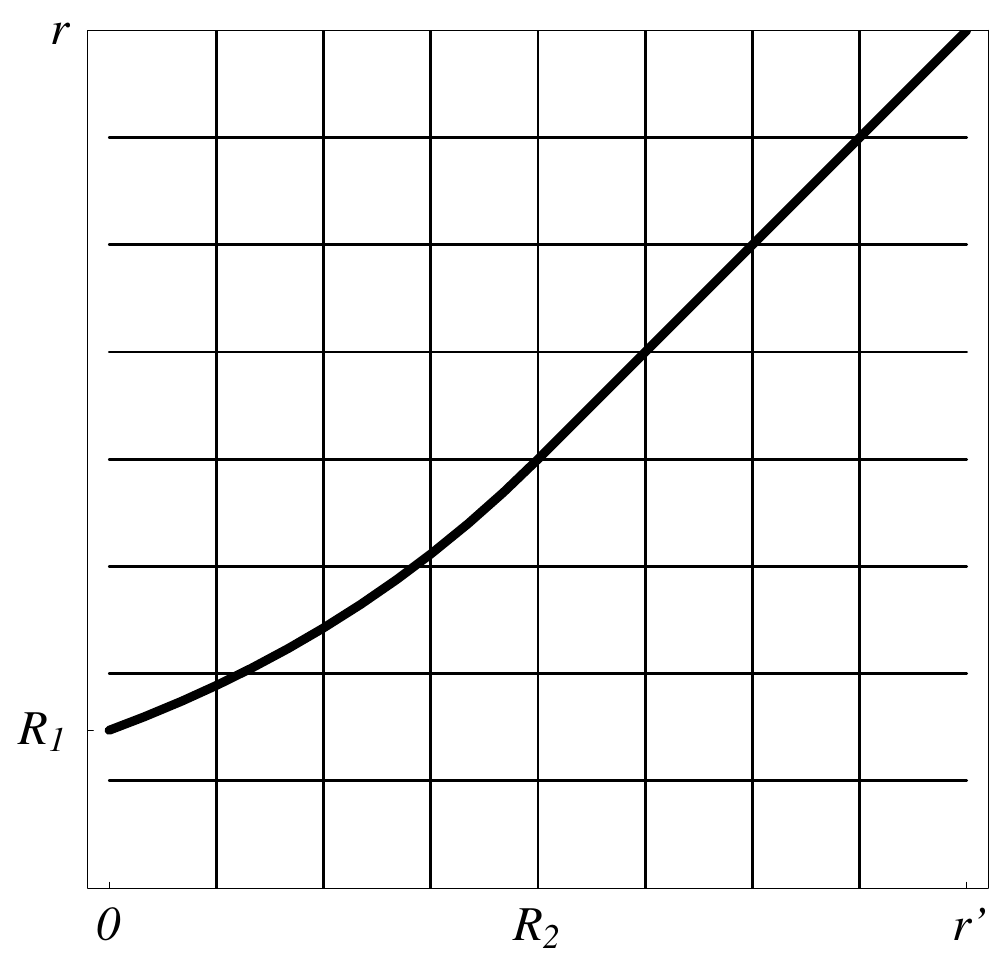}
\caption{
\small{
Cloaking transformation.
Radial transformation performed by a spherical cloaking device.
The radius $r'$ of spherical coordinates in electromagnetic space
is transformed to the radius $r$ in physical space, and vice versa.  
$R_1$ denotes the inner and $R_2$ the outer radius of the cloaking shell:
physical regions with radius $r<R_1$ are not reached by electromagnetic waves and for $r>R_2$ the electromagnetic coordinates agree 
with the coordinates of physical space.
}
\label{fig:cloakingradius}
}
\end{center}
\end{figure}

Transformation media that have holes in their coordinate grids in physical space act as perfect invisibility devices. Consider the following simple example in spherical coordinates \cite{PSS}. Suppose the device transforms the radius $r'$ in electromagnetic space, but does not affect the spherical angles,
\begin{equation}
r=r(r') \, ,  \quad \theta = \theta'  \, ,  \quad \phi = \phi'   .
\end{equation}
We obtain from Eq.\ (\ref{gsph}) the matrices $\Gamma =  \mathrm{diag} (1, r^2, r^2 \sin^2 \theta)$ and  $G' = \mathrm{diag} \, (1,$ $r'^2, r'^2 \sin^2 \theta)$, and, as in the previous subsection, the transformation matrix of the devices is $\Lambda = \mathrm{diag} \, (R, 1, 1)$ with $R = \mathrm{d}r / \mathrm{d}r'$.
Consequently,
\begin{equation}
\varepsilon \, \Gamma = \frac{r'^2}{r^2 R^2} \, \mathrm{diag} \left( R^2 \,,  \frac{r^2}{r'^2} \,,  \frac{r^2}{r'^2}  \right) =
\mathrm{diag} \left( \frac{r'^2}{r^2} \,,  \frac{1}{R^2}  \,,  \frac{1}{R^2}\right)   .
\end{equation}
Figure \ref{fig:cloakingradius} shows an example of a radial transformation when $r(r')$ reaches a finite value $R_1$ at the origin in electromagnetic space where $r'=0$. This means that in physical space the entire spherical shell of radius $R_1$ corresponds to a single point in electromagnetic space. Anything lurking inside that sphere has become excluded from the electromagnetic field: it has become invisible. 
Figure \ref{fig:cloakingradius} also shows that beyond the radius $R_2$ electromagnetic and physical coordinates agree --- $R_2$ describes the outer radius of the cloaking device, whereas $R_1$ is the inner radius.
If electromagnetic and physical coordinates agree beyond the outer layer of the cloak then outside of the device electromagnetic waves are indistinguishable from waves propagating through empty space. The device guides electromagnetic waves around the enclosed hidden object without causing disturbances.
So the object has not only disappeared from sight, but the act of hiding has become undetectable; the scenery behind the cloaking device would show no sign of the object nor of the device. In short, this transformation makes a perfect cloaking device.
This behavior has been verified using scattering theory
\cite{Ruan}
and the most appropriate boundary conditions for the inner layer
of the cloak have been identified \cite{GreenleafCoat}.

\begin{figure}[t]
\begin{center}
\includegraphics[width=30.0pc]{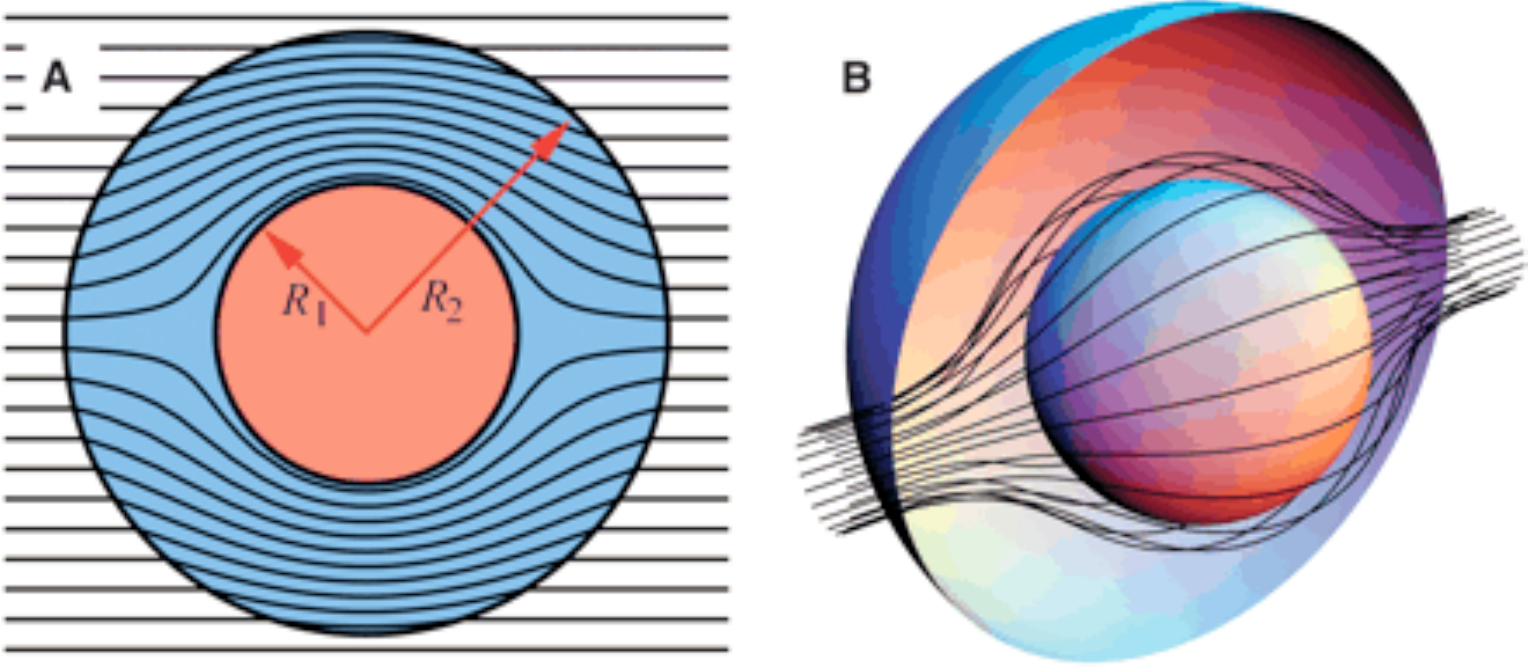}
\caption{
\small{
Spherical cloaking device. 
(From Ref.\ \cite{PSS}. Reprinted with permission from AAAS.)
The figure shows the trajectories of light rays. (A) two-dimensional cross section of rays striking the device, diverted within the annulus of cloaking material contained within $R_1 < r < R_2$ to emerge on the far side undeviated from their original course. (B) three-dimensional view of the same process.
}
\label{fig:invisibility}}
\end{center}
\end{figure}

Fortunately perhaps, perfect invisibility \cite{PSS} is severely limited \cite{GREE}. Figure \ref{fig:invisibility} shows the trajectories of light rays passing through the cloaking shell. The rays make a detour around the hidden core of the device, but they must arrive at the same time as if they were propagating through empty space. So there is only one option: inside the cloaking device the phase velocity of light must exceed $c$. This is possible in principle \cite{BG,MilonniLight}, but only in regions of the spectrum with narrow bandwidth that correspond to resonances in the material.
But, to make matters worse, light rays straddling the inner lining of the cloak must go around the inner core in precisely the same time it would take them to traverse a single point in electromagnetic space, in zero time. Light must propagate at infinite phase velocity! One can put these thoughts into precise mathematical terms \cite{GREE} by considering the product $\varepsilon_1 \varepsilon_2 \varepsilon_3$ of the three dielectric functions of the transformation medium. This product is the product of the eigenvalues of $\varepsilon \, \Gamma$, the determinant of $\varepsilon \, \Gamma$. We obtain from formula (\ref{eq:epsmu}) 
\begin{equation}
\varepsilon_1 \, \varepsilon_2 \, \varepsilon_3 = \mathrm{det} \left( \varepsilon \, \Gamma \right) =
\frac{\sqrt{\mathrm{det} \, G'}}{\sqrt{\mathrm{det} \, \Gamma}} \, \frac{1}{\mathrm{det} \, \Lambda}   .
\end{equation}
We showed in Sec.\ 3.2 that $\sqrt{\mathrm{det} \, G'}$ describes the volume element, here the volume element in electromagnetic space. The volume of a single point is zero; hence at least one of the eigenvalues of the dielectric tensor is zero --
the speed of light reaches infinity at the inner lining of the cloak. This inevitable feature restricts the performance of perfect invisibility devices to a single frequency set by the particular metamaterial used. Perfect invisibility devices are impractical, but imperfect devices inspired by these ideas may be possible.

\subsection{Perfect lenses}

The perfect lens, made by negative refraction  \cite{Pendry,Veselago1},
turns out to be an example of a transformation medium as well \cite{GREE}.
As in perfect invisibility devices, the unusual properties of the lens
are based on an unusual topology of the transformation.
For achieving invisibility, 
electromagnetic space does not
cover the entire physical space,
whereas for perfect lenses
electromagnetic space turns out to be multi-valued:
single points in electromagnetic space
are mapped to multiple points in physical space.

\begin{figure}[h]
\begin{center}
\includegraphics[width=16.0pc]{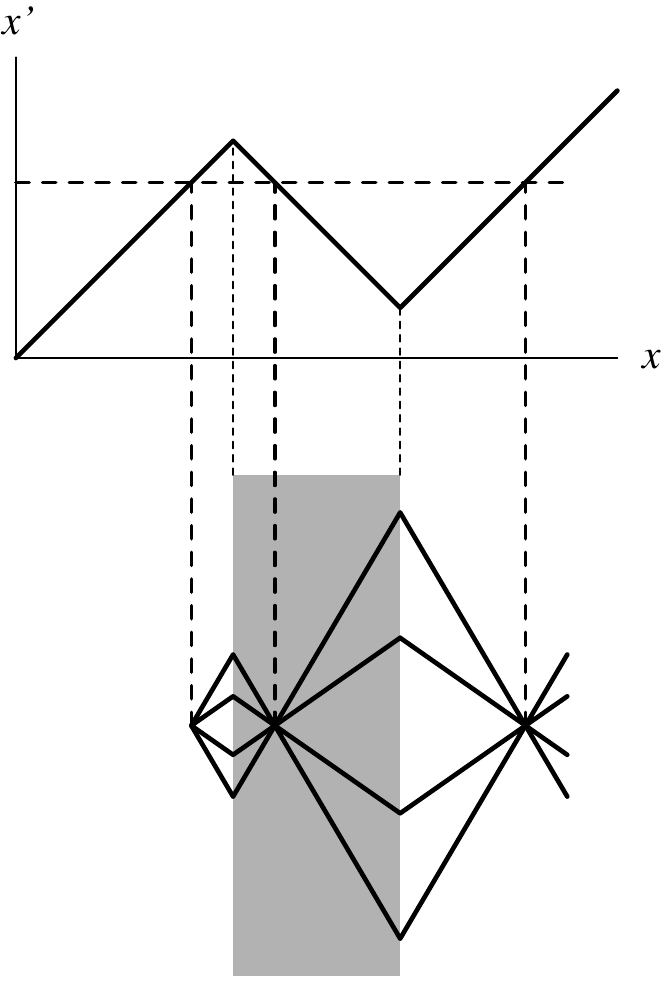}
\caption{
\small{
Perfect lens.
Negatively refracting perfect lenses employ transformation media.
The top figure shows a suitable coordinate transformation
from the physical $x$ axis to the electromagnetic $x'$,
the lower figure illustrates the corresponding device.
The inverse transformation from $x'$ to $x$ is either triple
or single-valued.
The triple-valued segment on the physical $x$ axis 
corresponds to the focal region of the lens: any source point
has two images, one inside the lens and one on the other side.
Since the device facilitates an exact coordinate transformation,
the images are perfect with a resolution below the
normal diffraction limit \cite{BornWolf}: 
the lens is perfect \cite{Pendry}.
In the device,
the transformation changes 
right-handed into left-handed coordinates.
Consequently, the medium employed here is left-handed, 
with negative refraction \cite{Veselago2}.
}
\label{fig:lens}}
\end{center}
\end{figure}

\begin{figure}[h]
\begin{center}
\includegraphics[width=16.0pc]{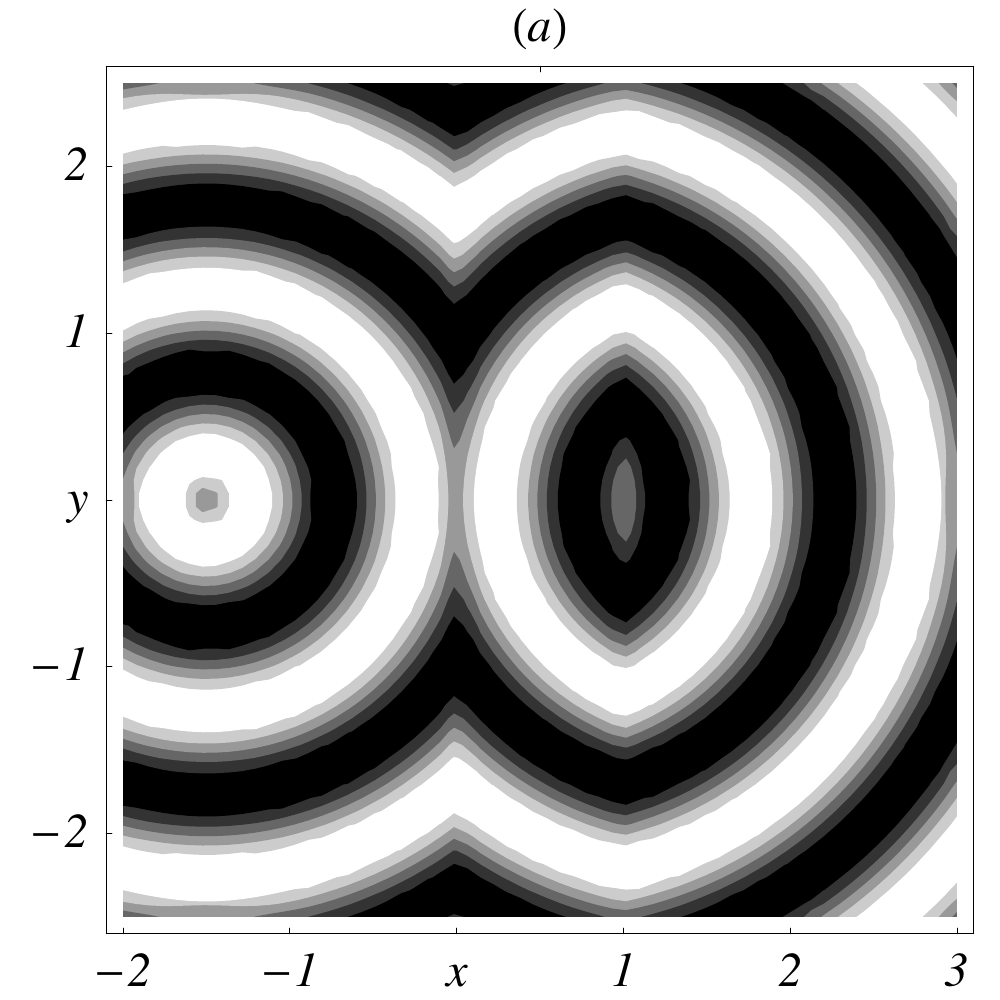}
\includegraphics[width=16.0pc]{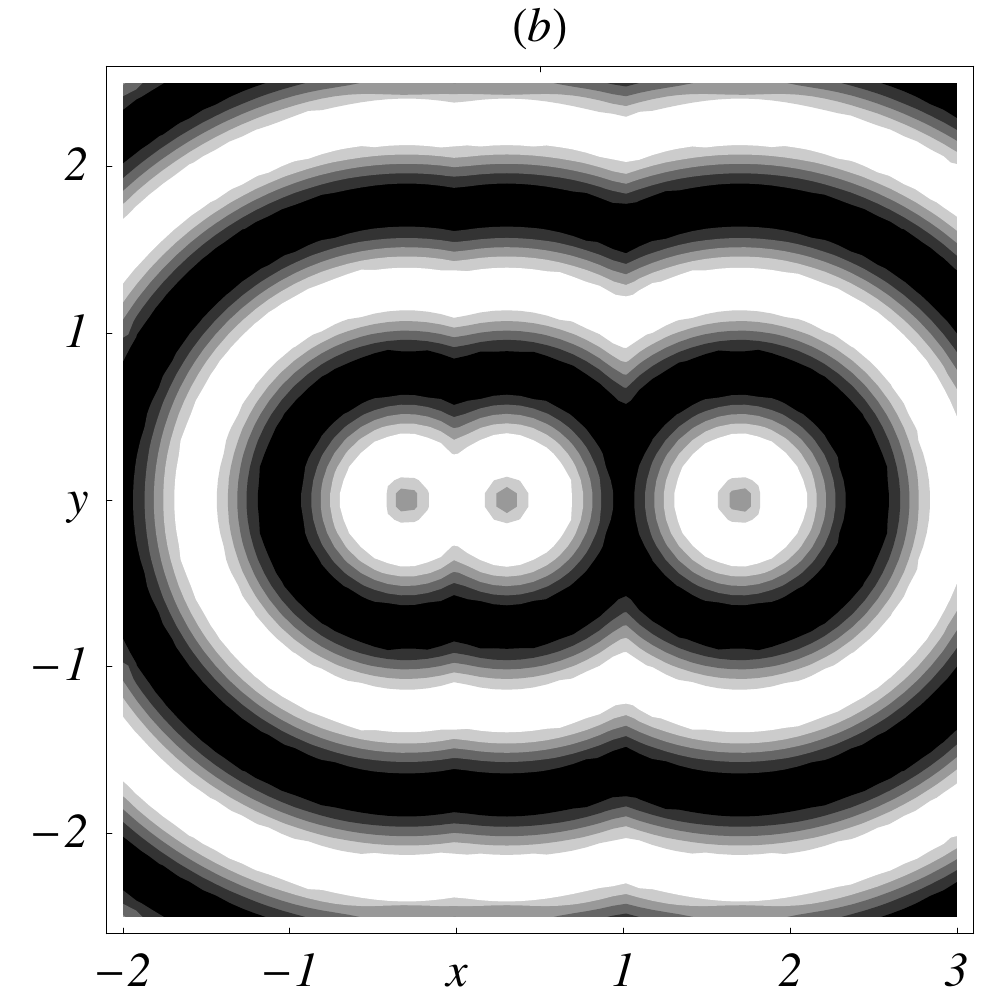}
\caption{
\small{
Propagation of electromagnetic waves in a perfect lens.
The lens facilitates the coordinate transformation
shown in Fig.\ \ref{fig:lens}.
Spherical waves in electromagnetic space 
are transformed into physical space. 
(a) The wave of Fig.\ \ref{fig:waves} (a) is emitted outside
the imaging range of the lens.
The wave is transformed by the lens, but the device is not
sufficiently thick to form an image.
(b) A wave is emitted inside the imaging range,
creating two images of the emission point, one inside the
device and one outside,
corresponding to the image points of Fig.\ \ref{fig:lens}. 
}
\label{fig:waveslens}}
\end{center}
\end{figure}

Consider in Cartesian coordinates
the transformation $x(x')$ illustrated 
in Fig.\ \ref{fig:lens}, 
whereas all other coordinates are not changed.
In the fold of the function $x(x')$
each point $x'$ in electromagnetic space 
has three faithful images in physical space.
Obviously, electromagnetic fields 
at one of those points are perfectly imaged onto
the others: the device is a perfect lens, see Fig.\ \ref{fig:waveslens}.
Perfect lensing was first analyzed \cite{Pendry}
as the imaging of evanescent waves
in a slab of negatively-refracting material,
waves that may carry images finer 
than the optical resolution limit \cite{BornWolf}.
Various aspects of this idea  \cite{Pendry}
have been subject to a considerable theoretical debate
\cite{Minkel,P1,P1Re,P2,P2Re}, 
but recent experiments
\cite{Fang,Grbic,Melville}
confirmed sub-resolution imaging.
Our pictorial argument leads to a simple intuitive
explanation of why such lenses are indeed perfect
\cite{GREE}.

We also immediately obtain from the theory of transformation media the reason 
why perfect lensing requires left-handed media with negative 
$\varepsilon$ and $\mu$ 
\cite{Marques,MilonniLight,Sarychev,Shelby,Veselago2}: 
inside the device, {\it i.e.} inside of the $x'$ fold,
the derivative of $x(x')$ becomes negative and
the coordinate system changes handedness.
The electromagnetic left-handedness of negative-index materials
appears in a left-handed coordinate system. 
We obtain from our recipe (\ref{eq:epsmu})
the compact result
\begin{equation}
\label{eq:lens}
\varepsilon = \mu = 
\mathrm{diag}
\left(\frac{\mathrm{d}x}{\mathrm{d}x'},
\frac{\mathrm{d}x'}{\mathrm{d}x},
\frac{\mathrm{d}x'}{\mathrm{d}x}\right)
\,.
\end{equation}
At the place where the transition between right-handed 
and left-handed coordinates occurs,
the derivative of $x(x')$ is undefined and so are
$\varepsilon$ and $\mu$,
but this has no adverse effect in practice
\cite{LeoNegative}.
It simply describes the discontinuity at the boundary 
of the medium where our transformation $x'(x)$
automatically gives the correct boundary conditions
\cite{Jackson},
because Maxwell's equations are satisfied 
arbitrarily close to the boundary.
Equation (\ref{eq:lens}) shows that
when $\mathrm{d}x'/\mathrm{d}x$
is $-1$ inside the device and $+1$ outside,
we obtain the standard perfect lens 
\cite{Pendry,Veselago1} based
on an isotropic left-handed material with  
$\varepsilon=\mu=-1$ inside.
Furthermore, we see that lenses with 
$\varepsilon=\mu=-1$ are
not the most general choice.
One could use an anisotropic medium to magnify
perfect images by embedding the
source or the image in transformation media with 
$|\mathrm{d}x'/\mathrm{d}x| \neq 1$.

\begin{figure}[h]
\begin{center}
\includegraphics[width=35.0pc]{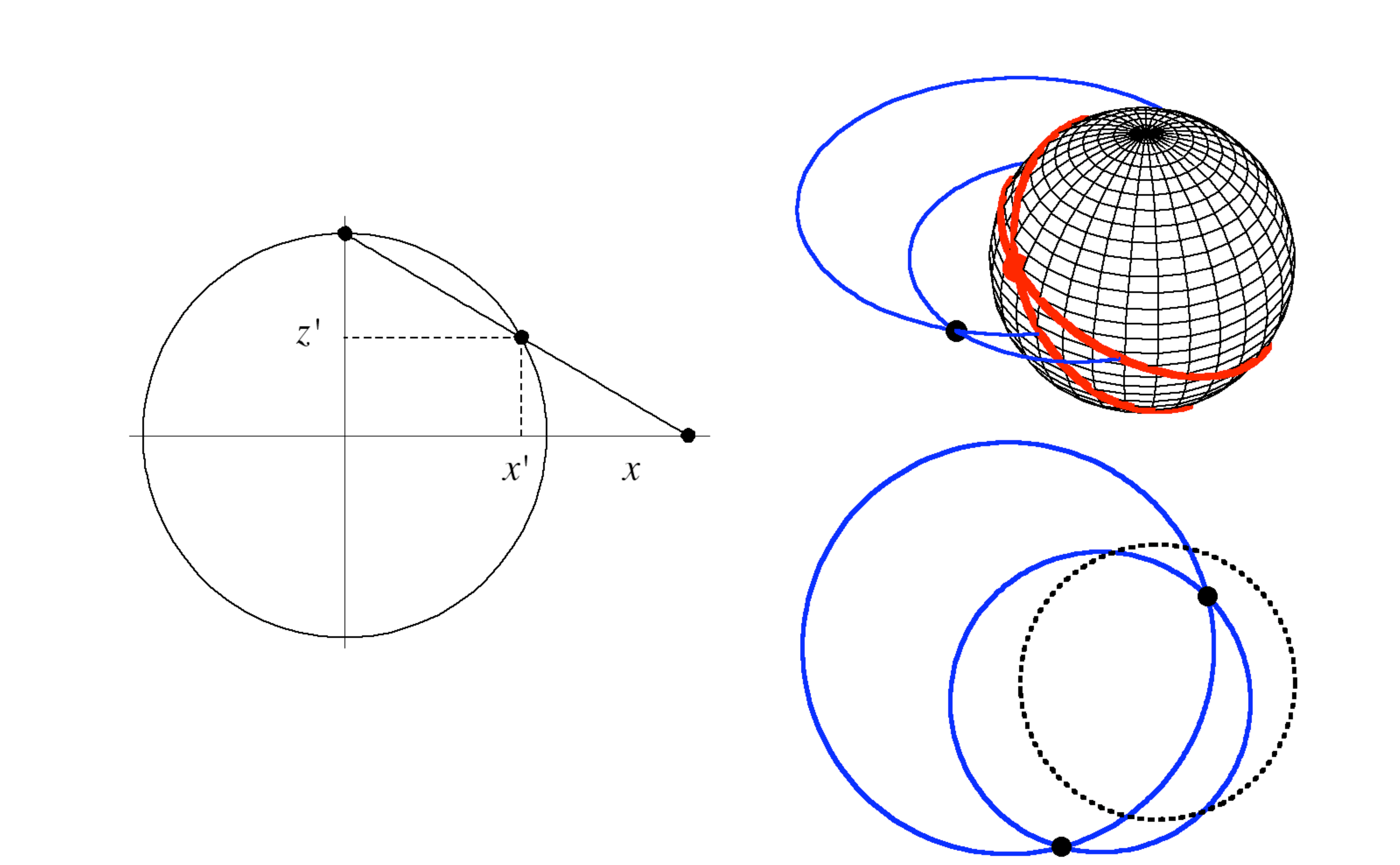}
\caption{
\small{
Maxwell's fish-eye.
The left figure illustrates the stereographic projection 
described in Eq.\ (\ref{eq:stereo}).
A line is drawn from the North Pole through a point on the sphere.
The image of this point is the intersection of this line with
the plane through the Equator. 
The top right figure shows light rays on the 
surface of the sphere and projected on to the Equatorial plane;
the bottom right figure shows the rays in the plane
where the dotted circle indicates the Equator.   
Light rays originating from one point meet again 
at the antipodal point:
Maxwell's fish-eye makes a perfect lens.
}
\label{fig:fisheye}}
\end{center}
\end{figure}

Note that the perfect lens made by negative refraction \cite{Pendry}
is not the only case of perfect imaging \cite{BornWolf}.
In 1854 Maxwell proposed a curious device called the fish-eye
\cite{Maxwell}.
In Maxwell's fish-eye light goes around in circles 
in such a way that any point in the device is imaged 
to a partner point with infinite precision. 
In turns out \cite{Luneburg} that Maxwell's fish-eye also is
an example of a transformation medium, 
but one with non-Euclidean geometry:
here physical space is not mapped to flat space, but
to our favorite example of a curved space, the surface of a sphere. 
In the following we use this visualization of the fish-eye 
to explain how it works.
Consider the stereographic projection
from the surface of a sphere with radius $r_0$ to the $x,y$ plane
as shown in Fig.\ \ref{fig:fisheye}.
In formulae,
\begin{equation}
x=\frac{x'}{1-z'/r_0} \,,\quad
y=\frac{y'}{1-z'/r_0} 
\quad\mbox{with}\quad
x'^2+y'^2+z'^2 = r_0^2 
\label{eq:stereo}
\end{equation}
and the inverse transformation
\begin{equation}
x'= \frac{2x}{1+(r/r_0)^2} \,,\quad
y'= \frac{2y}{1+(r/r_0)^2} \,,\quad
z' = r_0\frac{(r/r_0)^2-1}{(r/r_0)^2+1} \,,\quad
r^2 = x^2 +y^2
\,.
\label{eq:inverse}
\end{equation}
Imagine light propagating on the surface of the fictitious sphere
with uniform refractive index $n_0$.
In this case, the line element of electromagnetic space is
\begin{equation}
\mathrm{d}s^2
=n_0^2\left(\mathrm{d}x'^2+\mathrm{d}y'^2+\mathrm{d}z'^2\right)
\,.
\end{equation}
Similar to Eq.\ (\ref{eq:difftrans}) we express the differentials 
$\mathrm{d}x'$, $\mathrm{d}y'$ and $\mathrm{d}z'$
in terms of the differentials in physical space, and obtain 
\begin{equation}
\mathrm{d}s^2
=n^2\left(\mathrm{d}x^2 + \mathrm{d}y^2\right)
\,,\quad
n= \frac{2n_0r_0^2}{x^2+y^2+r_0^2}
\,.
\end{equation}
By the way, this result also proves that the stereographic projection 
is a conformal map of the surface of the sphere on 
the two-dimensional plane.\footnote{The stereographic projection, 
invented by Ptolemy, is an ingredient of the Mercator projection 
used in cartography; on the complex plane the Mercator projection is the logarithm of the stereographic projection \cite{Nehari}.}
We can easily extend this procedure from two-dimensional surfaces
to three-dimensional spaces by considering the stereographic 
projection from the hyper-surface of the four-dimensional sphere
to three-dimensional space or simply by rotating the $x, y$ plane 
around one axis, say the $y$ axis, while performing two-dimensional projections. Because of rotational symmetry, we obtain in 3D
\begin{equation}
\label{eq:fisheye}
\mathrm{d}s^2
=n^2\left(\mathrm{d}x^2 + \mathrm{d}y^2 + \mathrm{d}z^2\right)
\,,\quad
n= \frac{2n_0r_0^2}{x^2+y^2+z^2+r_0^2}
\,.
\end{equation}
Let us return to 2D where we can draw pictures, as shown
in Fig.\ \ref{fig:fisheye}.
On the surface of the sphere, light rays propagate along geodesics,
the great circles.
It is one of the remarkable properties of the stereographic projection
\cite{Nehari} that circles on the sphere are transformed into circles
on the plane. So, in physical space, light goes around in circles as well.
The great circles originating from one source point on the sphere 
meet again at the antipodal point.
In the stereographic projection, the image of the antipodal point
is the reflection of the source on a circle, the circle with the
radius $r_0$ of the sphere.  
Here the source is perfectly imaged:
Maxwell's fish-eye makes a perfect lens \cite{BornWolf}. 
However, this is quite an usual instrument: 
both the source and the image are embedded 
in the non-uniform refractive index profile
(\ref{eq:fisheye}) of the fish-eye.
In contrast, for the flat lens
made by negative refraction \cite{Pendry}
source and image are outside the device;
the lens acts across some distance.
The imaging range is the distance from the lens where 
multiple images are formed, where 
$x(x')$ is multivalued.
As figure \ref{fig:lens} shows, the imaging range
is equal to the thickness of the lens for 
the standard case where 
$\mathrm{d}x'/\mathrm{d}x =-1$ and, accordingly, 
$\epsilon=\mu=-1$.
Because of losses in the material, 
the imaging range has been very small 
in practice though \cite{Fang,Grbic,Melville}.

In another twist of the story, 
both the interior of a hollow sphere and a perfect flat lens
are predicted to cause unusual quantum effects,
they lead to repulsive Casimir forces 
\cite{Boyer68,LPCasimir}. 
The Casimir force is a force caused by, literally, ``nothing'' ---
by the zero-point energy of the quantum vacuum
\cite{LamoreauxReview,MandelWolf,Milonni}.
What is zero-point energy?
Imagine the electromagnetic field as a superposition of 
infinitely many stationary modes. 
Some of the mode coefficients may be zero,
but according to quantum physics being zero is a positive 
statement: the modes are in vacuum states.
Each mode oscillates like a harmonic oscillator 
with its eigenfrequency $\omega_\nu$. 
The vacuum state is the ground state of the harmonic oscillator,
but this state carries the energy $\hbar\omega_\nu/2$
\cite{LL3}. 
The sum of the ground-state energies of all the modes 
is called zero-point energy. 
So even in the absence of any electromagnetic fields,
in the complete quantum vacuum state,
the zero-point energy remains.
Moreover, this energy is infinite for infinitely many modes.
On the other hand, the mode frequencies $\omega_\nu$ 
depend on the boundary conditions.
Consider, as Casimir did \cite{Casimir}, 
quantum electromagnetism inside two parallel mirrors 
with distance $a$. 
Although the total zero-point energy is infinite,
the difference ${\cal U}$ between the zero-point energies per area
at $a$ and at $\infty$ turns out to be finite and negative
\cite{Casimir,DLP,Milonni}, 
\begin{equation}
{\cal U}(a)= -\frac{\hbar c \pi^2}{720 a^3} \,,
\label{eq:casimir}
\end{equation}
a mesmerizing combination of natural constants. 
Read ${\cal U}(a)$ as the potential for the force between the mirrors
with distance $a$. 
Since the potential falls with decreasing distance 
the mirrors are attracted to each other.
This attraction caused by the quantum vacuum has been observed 
\cite{Lamoreaux} and good quantitative agreement
with theoretical predictions was found \cite{Chan}. 
Normally, the Casimir force is weak, 
but it may become important for small distances $a$,
distances on the scale of nanomachines
\cite{Ball}.

Imagine that we place a perfect lens between the mirrors.
The lens should be sufficiently thick
such that the mirrors lie within the imaging range $b$ of the lens.
The lens maps physical space to electromagnetic space where 
the mirror distance is reduced to  
\begin{equation}
a' = 2b-a \,,\quad a<2b \,.
\end{equation}
Electromagnetic space is empty, apart from the mirrors; 
there they are attracted to each other with potential ${\cal U}(a')$,
but in physical space this attraction turns into repulsion, 
because when $a'$ decreases, $a$ must grow.
We obtain for the force per area
\begin{equation}
f  = -\frac{\mathrm{d} {\cal U}}{\mathrm{d} a} =
-\frac{\mathrm{d} {\cal U}(a')}{\mathrm{d} a'}
\frac{\mathrm{d} a'}{\mathrm{d} a} 
= \frac{\hbar c \pi^2}{240 a'^4} \,.
\end{equation}
This is not the only example of repulsive Casimir forces,
but so far it is the only one where repulsion is explained 
by a simple visual argument.  
The vacuum confined to the interior of a hollow 
conducting shell causes repulsion as well \cite{Boyer68};
the ``nothing'' would cause the shell to explode 
if the shell is sufficiently small such that the Casimir force is strong.
Another example is a miniature sandwich of different dielectric media,
silicon, ethanol and gold \cite{Cap07}.
Also in this case the Casimir force is repulsive for reasons
that are far less intuitive \cite{DLP}.
It turns out \cite{LPCasimir} that the perfect lens needs to have gain   
for sustaining a repulsive Casimir force,
but, in principle, it
represents a simple case where two mirrors repel each other
across some distance, 
without any adjacent mediator (such as the ethanol in a layer of 
silicon, ethanol and gold).
The Casimir repulsion could be sufficiently strong such that
one of the mirrors would overcome gravity, levitating
on zero-point energy.


\subsection{Moving media}

Perfect invisibility devices and perfect lenses are made of spatial transformation media -- metamaterials that facilitate transformations in space, but that do not alter the measure of time. Here we develop a physical interpretation of electromagnetism under space-time transformations. We show that they correspond to moving media \cite{Stor,LeoGeometry} (or media with moving dielectric properties \cite{Philbin}). 
For this we translate Fermat's principle (\ref{eq:fermat}) into a space-time geometry for moving media with isotropic and equal $\varepsilon = \mu = n$. We describe the electrodynamics of moving media in an inertial frame in Cartesian coordinates,  
the laboratory frame, but as an intermediate step we take the liberty of performing transformations to other inertial frames of reference.

Imagine the medium consists of infinitesimally small cells moving with velocities $\bm u$. The velocity profile may vary in space and time, but for a given space-time point we can always construct a locally co-moving inertial frame. In such a frame, the medium is not moving, at least at the specific point we are considering, and so we can apply here the standard theory of electromagnetism in media at rest. In particular, Fermat's principle (\ref{eq:fermat}) suggests that the electromagnetic field perceives here a spatial geometry with the line element 
$\mathrm{d}l^2 = n^2 (\mathrm{d}x^2+\mathrm{d}y^2+\mathrm{d}z^2)$. We translate the spatial geometry into a space-time geometry with the space-time line element 
$\mathrm{d}s^2 = c^2 \mathrm{d}t^2 - \mathrm{d}l^2$, and hence the metric tensor 
$g_{\alpha \beta} = \mathrm{diag} \, (1, -n^2, -n^2, -n^2)$ with the determinant $-n^6$ and the inverse matrix $g^{\alpha \beta} = \mathrm{diag} \, (1, -n^{-2}, -n^{-2}, -n^{-2})$. The reader easily verifies from Eq.\ (\ref{eq:constrel}) that this geometry indeed describes a medium at rest with $\varepsilon = \mu = n$.

The local geometry of the medium is still purely spatial; it influences the spatial part of the metric, but not the measure of time. We can change this using the conformal invariance of Maxwell's equations (in the absence of external charges and currents). We may multiply the space-time metric by an arbitrary factor, called conformal factor, and still obtain exactly the same constitutive equations (\ref{eq:constitution}) as before. Multiplying $\mathrm{diag} \, (1, -n^2, -n^2, -n^2)$ by $n^{-2}$ we obtain the new metric tensor $g_{\alpha \beta} = \mathrm{diag} \, (n^{-2}, -1, -1, -1)$ with determinant $-n^{-2}$ and inverse matrix $g^{\alpha \beta} = \mathrm{diag} \, (n^2, -1, -1, -1)$.
This geometry influences the measure of time (in the metric element $g_{00}$) but not the spatial part of the metric. We write the metric tensor as
\begin{eqnarray}
g_{\alpha \beta} & = & \eta_{\alpha \beta} + (n^{-2} -1) \, u_\alpha \, u_\beta 
\,  ,   \quad
g^{\alpha \beta}   =   \eta^{\alpha \beta} + (n^2 -1) \, u^\alpha \, u^\beta 
\label{eq:gordon} \\
\eta_{\alpha \beta} & = & \mathrm{diag} \, (1, -1, -1, -1)   =   \eta^{\alpha \beta}
\label{eq:minkovski}
\end{eqnarray}
and $u^\alpha = (1, 0, 0, 0) = u_\alpha$.
Then we use a key insight from special relativity \cite{LL2}: $\eta_{\alpha \beta}$ describes the metric tensor (\ref{mink}) of flat space-time in an inertial frame (the Minkowski metric). Since all inertial frames are equivalent, $\eta_{\alpha \beta}$ does not change in transformations from one inertial frame to another, for example in the transformation from the laboratory frame to the locally co-moving frame and vice versa. Using this insight from relativity, we also develop a physical interpretation for $u^\alpha$ and $u_\alpha$ that allows us to translate the  $g^{\alpha \beta}$ and $g_{\alpha \beta}$ back to the laboratory frame: we regard $u^\alpha$ as the local four-velocity of the medium,
\begin{equation}
u^\alpha = \eta^{\alpha \beta} \, u_\beta = \frac{\mathrm{d}x^\alpha}{\mathrm{d}s}
  .
\label{eq:ufour}
\end{equation}
Here $\mathrm{d}s$ is the line element (\ref{mink}) in flat space-time,
\begin{equation}
\mathrm{d}s^2 = \eta_{\alpha \beta}   \mathrm{d}x^\alpha \, \mathrm{d}x^\beta = 
 c^2 \mathrm{d}t^2 - \mathrm{d}x^2 - \mathrm{d}y^2 - \mathrm{d}z^2 = 
 \left( 1 - \frac{u^2}{c^2} \right) \, \mathrm{d}t^2
\label{eq:ds}
\end{equation}
expressed in terms of the three-dimensional velocity vector of the medium
\begin{equation}
{\bm u} = \frac{\mathrm{d} {\bm x}}{\mathrm{d}t}   .
\label{eq:u}
\end{equation}
For zero velocity, {\it i.e.} in the locally co-moving frame, we get $u^\alpha = u_\alpha = (1, 0, 0, 0)$. Hence we can indeed interpret the $u^\alpha$ and $u_\alpha$ in the metric (\ref{eq:gordon}) as the local four-velocities. The important point is that 
$u^\alpha$ transforms like the four-dimensional coordinate vector $x^\alpha$, as the position of the index in $u^\alpha$ indicates, because $\mathrm{d}s$ is invariant.
Consequently, the expressions (\ref{eq:gordon}) are not only true in locally co-moving frames, but in the laboratory frame as well. All we have to do is to calculate $u^\alpha$ and $u_\alpha$ for non-zero velocities (the local velocities of the moving medium).
We obtain from Eqs.\ (\ref{eq:ufour}), (\ref{eq:ds}) and (\ref{eq:u}) the expressions
\begin{equation}
u^\alpha = \frac{(1, {\bm u}/c)}{\sqrt{1 - u^2/c^2}} \,, \quad u_\alpha = \frac{(1, - {\bm u}/c)}{\sqrt{1 - u^2/c^2}}   .
\end{equation}
In this way we have established the space-time geometry of moving media discovered in 1923 by Walter Gordon (and independently rediscovered several times \cite{Stor,PMQ1,PMQ2}). In matrix form, the metric tensors appear as 
\begin{eqnarray}
g^{\alpha \beta} & = & \left( 
\begin{array}{l@{\quad}l}
\displaystyle \frac{c^2 n^2 - u^2}{c^2 - u^2} & \displaystyle \frac{n^2 - 1}{c^2 -u^2} \, c \, {\bm u} \\
\displaystyle \frac{n^2 - 1}{c^2 -u^2} \, c \, {\bm u} & 
- \mathds{1} + \displaystyle \frac{n^2 - 1}{c^2 -u^2} \, {\bm u} \otimes {\bm u} 
\end{array}
\right)   , 
\label{eq:gordon2} \\
g & = &  \mathrm{det} \left( g_{\alpha \beta} \right)   =   - n^{-2}   , \\
g_{\alpha \beta} & = & \left( 
\begin{array}{l@{\quad}l}
\displaystyle \frac{c^2 n^{-2} - u^2}{c^2 - u^2} & \displaystyle \frac{1 - n^{-2}}{c^2 -u^2} \, c \, {\bm u} \\
\displaystyle \frac{1 - n^{-2}}{c^2 -u^2} \, c \, {\bm u} & 
- \mathds{1} + \displaystyle \frac{n^{-2} - 1}{c^2 -u^2} \, {\bm u} \otimes {\bm u} 
\end{array}
\right)   .
\end{eqnarray}
We obtain from Eq.\ (\ref{eq:const}) the dielectric tensors of moving media (that are locally isotropic and impedance matched),
\begin{equation}
\varepsilon = \mu = \frac{n}{1 - u^2 n^2 / c^2} \left( \left( 1 - \frac{u^2}{c^2} \right) \, \mathds{1}  +  \left( 1 - n^{-2} \right) \frac{{\bm u} \otimes {\bm u}}{c^2}
\right) \approx  n \, \mathds{1}
\end{equation}
in the limit of low velocity, $u/c \ll 1$.
Furthermore, since for moving media $g_{\alpha \beta}$ contains mixed space-time elements $g_{0i}$, we also obtain the vector
\begin{equation}
{\bm w} = \frac{n^2 - 1}{1 - u^2 n^2 / c^2} \, {\bm u} \approx
(n^2 -1) \, {\bm u}   .
\label{eq:low}
\end{equation}
In this way we have found a possible physical interpretation for the bi-anisotropy vector $\bm w$ of electromagnetism in space-time geometries: 
$\bm w$ may appear as the velocity of a moving medium.

\subsection{Optical Aharonov-Bohm effect}

Perfect invisibility devices and perfect lenses 
exploit transformations to non-trivial topologies in space -- excluded regions in physical space or folds in electromagnetic space.
Here we study a simple example, the propagation of light through a vortex, that turns out to represent a transformation medium with multi-valued space-time geometry. 
Consider a vortex in a fluid, say in water flowing down a plug hole. A vortex concentrates the vorticity $ \bm{\nabla} \times \bm u$ of the swirling flow $\bm u$ in one line, the vortex core. In Cartesian coordinates the velocity profile of a straight vortex is given by the expression
\begin{equation}
{\bm u} =  \frac{\cal W}{x^2+y^2} \, \left(
\begin{array}{c}
-y  \\ x \\ 0
\end{array}
\right)   ,
\label{eq:vortex}
\end{equation}
because for this profile $ \bm{\nabla} \times \bm u$ vanishes, except at the $z$ axis where $\bm u$ diverges, but the circulation $\oint {\bm u} \cdot  \mathrm{d} {\bm r}$ gives the finite value $2 \pi \cal W$. All vorticity is concentrated along the vortex line. 

It might be a useful exercise for the reader to express the velocity profile (\ref{eq:vortex}) of the vortex in cylindrical coordinates (the most natural coordinates for this situation). According to Eqs.\ (\ref{etrans}) and (\ref{Lambdas}) the coordinate-basis vectors of the cylindrical coordinates are
\begin{equation}
\bm{e}_r=\cos\phi\,\bm{i}+\sin\phi\,\bm{j}, \quad
\bm{e}_\theta=-r\sin\phi\,\bm{i}+r\cos\phi\,\bm{j}, \quad
\bm{e}_z=\bm{k}. \label{ec}
\end{equation}
The metric tensor is the familiar $\gamma_{ij}=\mathrm{diag}(1,r^2,1)$. One obtains for the velocity profile of the vortex
\begin{equation}
u^i =  \frac{\cal W}{r^2} \, \left(
\begin{array}{c}
0  \\ 1 \\ 0
\end{array}
\right) 
 ,\quad
u_i=\left(0,{\cal W},0\right)
.
\label{eq:vortexc}
\end{equation}
We immediately see from the expressions (\ref{curlcpts}) of the curl in arbitrary coordinates that $ \bm{\nabla}\times\bm{u}$ vanishes, because $u_i$ is constant. The vorticity is concentrated in the $z$ axis that is excluded from the cylindrical coordinates. 

Imagine the vortex is illuminated from the side. Any moving medium drags light, as Fresnel \cite{Fresnel} discovered in 1818 (deducing the correct result without knowing special relativity) and as Fizeau \cite{Fizeau} observed for water in uniform motion in 1851 (in interferometric measurements without lasers). The basics of Fresnel's drag are very simple: light propagating with the flow is advanced, whereas light propagating against the current lags behind. (The degree of dragging, Fresnel's dragging coefficient, depends on special relativity though.)
Dragged by the moving medium, light rays experience phase shifts and, if the phase fronts are deformed, light rays are deflected. For flow speeds $u$ much smaller than the speed of light in the medium, $c/n$, we expect that the phase shift is proportional to the integral of $\bm u$ along the propagation of light. For regions with vanishing vorticity, we can deform the contours of the phase integrals $\int {\bm u} \cdot  \mathrm{d} {\bm r}$ and so the phase difference between light waves that have passed the vortex on different sides is proportional to the circulation  $\oint {\bm u} \cdot  \mathrm{d} {\bm r}$, a constant. Hence we expect that the water vortex does not deflect light, but imprints a characteristic phase slip, in analogy to the Aharonov-Bohm effect \cite{AB,Peshkin,Tonomura}.
In the Aharonov-Bohm effect
charged matter waves, electrons for example, are not deflected by a vortex in the magnetic vector potential, but experience a characteristic phase shift. 
The optical analogue of the Aharonov-Bohm effect was first described by Jon Hannay in his PhD thesis \cite{Hannay} and has also been independently rediscovered \cite{CFM}. The effect could be used to detect quantum vortices with slow light \cite{Liten} and it is related to the Aharonov-Bohm effect with surface waves \cite{Berryetal,Roux,Vivanco} and to the the gravitational Aharonov-Bohm effect \cite{Stachel}.

Assuming that in the optical Aharonov-Bohm effect the phase modulation is proportional to the integral of $\int {\bm u} \cdot  \mathrm{d} {\bm r}$ we guess a space-time coordinate transformation that should describe the wave propagation. Then we use our formalism to verify that this guess is correct. The integral of the velocity profile (\ref{eq:vortex}) gives
\begin{equation}
\int {\bm u} \cdot  \mathrm{d} {\bm r} = {\cal W} \int \frac{-y \, \mathrm{d}x + x \, \mathrm{d}y}{x^2+y^2} = {\cal W} \int \mathrm{d} \left( \arctan \frac{y}{x} \right) = {\cal W} \phi  .
\end{equation}
This phase should modulate the time evolution of the wave. The water has a uniform refractive index $n$ that simply rescales the wavelength. Therefore we assume in physical space, using cylindrical coordinates, 
\begin{equation}
 c \, t = c \, t' - a \, \phi' \, , \quad  r = \frac{r'}{n} \, , \quad  \phi = \phi' \, , \quad  z = \frac{z'}{n} 
\label{eq:abtrans}
\end{equation}
with a constant $a$ to be determined later. In electromagnetic space we obtain 
\begin{equation}
 c \, t' = c \, t + a \, \phi \, ,  \quad r' = n \, r \, , \quad  \phi' = \phi \, , \quad  z' = n \, z   . 
\label{eq:ab2trans} 
\end{equation}
Electromagnetic space is flat and empty with the Minkowski line element (\ref{eq:ds}) in primed coordinates expressed in physical coordinates,
\begin{equation}
\mathrm{d} s^2 = ( c \, \mathrm{d}t + a \, \mathrm{d}\phi )^2 -
n^2 ( \mathrm{d}r^2 + r^2 \mathrm{d}\phi^2) - n^2 \, \mathrm{d}z^2   . 
\end{equation}
So the metric tensors are
\begin{eqnarray}
g_{\alpha\beta} & = &  \left(
\begin{array}{c@{\quad}c@{\quad}c@{\quad}c}
1 & 0 & a & 0 \\
0 & -n^2 & 0 & 0 \\
a & 0 & a^2-n^2r^2 & 0 \\
0 & 0 & 0 & -n^2
\end{array} 
\right)   ,\\
g & = & \mathrm{det} \, \left( g_{\alpha\beta} \right)  = - r^2 n^6   ,\\
g^{\alpha\beta} & = & \left( 
\begin{array}{c@{\quad}c@{\quad}c@{\quad}c}
1 - \displaystyle \frac{a^2}{n^2 r^2} & 0 & \displaystyle \frac{a}{n^2 r^2} & 0 \\
0 & - \displaystyle \frac{1}{n^2} & 0 & 0 \\
\displaystyle \frac{a}{n^2 r^2} & 0 & - \displaystyle \frac{1}{n^2r^2} & 0 \\
0 & 0 & 0 & - \displaystyle \frac{1}{n^2}
\end{array} 
\right)   .
\end{eqnarray}
We recall that in cylindrical coordinates the spatial metric is described by the matrix $\Gamma = \mathrm{diag} \, (1, r^2, 1)$ with determinant $\gamma=r^2$ and obtain from the constitutive equations (\ref{eq:const}) the dielectric tensors
\begin{equation}
\varepsilon = \mu = n \, \mathrm{diag} \, (1, r^{-2}, 1)
\end{equation}
and hence 
\begin{equation}
\varepsilon \, \Gamma = \mu \, \Gamma = n \, \mathds{1}    ;
\end{equation}
the medium is isotropic with refractive index $n$. We also get from the constitutive equations (\ref{eq:const}) the bi-anisotropy vector in cylindrical coordinates 
\begin{equation}
w_i = \left( 0 , a , 0 \right)
\end{equation}
that corresponds to the velocity profile (\ref{eq:vortexc}) in the low-velocity limit (\ref{eq:low}) for 
\begin{equation}
a = ( n^2 - 1 ) \, {\cal W}   .
\end{equation}
These results prove that the space-time transformation (\ref{eq:ab2trans}) 
generates from electromagnetic waves in empty space solutions of Maxwell's equations for the vortex (\ref{eq:vortex}) in the limit of low velocities.

\begin{figure}[t]
\begin{center}
\includegraphics[width=30.0pc]{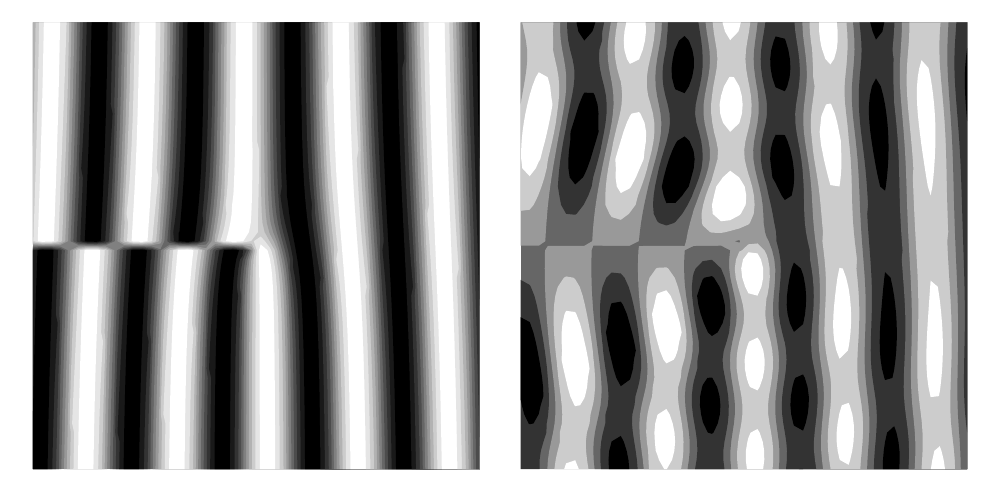}
\caption{
\small{
Aharonov-Bohm effect.
A fluid vortex generates the optical Aharonov-Bohm effect
described by the coordinate transformation (\ref{eq:abtrans}).
Light, incident from the right, 
is Fresnel-dragged by the moving medium:
light propagating with the flow is advanced,
whereas light propagating against the current is retarded.
The wave should develop the phase slip shown in the left figure.
However, although the transformation (\ref{eq:abtrans}) is exact,
physical space-time would be described in multi-valued 
coordinates here.
Instead of the simple phase slip,
the light turns out to exhibit the characteristic interference
pattern illustrated in the right figure
\cite{AB}.
}
\label{fig:ab}}
\end{center}
\end{figure}

However, although this solution describes the dominant features of the wave propagation through the vortex, it does not have the right topology in general; the solution is not periodic in the angle $\phi$.
For example, for a monochromatic wave of frequency $\omega$ the phase grows by $2 \pi \, a \, \omega$ after completing one circle, which is not an integer multiple of $2 \pi$ in general.
If we take the coordinate transformation (\ref{eq:abtrans}) literally, physical space has become multi-valued, with a branch cut in the direction of incidence. 
This branch cut describes the phase slip of light after passing through the vortex, but it cannot possibly be mathematically rigorous. Figure \ref{fig:ab} shows the correct solution due to Aharonov and Bohm \cite{AB} when additional scattering resolves the topological dilemma of the coordinate transformation (\ref{eq:abtrans}).
But this is a subject beyond the scope of this article, and so is light propagation in rapidly spinning media \cite{Stor} that describes how light may get sucked into the vortex. 

\subsection{Analogue of the event horizon}

\begin{figure}[t]
\begin{center}
\includegraphics[width=25.0pc]{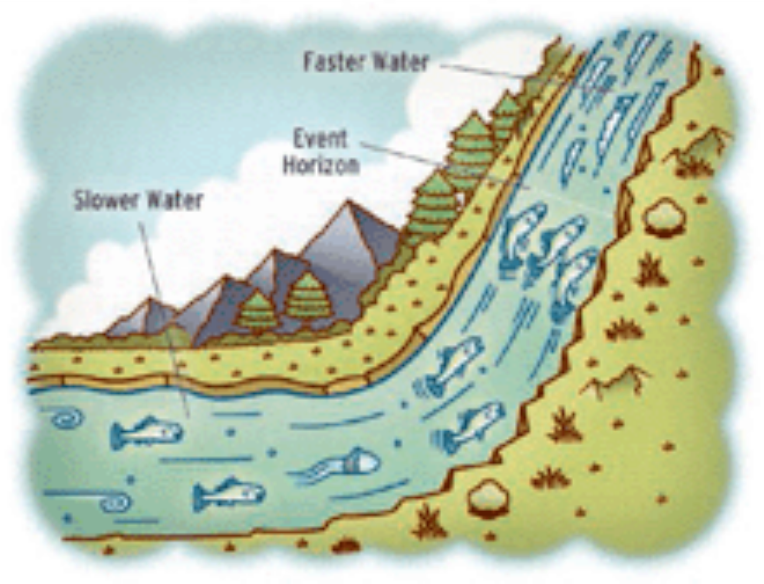}
\caption{
\small{
Aquatic analogue of the event horizon. 
(From Ref.\ \cite{Cho}. Reprinted with permission of Peter Hoey.)
Current can stop fish moving upstream and mimic an event horizon. A moving optical medium captures the physics at the event horizon, too. A pulse in an optical fibre creates such a moving medium \cite{Philbin}.
}
\label{fig:fish}}
\end{center}
\end{figure}

Transformation media are at the heart of invisibility devices, perfect lenses and the optical Aharonov-Bohm effect; here we explain that they also describe much of the essential physics at the event horizon of a black hole.
In 1972 William Unruh \cite{Unruh2008} invented a simple analogy of the event horizon as an illustration for a colloquium at Oxford: black holes resemble rivers. 
Imagine a river is flowing towards a waterfall. Suppose that the river carries waves that propagate with speed $c'$ relative to the water, but the water is rapidly moving with velocity $u$ that at some point exceeds $c'$. Beyond this point, waves are swept away towards the waterfall. The point of no return is the horizon of the black hole. 
Imagine another situation: a river flowing out into the sea, getting slower. Waves coming from the sea are blocked at the line where the flow exceeds the wave speed \cite{Suastika1,Suastika2}. Here the river establishes the analogue of the white hole, an object that nothing can enter. Such analogies occur also in many other situations and they illustrate some essential features of the event horizon, see Fig.\ \ref{fig:fish}. Horizons are perfect traps: one would not notice anything suspicious while passing the horizon with the flow, but to return is impossible. In the other direction, trying to escape against the current, waves get stuck at the horizon; they freeze with dramatically shrinking wavelengths. These analogues are not mere analogies, they are mathematically equivalent to wave propagation in general relativity (as long as higher-order dispersion is irrelevant \cite{BroutTrans,Jacobson,UnruhTrans}). 
For example, sound waves in moving fluids behave like waves in a certain space-time geometry that depends on the flow and the local velocity of sound \cite{Moncrief,Unruh,Visser}. Or, as we have already explained, light experiences a moving medium as the effective space-time geometry (\ref{eq:gordon}). At the horizon, where the flow $u$ reaches the speed of light $c/n$, the measure of time $g_{00}$ in Gordon's metric (\ref{eq:gordon2}) vanishes. Time comes to a standstill. Close to the horizon the wavelength is dramatically reduced (until due to optical dispersion the refractive index changes, tuning the light out of the grip of the horizon.) For short wavelengths the lateral dimensions of the horizon are irrelevant. Therefore, the optics at the horizon is essentially one-dimensional. 

Consider a one-dimensional model: both medium and light move along the $z$ axis, the electromagnetic field vectors are pointing orthogonal to $z$ in the $x,y$ plane. In the following we show that this situation corresponds to a space-time transformation medium. We write down a coordinate transformation that maps one-dimensional wave propagation in moving media to propagation in empty electromagnetic space. How to find this transformation?  In an empty one-dimensional space light would freely propagate with the speed $c$ as a superposition of wavepackets that move either in positive or negative direction while maintaining their shapes. So these two wavepackets depend only on   
\begin{equation}
t_\pm = t' \mp \frac{z'}{c}    .
\end{equation}
In physical space, the wavepackets do not propagate with the speed of light in vacuum, but with the velocity $v_\pm$ that is given by the relativistic addition \cite{LL2} of the flow speed $u$ and the speed of light $\pm c/n$ in positive or negative direction,
\begin{equation}
t_\pm = t - \int \frac{\mathrm{d} z}{v_\pm}  \, , \quad  v_\pm = \frac{u \pm \displaystyle \frac{c}{n}}{1 \pm \displaystyle \frac{u}{c \, n}}   .
\label{eq:pm}
\end{equation}
So we expect that the one-dimensional wave propagation in the moving medium corresponds to waves in empty space by the transformation
\begin{equation}
c \, t' = \frac{c}{2}\, (t_- + t_+) \, , \quad x' = x \,, \quad y' = y \,,  \quad
z' = \frac{c}{2}\, (t_- - t_+)   .
\label{eq:wtrans}
\end{equation}
To prove this assertion we apply the theory of transformation media. For ${\bm u} = (0, 0, u)$ the effective space-time geometry of the moving medium is described by the metric (\ref{eq:gordon2}) with 
\begin{equation}
g^{\alpha\beta} = \left( 
\begin{array}{c@{\quad}c@{\quad}c@{\quad}c}
\displaystyle \frac{c^2 n^2 - u^2}{c^2 - u^2} & 0 & 0 & \displaystyle \frac{(n^2-1)\, c \, u}{c^2 - u^2} \\
0 & - 1 & 0 & 0 \\
0 & 0 & - 1 & 0 \\
\displaystyle \frac{(n^2-1)\, c \, u}{c^2 - u^2} & 0 & 0 & \displaystyle \frac{n^2 u^2 - c^2}{c^2 - u^2}
\end{array} 
\right)   .
\end{equation}
We transform $g^{\alpha\beta}$ to electromagnetic space,
\begin{equation}
g^{\alpha'\beta'} = {\Lambda^{\alpha'}}_\alpha \, g^{\alpha\beta} \, {\Lambda^{\beta}}_{\beta'}
\end{equation}
with the matrix ${\Lambda^{\alpha'}}_\alpha$ that we obtain by differentiating the new coordinates $x^{\alpha'}$ with respect to the old coordinates $x^\alpha$ of physical space
\begin{equation}
{\Lambda^{\alpha'}}_\alpha = \left( 
\begin{array}{c@{\quad}c@{\quad}c@{\quad}c}
1 & 0 & 0 & \displaystyle \frac{(n^2-1)\, c \, u}{c^2 - n^2 u^2} \\
0 & 1 & 0 & 0 \\
0 & 0 & 1 & 0 \\
0 & 0 & 0 & \displaystyle \frac{n ( c^2 - u^2 )}{c^2 - n^2 u^2} 
\end{array} 
\right)   .
\end{equation}
The result is the diagonal matrix
\begin{equation}
g^{\alpha'\beta'} = \mathrm{diag} \, \left( \frac{n^2 ( c^2 - u^2 )}{c^2 - n^2 u^2} \,,  - \frac{n^2 ( c^2 - u^2 )}{c^2 - n^2 u^2} \,,  -1 \,,  -1  \right)
\end{equation}
with the inverse 
\begin{equation}
g_{\alpha'\beta'} = \mathrm{diag} \, \left( \frac{ c^2 - n^2 u^2}{
n^2 ( c^2 - u^2 )} \,,  - \frac{ c^2 - n^2 u^2}{
n^2 ( c^2 - u^2 )} \,,  -1 \,,  -1  \right)
\end{equation}
and the determinant of the metric 
\begin{equation}
g' = - \, \frac{(c^2 - n^2 u^2)^2}{n^4 ( c^2 - u^2 )^2 }    .
\end{equation}
The matrix $g_{\alpha'\beta'}$ describes the geometry in electromagnetic space. To find out how this geometry appears as a medium we use the constitutive equations (\ref{eq:constitution}) in electromagnetic space, with primed instead of unprimed metric tensors. Since $g_{\alpha'\beta'}$ is diagonal, the bi-anisotropy vector $\bm w'$ vanishes: in electromagnetic space the medium is at rest. We also obtain
\begin{equation}
\varepsilon'_x=\mu'_x=\varepsilon'_y=\mu'_y=1   .
\end{equation}
Since electromagnetic waves propagating in the $z$ direction are polarized in the $x,y$ plane their electromagnetic fields only experience the $x$ and $y$ components of the dielectric tensors. Consequently, for one-dimensional wave propagation electromagnetic space is empty, waves are free here, whereas in physical space they appear as modulated wavepackets according to the transformations (\ref{eq:pm}) and (\ref{eq:wtrans}).

We mapped one-dimensional electromagnetic waves in moving media onto waves in empty space, but what happens at the horizon? Why are horizons special? Suppose, without loss of generality, that the medium moves in the negative $z$ direction and develops a black-hole horizon at $z=0$: for positive $z$ the flow speed $|u|$ lies below $c/n$ and for negative $z$ the flow exceeds $c/n$. For the waves that propagate against the current in the positive $z$ direction the transformation (\ref{eq:pm}) develops a pole at the horizon. We linearize $u(z)$ here and obtain
\begin{equation}
t_+ = t - \frac{\ln |z|}{\alpha}  \, , \quad  \alpha = \frac{1}{1-n^{-2}} \, 
\left. \frac{\mathrm{d} u}{\mathrm{d} z} \right|_0     .
\label{eq:ln}
\end{equation}
For any given time $t$, the entire range of $t_+$ is filled for $z>0$ or $z<0$, for either side of the horizon. So electromagnetic space has become multi-valued and physical space is cut into two distinct regions, because waves confined to either one of the two sides would never interact with each other. But this is not entirely true \cite{Hawking}. Picture a wavepacket that, after having barely escaped from the horizon, propagates in a region where the medium moves at uniform speed. The wavepacket is purely forward-propagating; so it consists of a superpostion of plane waves with purely positive wavenumbers $k$ and corresponding positive frequencies $\omega$. On the other hand, recall the result from complex analysis \cite{Ablowitz} that Fourier transforms with positive spectra are analytic on the upper half plane (because integrals over $k$ containing the analytic function $\exp (\mathrm{i} k z)$ converge and are continuous for positive $k$ and positive $\mathrm{Im} \, z$). Consequently, if we analytically continue the wave packet on the complex $z$ plane it should be analytic for $\mathrm{Im} \, z > 0$, even if we trace its evolution back to the horizon. Analytic functions are continuous; so the escaping wavepacket must have partially originated from beyond the horizon --- some part of the wave must have tunneled through.
In order to calculate the tunnel amplitude we use the following trick \cite{DR}: we combine two waves $\exp (-\mathrm{i} \omega t_+)$ localized on either side of the horizon such that they seamlessly form an analytic function on the upper half $z$ plane. From 
\begin{equation}
\exp \left( - \pi \frac{\omega}{\alpha} + \mathrm{i} \ln (-z) \right) = 
\exp \left( \mathrm{i} \frac{\omega}{\alpha} \ln (z) \right)
\end{equation}
follows that we should give the waves on the left side of the horizon the prefactor $\exp (-\pi \omega/\alpha)$. Then the two waves combined make an analytic function. On the left side the medium moves faster than the speed of light in the medium. So here the phase of the wave moves backwards. In locally co-moving frames the wave would oscillate with negative frequencies. Such negative frequencies have recently been observed in water waves \cite{Rousseaux}. The negative-frequency wave makes a negative contribution to the energy;
so we ought to subtract the modulus square of its amplitude in the total intensity of the wave (a technically more involved argument \cite{LeoInstruments,PhilbinOnline} confirms this reasoning). Therefore the ratio $N$ of the tunneled to the total spectral intensity is
\begin{equation}
N = \frac{\exp \left( - 2 \pi \displaystyle \frac{\omega}{\alpha} \right)}{1 - \exp \left( - 2 \pi \displaystyle \frac{\omega}{\alpha} \right)} 
= \left[\exp \left( \frac{ 2 \pi \omega}{\alpha} \right) - 1 \right]^{-1}   .
\label{eq:tunnel}
\end{equation}
According to the quantum field theory at horizons \cite{BD,Brout,LeoInstruments,PhilbinOnline}, positive- and negative-frequency photon pairs are spontaneously created from the quantum vacuum, because they do not cost any energy. They are emitted at the tunneling rate (\ref{eq:tunnel}) that one can also interpret as the Planck spectrum \cite{MandelWolf}
\begin{equation}\label{eq:planck}
N = \left[\exp \left( \frac{ \hbar \omega}{k_B T} \right) - 1 \right]^{-1}
\end{equation}
with temperature ($k_B$ denotes Boltzmann's constant)
\begin{equation}\label{eq:hawking}
k_B T = \frac{\hbar \alpha}{2 \pi}    .
\end{equation}
Black holes are not black after all. The horizon emits light with a Planck spectrum (\ref{eq:planck}) and  temperature (\ref{eq:hawking}). The light consists of entangled photon pairs where each pair contains one photon with positive and one with negative frequency, one emitted outside and the other inside the horizon. Seen from one side of the horizon, the black hole is a black-body radiator. This flash of insight, due to Stephen Hawking in 1974 \cite{Hawking}, has been one of the most influential predictions of modern theoretical physics, because it illuminated a vastly unexplored intellectual landscape: Hawking's theory supports Bekenstein's idea \cite{Bekenstein} that the horizon carries entropy that is proportional to its area and it is a quantum phenomenon of a space-time geometry. Hawking's effect thus connects three vastly different areas of physics --- thermodynamics, quantum mechanics and general relativity. In modern attempts for finding a quantum theory of gravity, such as loop quantum gravity \cite{Rovelli} and superstring theory \cite{GSW}, the correct prediction of the Bekenstein-Hawking entropy has been used as a benchmark. However, the Hawking temperature of solar-mass black holes lies about eight orders of magnitude below the temperature of the cosmic microwave background; so, most probably, there is no chance of directly observing Hawking radiation in astrophysics. The benchmark of some of the most advanced theories of physics seems destined to remain theory.

On the other hand, 
laboratory analogues
of the event horizon 
may demonstrate the physics behind Hawking radiation
\cite{Novello,Philbin,SUbook,Volovik}.
For example, one could perhaps generate a detectable amount 
of Hawking radiation using few-cycle light pulses \cite{BK,Kartner}
in photonic-crystal fibres \cite{Russell}. 
According to nonlinear fibre optics \cite{Agrawal} the pulses behave, for all practical purposes, like one-dimensional moving media \cite{Idea,Philbin}: due to the optical Kerr effect they create additional contributions to the refractive index that move with the light pulses --- media that move at the speed of light.  Apart from optical dispersion \cite{BornWolf}, such moving media resemble the case considered in this article. How Hawking radiation emerges from elementary processes in gravity has remained largely a mystery, but laboratory analogues have a chance to, quite literally, shed light on one of the most fascinating pieces of physics, the creation of light at horizons. 


\section{Summary}

Optical media appear as geometries and geometries appear as media. 
Or, to be more precise, 
dielectric media appear to light 
as if they were changing the geometry of space
(or of space-time if the media are moving or are bi-anisotropic). 
The modified and sometimes distorted geometry perceived by light
is the cause of many optical illusions, including the ultimate illusion:
invisibility \cite{LeoConform,PSS}.
In order to deliberately create such an illusion one may design the 
desired geometry first and then construct a dielectric material 
for implementing it in practice. 
The simplest geometries are created by coordinate transformations
where light propagation in physical space appears to be curved,
but where, in a transformed space, light propagates along straight lines. 
Media that facilitate coordinate transformations are known as 
transformation media. 

Connections between geometry and optics have a long and distinguished
history: Fermat's principle has inspired the principle of least action in classical mechanics, the connection between geometrical optics and wave optics motivated quantum mechanics and path integrals in quantum field theory and statistical mechanics, and the principle of least action inspired the idea of inertial motion along geodesics in general relativity. But transformation optics is probably the first case where geometrical ideas that typically belong to the lofty sphere of general relativity have become practically useful in rather down-to-earth engineering applications.  Both sides benefit from this connection: engineers get motivated to learn and appreciate the tools of geometry, because applications are a great incentive to learn, and applied mathematicians or relativists  are confronted with and often delighted by the way in which these ideas appear in the real world, both in technical applications and also in some natural phenomena on Earth.  Connections between theory and experiment and between applied and fundamental science have been the greatest strength of science. The subject of this article, transformation optics, is an example of such connections. We hope to have written it in a form such that both practical-minded engineers and  theoretical-minded physicists and mathematicians find some treasure (and pleasure) in it.

This article is a primer, an introduction to the geometry of light and the concepts of transformation optics. We focused on the principal ideas and introduced them without assuming much prior knowledge, only standard mathematics, some basic physics and sufficient stamina of the reader. Being a primer, the article necessarily omits many aspects of this field.  
We analyzed only four examples of transformation media ---  
cloaking devices, perfect lenses, vortices and horizons ---
but these four cases both illustrate characteristic non-trivial topologies, 
each one with different physics,
and they have been experimentally verified, at least to some extent.
We did not discuss transformation media that re-scale space without 
changing the topology, see for example Refs.\ \cite{Kildishev,SPS2}, nor did we describe intriguing ideas that have not been implemented yet such as electromagnetic wormholes \cite{GreenleafWorm}. We omitted the analysis of active cloaking devices \cite{Miller} or plasmonic coverings \cite{Alu,MN}, because they are not directly related to transformation media (but perhaps the latter may appear as transformation media in disguise). We also did not discuss cloaking devices that combine coordinate transformations with refractive-index profiles that cannot be transformed away, first ideas of non-Euclidian cloaking \cite{LeoNotes,Ochiai}. Furthermore, we did not explain in detail how transformation media are implemented in practice, where the properties of metamaterials come from and how they are made, because this is a subject that fills entire books \cite{Marques,Milton,Sarychev}. 
We focused on the main ideas and some connections between optics, in particular transformation optics, and other areas of physics and mathematics.  It is quite remarkable how many different ideas optics combines and connects, but also here we were forced to make omissions. For example, we did not discuss applications of transformation media in acoustics, see for instance Refs.\ \cite{Chen,Cummer,Fang2,MBW}, and in quantum mechanics \cite{LeoConform}. We also focused primarily on the classical optics of transformation media and only mentioned their quantum optics \cite{LeoInstruments,LPCasimir,LPQuant,PhilbinOnline} without going into detail. Last, and least, we did not even attempt to represent the recent literature related to transformation optics, because this literature is too recent and rapidly growing --- had we included it this article would be outdated in a very short time. We hope to have compensated for these shortcomings and omissions by being clear and pedagogical in the main ideas and by focusing on the ``new things in old things'' and explaining ``old things in new things'', by telling the aspects of the story that we believe are already guaranteed to last and to remain inspiring for a long time. 

\section*{Acknowledgments}

We are privileged for having benefited from many inspiring conversations about ``geometry, light and a wee bit of magic''. 
In particular, we would like to thank
John Allen,
Sir Michael Berry,
Leda Boussiakou,
Luciana Davila-Romero,
Mark Dennis,
Malcolm Dunn,
Ildar Gabitov,
Greg Gbur,
Andrew Green,
Awatif Hendi,
Julian Henn,
Chris Hooley,
Sir Peter Knight,
Natalia Korolkova,
Irina Leonhardt,
Renaud Parentani, 
Harry Paul,
Paul Piwnicki,
Sir John Pendry,
Wolfgang Schleich,
David Smith,
Stig Stenholm,
Tom\'a\v{s} Tyc and
Grigori Volovik.
Our work was supported by the Leverhulme Trust,
the Engineering and Physical Sciences Research Council,
the Max Planck Society, 
and a Royal Society Wolfson Research Merit Award.

\newpage


\renewcommand{\theequation}{A\arabic{equation}}
\setcounter{equation}{0}

\section*{Appendix A}
The physical principle underlying general relativity is that all the laws of physics should be expressible as tensor equations in four-dimensional (possibly curved) space-time \cite{Telephone}. As a consequence of the transformation properties of tensors, this means that the laws of physics will have the same form in any space-time coordinate system. This property of tensor equations is called {\it general covariance}. In this appendix we show that the free-space Maxwell equations 
in generally covariant form are equivalent 
to Maxwell's equations in a material medium 
with constitutive equations 
(\ref{eq:constitution})-(\ref{eq:constrel}). 

The space-time form of Maxwell's equations is written in terms of the electromagnetic field tensor $F_{\mu\nu}$, constructed from the 
the $\bm{E}$ and $\bm{B}$ fields; in a right-handed coordinate system 
\begin{equation} \label{aF}
 F_{\mu\nu}=  \begin{pmatrix}
       0 &\  \ -E_1 & \ \ -E_2 & \ \ -E_3 \\
      E_1 & \ \ 0 & \ \ cB_3 & \ \ -cB_2 \\
      E_2 & \ \ -cB_3 & \ \ 0 & \ \ cB_1 \\
      E_3 & \ \ cB_2 & \ \ -cB_1 & \ \ 0  
   \end{pmatrix}.
\end{equation}
The generally covariant Maxwell equations are 
\cite{Telephone}
\begin{equation}
F_{[\mu\nu;\lambda]}=F_{[\mu\nu,\lambda]}=0,\quad
\varepsilon_0F^{\mu\nu}_{\ \ \ ;\nu}=
\frac{\varepsilon_0}{\sqrt{-g}}
\left(\sqrt{-g}F^{\mu\nu}\right)_{,\nu}=j^\mu, 
\label{amax} 
\end{equation}
where  
$j^\mu=(\rho,j^i/c)$ is the four-current, and
the square brackets denote antisymmetrization. This last operation produces a completely antisymmetric tensor with three indices. In Eq.\ (\ref{amax}) $F_{[\mu\nu,\lambda]}$ is explicitly
\begin{equation}
F_{[\mu\nu,\lambda]}=F_{\mu\nu,\lambda}+F_{\nu\lambda,\mu}+F_{\lambda\mu,\nu},
\end{equation}
since the electromagnetic field tensor already is antisymmetric: $F_{\mu\nu}=-F_{\nu\mu}$. The forms of (\ref{amax}) containing partial rather than covariant derivatives are obtained by using the antisymmetry of $F_{\mu\nu}$ and $F^{\mu\nu}$. In a left-handed coordinate system the relation between $F_{\mu\nu}$ and the magnetic field differs by a sign from that in Eq.\ (\ref{aF}). In this Appendix we will confine ourselves to right-handed systems; this issue of handedness is dealt with in \S 4.
We define a quantity $H^{\mu\nu}$ by
\begin{gather} 
H^{\mu\nu}=\varepsilon_0\sqrt{-g}F^{\mu\nu}=
\varepsilon_0\sqrt{-g}g^{\mu\lambda}g^{\nu\rho}F_{\lambda\rho}  
\label{acons1}  \\[8pt]
\Longrightarrow\quad F_{\mu\nu}=
\frac{1}{\varepsilon_0
\sqrt{-g}}g_{\mu\lambda}g_{\nu\rho}H^{\lambda\rho}  
\label{acons2}
\end{gather}
and regard $H^{\mu\nu}$ 
as being constructed 
from $\bm{D}$ and $\bm{H}$ fields as follows:
\begin{equation} \label{aH}
 H^{\mu\nu}=  \begin{pmatrix}
       0 &\  \ D^1 & \ \ D^2 & \ \ D^3 \\
      -D^1 & \ \ 0 & \ \ H^3/c & \ \ -H^2/c \\
      -D^2 & \ \ -H^3/c & \ \ 0 & \ \ H^1/c \\
      -D^3 & \ \ H^2/c & \ \ -H^1/c & \ \ 0  
   \end{pmatrix}.
\end{equation}
Then, introducing a new four-current $J^\mu=\sqrt{-g}j^\mu$, 
the free-space equations (\ref{amax}) can be written as
\begin{equation}
F_{[\mu\nu,\lambda]}=0, 
\qquad  H^{\mu\nu}_{\ \ \ ,\nu}=J^\mu,
\end{equation}
which are the macroscopic Maxwell equations (\ref{mac2}) in right-handed Cartesian coordinates. The constitutive equations are given by Eqs.\ (\ref{aF}) and (\ref{acons1})-(\ref{aH}).
To obtain relations between the vector fields 
$\bm{D}$, $\bm{H}$ and 
$\bm{E}$, $\bm{B}$ 
consider first the components $F_{0i}$; 
from Eqs.\ (\ref{aF}), (\ref{acons2}) and (\ref{aH}) 
one obtains
\begin{equation} \label{aecon}
E_i=\frac{1}{\varepsilon_0\sqrt{-g}}
\left(g_{i0}g_{j0}-g_{ij}g_{00}\right)D^j-
\frac{1}{\varepsilon_0c\sqrt{-g}}[jkl]g_{0j}g_{ik}H^l.
\end{equation}
We simplify this result as follows.
The identity
\begin{equation}
g_{\mu\lambda}g^{\lambda\nu}=\delta_{\mu}^{\nu}
\end{equation}
gives
\begin{gather}
g_{i\lambda}g^{\lambda 0}=0 
\quad \Longrightarrow \quad  
g_{i0}=-\frac{1}{g^{00}}\,g_{ij}g^{j0}, 
\label{agid1} \\[8pt]
g_{0\lambda}g^{\lambda i}=0 
\quad \Longrightarrow \quad  
g^{i0}=-\frac{1}{g_{00}}\,g^{ij}g_{j0}, 
\label{agid2} \\[8pt]
g_{j\lambda}g^{\lambda i}=
g_{j0}g^{0 i}+g_{jk}g^{ki}=\delta^i_j.  
\label{agid3}
\end{gather}
Use of Eqs.\ (\ref{agid1}) or (\ref{agid2}) 
in Eq.\ (\ref{agid3}) produces the two relations
\begin{equation}
\left(g^{ij}-\frac{1}{g^{00}}g^{i0}g^{k0}\right)g_{kj}=
\delta^i_j,\quad
g^{ik}\left(g_{kj}-\frac{1}{g_{00}}g_{k0}g_{j0}\right)=
\delta^i_j, \label{agid5}
\end{equation}
which reveal inverse-related $3\times 3$ matrices. 
In view of Eqs.\ (\ref{agid3}) and (\ref{agid5}),  
multiplying Eq.\ (\ref{aecon}) by $g^{li}$ and contracting on the index $i$ yields
\begin{equation} \label{aDcons}
D_i=-\frac{\varepsilon_0
\sqrt{-g}}{g_{00}}g^{ij}E^j+
\frac{1}{cg_{00}}[ijk]g_{j0}H^k,
\end{equation}
the first of the constitutive equations 
(\ref{eq:constitution}) with (\ref{eq:constrel}).

To obtain the second constitutive relation, 
we employ the tensors dual to $F_{\mu\nu}$ and $H^{\mu\nu}$. 
This requires use of the 4D Levi-Civita tensor, which in a right-handed system is
given by
\begin{equation}
\epsilon_{\mu\nu\lambda\rho}=\sqrt{-g}\,[\mu\nu\lambda\rho],\quad 
\epsilon^{\mu\nu\lambda\rho}=
-\frac{1}{\sqrt{-g}}\,[\mu\nu\lambda\rho],\quad 
[0123]=+1.
\end{equation}
The dual tensors $^*\!F^{\mu\nu}$ and $^*\!H_{\mu\nu}$ 
are defined by~\cite{Telephone}
\begin{gather}
^*\!F^{\mu\nu}=
\frac{1}{2}\epsilon^{\mu\nu\lambda\rho}F_{\lambda\rho} 
\qquad \Longrightarrow \qquad
F_{\mu\nu}=
\frac{1}{2}\epsilon_{\mu\nu\lambda\rho}\,^*\!F^{\lambda\rho}, 
\\[8pt]
^*\!H_{\mu\nu}=
\frac{1}{2}\epsilon_{\mu\nu\lambda\rho}H^{\lambda\rho} 
\qquad \Longrightarrow \qquad
H^{\mu\nu}=
\frac{1}{2}\epsilon_{\mu\nu\lambda\rho}\,^*\!H^{\lambda\rho},
\end{gather}
so they have components
\begin{gather} 
 ^*\!F^{\mu\nu}=  \frac{1}{\sqrt{-g}}\begin{pmatrix}
       0 &\  \ cB_1 & \ \ cB_2 & \ \ cB_3 \\
      -cB_1 & \ \ 0 & \ \ -E_3 & \ \ E_2 \\
      -cB_2 & \ \ E_3 & \ \ 0 & \ \ -E_1 \\
      -cB_3 & \ \ -E_2 & \ \ E_1 & \ \ 0  
   \end{pmatrix}, \label{adual1} \\[8pt]
 ^*\!H_{\mu\nu}=  \sqrt{-g}\begin{pmatrix}
       0 &\  \ -H^1/c & \ \ -H^2/c & \ \ -H^3/c \\
      H^1/c & \ \ 0 & \ \ -D^3 & \ \ D^2 \\
      H^2/c & \ \ D^3 & \ \ 0 & \ \ -D^1 \\
      H^3/c & \ \ -D^2 & \ \ D^1 & \ \ 0  
   \end{pmatrix}. \label{adual2}
\end{gather}
Re-expressed in terms of the dual tensors, 
the constitutive equations (\ref{acons1})-(\ref{acons2}) read
\begin{gather}
^*\!H_{\mu\nu}=
\varepsilon_0\sqrt{-g}g_{\mu\lambda}g_{\nu\rho}
\,^*\!F^{\lambda\rho},  \label{acons3}  \\[8pt]
^*\!F^{\mu\nu}=
\frac{1}{\varepsilon_0\sqrt{-g}}
g^{\mu\lambda}g^{\nu\rho}\,^*\!H_{\lambda\rho},
\label{acons4}
\end{gather}
where we applied the four-dimensional version of the double vector product (\ref{baccabgen})
\begin{equation}
\epsilon^{\mu\nu\lambda\rho}\epsilon_{\lambda\rho\sigma\tau}=-2(\delta^\mu_{\ \sigma}\delta^\nu_{\ \tau}-\delta^\mu_{\ \tau}\delta^\nu_{\ \sigma}).
\end{equation}
Writing out $^*\!H_{0i}$ using 
Eqs.\ (\ref{adual1})-(\ref{acons3}) 
one finds
\begin{equation}  \label{aHEB}
H^i=-\frac{\varepsilon_0c^2}{\sqrt{-g}}
\left(g_{00}g_{ij}-g_{i0}g_{j0}\right)B_j+
\frac{\varepsilon_0c}{\sqrt{-g}}[jkl]g_{j0}g_{ik}E_l \,.
\end{equation}
Comparison of this with Eqs.\ (\ref{aecon}) and (\ref{aDcons}) 
shows that
\begin{equation}
B_i=-\frac{\sqrt{-g}}{\epsilon_0c^2g_{00}}g^{ij}H^j-
\frac{1}{cg_{00}}[ijk]g_{j0}E_k,
\end{equation}
which is the second of 
the constitutive equations 
(\ref{eq:constitution}) with (\ref{eq:constrel}). 

In this way we derived Plebanski's constitutive equations (\ref{eq:constitution})-(\ref{eq:constrel}) \cite{Plebanski}. 
Note that several other relations between  
$\bm{D}$, $\bm{H}$, $\bm{E}$ and $\bm{B}$ 
are contained in Eqs.\ (\ref{acons1})-(\ref{acons2}) and 
(\ref{acons3})-(\ref{acons4}). 
For example, to express $\bm{D}$ and $\bm{H}$ 
in terms of $\bm{E}$ and $\bm{B}$ we need only take the time-space components 
of (\ref{acons1}), obtaining
\begin{equation}  \label{aDEB}
D^i=\varepsilon_0\sqrt{-g}\left(
g^{i0}g^{j0}-g^{ij}g^{00}\right)E_j-
\varepsilon_0c\sqrt{-g}[jkl]g^{k0}g^{ij}B_l,
\end{equation}
and the required formulae are 
Eqs.\ (\ref{aHEB}) and (\ref{aDEB}).


\end{document}